\documentclass[aps,prb,reprint,nofootinbib,superscriptaddress,twocolumn]{revtex4-2}
\usepackage[T1]{fontenc}
\usepackage[utf8]{inputenc}
\usepackage{amsmath}
\usepackage{amsfonts}
\usepackage{amssymb}
\usepackage{newtxtext,newtxmath}
\usepackage{graphicx}
\usepackage{xcolor}
\usepackage{hyperref}
\hypersetup{colorlinks=true,linkcolor=blue,urlcolor=blue,citecolor=blue}
\usepackage{sublabel}

\usepackage{soul}

\begin{document}

\title{Quantum geometric tensor and quantum phase transitions in the Lipkin-Meshkov-Glick model}

\author{Daniel Guti\'errez-Ruiz}
\email{daniel.gutierrez@correo.nucleares.unam.mx}
\affiliation{Instituto de Ciencias Nucleares, Universidad Nacional Aut\'onoma de M\'exico, Apartado Postal 70-543, Ciudad de M\'exico, 04510, M\'exico}

\author{Diego Gonzalez}
\email{dgonzalez@fis.cinvestav.mx}
\affiliation{Instituto de Ciencias Nucleares, Universidad Nacional Aut\'onoma de M\'exico, Apartado Postal 70-543, Ciudad de M\'exico, 04510, M\'exico}
\affiliation{Departamento de F\'isica, Cinvestav, Avenida Instituto Polit\'ecnico Nacional 2508, San Pedro Zacatenco, 07360, Gustavo A. Madero, Ciudad de M\'exico, M\'exico}

\author{Jorge Ch\'avez-Carlos}
\email{jchavez@icf.unam.mx}
\affiliation{Instituto de Ciencias F\'isicas, Universidad Nacional Aut\'onoma de M\'exico, Cuernavaca, Morelos, 62210, M\'exico}

\author{Jorge G. Hirsch}
\email{hirsch@nucleares.unam.mx}

\author{J. David Vergara}
\email{vergara@nucleares.unam.mx}

\affiliation{Instituto de Ciencias Nucleares, Universidad Nacional Aut\'onoma de M\'exico, Apartado Postal 70-543, Ciudad de M\'exico, 04510, M\'exico}

\date{\today}

\begin{abstract}
	
We study the quantum metric tensor and its scalar curvature for a particular version of the Lipkin-Meshkov-Glick model. We build the classical Hamiltonian using Bloch coherent states and find its stationary points. They exhibit the presence of a ground state quantum phase transition, where a bifurcation occurs, showing a change of stability associated with an excited state quantum phase transition. Symmetrically, for a sign change in one Hamiltonian parameter, the same phenomenon is observed in the highest energy state. Employing  the Holstein-Primakoff approximation, we derive analytic expressions for the quantum metric tensor and compute the scalar and Berry curvatures. We contrast the analytic results with their finite-size counterparts obtained through exact numerical diagonalization and find an excellent agreement between them for large sizes of the system in a wide region of the parameter space, except in points near the phase transition where the Holstein-Primakoff approximation ceases to be valid.

\end{abstract} 

\maketitle

\section{Introduction}

The geometry of quantum phase transitions is a vast and exciting subject (see Ref.~\cite{Carollo2020} for a comprehensive and recent review). A fundamental structure that characterizes a quantum system's geometry is the {\em quantum geometric tensor} (QGT). This quantity, introduced by Provost and Valle~\cite{Provost}, is useful to measure the distance in the parameter space between two quantum states with infinitesimally shifted parameters. Experimental protocols to measure its components have been proposed in~\cite{Ozawa2018} and~\cite{Bleu2018} and were carried out in~\cite{Ozawa2020} by exploiting the relation between the QGT and the response of quantum systems under periodic modulations.  In the context of topological phase transitions and topological invariants, the quantum geometric tensor plays an important role and has been measured~\cite{TanPRL,TanPRLErratum}. The critical scaling behavior of its components has been analyzed in~\cite{ZanardiPRL20072,SarkarCritical}. 

From its definition, it has been recognized that the QGT has a divergent behavior at points in parameter space that induce level-crossings, and therefore, that is useful to detect quantum phase transitions (QPTs) ~\cite{ZanardiPRL2007}. QPTs occur at zero temperature and are characterized by a change in the ground state's analytic properties induced by quantum fluctuations. From a theoretical standpoint, the QGT can be obtained through a second-order expansion of the distance induced by the {\em fidelity}, a well-known tool in information theory.  This quantity has an abrupt change in a quantum phase transition, and it has been used extensively (see Refs.~\cite{ZanardiPRE2006,Gonzalez2020Wigner,Gu} and references therein). Moreover, alternative definitions of the fidelity and machine learning approaches have recently been used to study phase transitions on spin models with Hamiltonians depending on parameters \cite{Defenu_2020}. Mathematically, the fidelity emerges from the Fubini-Study distance, which is defined in the projective Hilbert space and is a measure of statistical indistinguishability between pure or mixed quantum states~\cite{Wooters,tomczak2021probing}.

In the general formulation of Ref.~\cite{Provost}, the QGT includes in its real part the quantum metric tensor (QMT), and in its imaginary part the Berry curvature. The study of the geometric structures that emerge from the QMT has been carried out in various systems. Geodesics were studied in~\cite{SarkarGeodesics,SarkarCritical,Polkovnikov}, and an analysis of singularities and their relation with topology for the ground state was made in~\cite{KolodrubetzPRB,KolodrubetzRep}. 

A specific geometric structure that stands out is the {\em scalar curvature (or Ricci scalar)}, a local invariant that quantifies the deviation of the space of parameters from being  Euclidean. Consequently, the value of the scalar curvature at a point is independent of the parameter space coordinates that were employed to calculate the metric and hence can be used to identify whether or not a singularity in the metric is genuine, or coordinate dependent.  In quantum mechanics, one can naturally compute the scalar curvature of the QMT and observe its behavior as a function of the parameters, particularly at a region in parameter space where a QPT occurs. In the thermodynamic limit of a transverse $XY$ spin chain it was found that the scalar curvature is an indicator of the QPT~\cite{ZanardiPRL2007}, while in the Dicke model of quantum optics  the scalar curvature does not show any particular behavior, but the QMT diverges at the QPT ~\cite{SarkarPRE2012}.
This contrasts strongly with classical (thermodynamic) phase transitions, where a Riemannian metric can also be defined~\cite{Weinhold,Ruppeiner}, and its corresponding scalar curvature shows a singularity in the thermodynamic space across the phase transition~\cite{Ruppeiner1,Ruppeiner2}. It is also at variance with the behavior of the fidelity susceptibility, which is a component of the QMT and exhibits a peak at the QPT in the Dicke model for a finite number of particles \cite{Castanos2012a,Bastarrachea2014}  and in the vicinity the Berezinskii-Kosterlitz-Thouless QPT in one-dimensional systems for finite sizes \cite{Sun2015} and in the thermodynamic limit  \cite{Cincio2019,tomczak2021probing}.

In what follows, we consider the Lipkin-Meshkov-Glick (LMG) model~\cite{Lipkin1,Lipkin2,Lipkin3} and analyze in detail the behavior of the QMT and its scalar curvature, in the thermodynamic limit and also for finite sizes of the system. Some of the fidelity susceptibility components at the quantum stationary points have been studied varying one of the Hamiltonian parameters \cite{Kwok2008,Gu,Quan2009,Castanos2012}.  In ~\cite{SarkarPRE2012,Scherer2009} it was found that, for the standard version of the LMG model, the fidelity metric is ill-defined at zero temperature.  We extend these studies in the present work by including in the LMG Hamiltonian two terms linear in the angular momentum components. This extended model has a well defined QMT, whose characteristics and relationship with the QPT are studied in detail employing both analytical and numerical methods.  In particular, we perform a meticulous analysis of the scalar curvature, and we show some of the exciting properties that this geometric invariant has in the QPT.

The paper is organized as follows.  In Sec. II, we introduce the mathematical tools employed in the study of the geometry of the Hamiltonian parameter space. In Sec. III, we present the LMG model, the Hamiltonian, its semi-classical limit, stationary points, and quantum phases employing $SU(2)$ coherent states. The geometry of the parameter space of the LMG model is studied in Sec. IV for the ground state and the highest energy state, including the quantum metric tensor and the scalar curvature. Analytic expressions derived employing the truncated Holstein-Primakoff expansion are compared with numerical calculations for various system dimensions. Conclusions are discussed in Sec. V.  Appendix A shows the difficulties found in evaluating the scalar curvature employing coherent states.

\section{Geometry of the parameter space}

Consider a quantum system whose Hamiltonian $\hat{H}(x)$ depends smoothly on a set of $m$ real adiabatic parameters denoted by $x=\{x^{i}\}$ $(i=1,...,m)$, which parameterize some $m$-dimensional parameter manifold  $\mathcal{M}$. For an orthonormal eigenvector $|n(x)\rangle$ of the system with nondegenerate eigenvalue $E_n(x)$,  the quantum geometric tensor (QGT) is defined by~\cite{Provost}
 \begin{equation}
 Q_{ij}^{(n)}:=\langle\partial_{i}n|\partial_{j}n\rangle-\langle\partial_{i}n|n\rangle\langle n|\partial_{j}n\rangle,\label{QGTProvost}
 \end{equation}
where $\partial_{i}:=\frac{\partial}{\partial x^{i}}$. The QGT transforms as a covariant tensor of rank two and  is invariant under the $U(1)$ gauge transformation $|n(x)\rangle\rightarrow e^{i\alpha_{n}(x)}|n(x)\rangle$ with $\alpha_{n}(x)$ a smooth function of the parameters. The (symmetric) real part of the QGT yields the QMT~\cite{Provost}
\begin{equation}\label{QMT}
g^{(n)}_{ij} = {\rm Re} \, Q^{(n)}_{ij},
\end{equation}
which is a Riemannian metric and provides the distance  $\delta\ell^{2}=g_{ij}^{(n)}(x)\delta x^{i}\delta x^{j}$ between the quantum states $|n(x)\rangle$ and $|n(x+\delta x)\rangle$, corresponding to infinitesimally different parameters. The (antisymmetric) imaginary part of the QGT encodes the Berry curvature~\cite{Berry1985}
\begin{equation}\label{BerryC}
F^{(n)}_{ij}=-2 \, {\rm Im} \, Q^{(n)}_{ij},
\end{equation}
which, after being integrated over a surface subtended by a closed path in the parameter space, gives rise to the geometric Berry phase~\cite{Berry1985}.

An important aspect of the QGT is that its singularities are associated with QPTs~\cite{ZanardiPRL2007,ZanardiPRL20072}. To gain a better understanding of this aspect, let us write the QGT in a perturbative form.  Inserting the identity operator $\mathbb{I}=\sum_{m}|m\rangle\langle m|$ in the first term of Eq.~(\ref{QGTProvost}) and using 
\begin{equation}
\langle m|\partial_{i}n\rangle = \frac{\langle m|\partial_{i}\hat{H}|n\rangle}{E_{n}-E_{m}} \qquad {\rm for} \qquad m\neq n,
\end{equation}
which follows from the eigenvalue equation $\hat{H}|n\rangle=E_n|n\rangle$, the QGT takes the form~\cite{Gu}
\begin{equation}
  Q_{ij}^{(n)}=\sum_{m\neq n}\frac{\langle n|\partial_{i}\hat{H}|m\rangle\langle m|\partial_{j}\hat{H}|n\rangle}{(E_{m}-E_{n})^{2}}.\label{QGT}
\end{equation}
This expression shows that at the stationary points of the QPT, which are characterized by the ground-state level crossing, the components of the QGT (and hence also the components of the QMT and the Berry curvature) are singular. In general, it is clear from Eq~(\ref{QGT}) that the components of the QGT are singular at the points where the parameters take a value $x^{*}\in \mathcal{M}$ such that $E_{n}(x^{*})=E_{m}(x^{*})$.

To determine whether or not the QMT has a genuine singularity, i.e., a singularity that cannot be removed by changing coordinates, we can use the scalar curvature, which is independent of the coordinates that were chosen to parameterize the manifold $\mathcal{M}$. For a two-dimensional manifold $\mathcal{M}$ endowed with the Riemannian QMT, which is the case of interest in this paper, the scalar curvature $R$ can be computed using the expression:
\begin{equation}
R=\frac{1}{\sqrt{|g|}}({\cal A}+{\cal B}),\label{scalar}
\end{equation}
where  $g=\det[g_{ij}]$ and we have defined the quantities ${\cal A}$ and ${\cal B}$ by:
\begin{align}
{\cal A} & :=\partial_{1}\left(\frac{g_{12}}{g_{11}\sqrt{|g|}}\partial_{2}g_{11}-\frac{1}{\sqrt{|g|}}\partial_{1}g_{22}\right),\nonumber \\
{\cal B} & :=\partial_{2}\left(\frac{2}{\sqrt{|g|}}\partial_{1}g_{12}-\frac{1}{\sqrt{|g|}}\partial_{2}g_{11}-\frac{g_{12}}{g_{11}\sqrt{|g|}}\partial_{1}g_{11}\right).
\end{align} 

In what follows, we analyze the Riemannian curvature of the LMG model, which in this case is contained in the scalar curvature, and how it changes across a quantum phase transition and the information it offers.  

\section{Lipkin Meshkov Glick (LMG) model}

The LMG model ~\cite{Lipkin1,Lipkin2,Lipkin3} was originally employed as a tool to describe the shape phase transition in nuclei and, at the same time, to explore the limits of different approximate methods \cite{Dreiss1969,Schuck1973,Catara1994,Wahlen2017}, taking advantage of its simplicity. It provides a description of the collective modes of systems with two degrees of freedom, like a single particle evolving in a double-well potential or an interacting two-level boson system, and includes one- and two-body interactions, which can be mapped to a quantum top in a constant magnetic field \cite{Turbiner1988,Ulyanov1992}. It has been useful in the description of a wide variety of many body systems: spin systems with infinite-range interactions \cite{Botet1983},  quantum superpositions of Bose-Einstein condensates \cite{Cirac1998}, and  single molecular magnets \cite{Gatteschi2006,Bogani2008}. In recent years this fully connected model provided the testing ground to explore novel properties like time-translation symmetry breaking and time crystals \cite{Russomanno2017}, and the dynamics around the ground state and the excited quantum phase transitions, including quantum driving and control \cite{Campbell2015},  dynamical scaling behavior across adiabatic quantum phase transitions  \cite{Acevedo2014,Defenu2018}, quench dynamics \cite{Puebla2020} and nonadiabatic dynamics \cite{Kopylov2017}, which were experimentally studied employing trapped atoms \cite{Makhalov2019}. 

The scaling properties of the entanglement at quantum phase transition in the LMG model were studied in detail by Vidal et al.~\cite{Dusuel2004,Vidal2004,Latorre2005,Barthel2006}. The LMG model is part of an extended family of exactly solvable Hamiltonians, which can be built using a complete set of mutually commuting quantum invariants of a generalized Gaudin Lie algebra  \cite{Ortiz2005}. Employing spin coherent states, detailed analysis of the quantum phases  \cite{Gilmore1978,Kuriyama2003} and the order of the respective phase transitions have been performed \cite{Castanos2005}. Having a finite Hilbert space, the complete energy spectrum of the LMG model can be obtained through a numerical diagonalization for a large number of individual spins \cite{Castanos2006}. While the ground state has two quantum phases,  there is a divergence of the density of states at the threshold energy \cite{Leyvraz2005,Heiss2005} and the complete spectrum has four qualitatively different regions in the parameter space \cite{Ribeiro2007}.  This divergence is a paradigmatic example of an excited state quantum phase transition (ESQPT) \cite{Cejnar2006,Caprio2008,Cejnar2010,Perez2011,Brandes2013,Santos2015,Santos2016}. At this energy, the LMG has a positive Lyapunov exponent, which is not a signature of chaos but an instability, where the quantum out-of-time-order-correlator (OTOC) also exhibits an exponential growth \cite{Pilatowsky2020}.

The quantum criticality of the LMG model has been studied, varying one of the Hamiltonian parameters employing the fidelity susceptibility \cite{Kwok2008,Gu,Quan2009,Castanos2012}. The relationship between the information geometry and quantum phase transitions was explored in the Dicke model, showing that the scalar curvature is continuous across the phase transition boundary ~\cite{SarkarPRE2012}. In the same work, it was found that for the standard version of the LMG model, the metric's determinant is zero, signaling that information geometry is ill defined in this model. In Ref. \cite{Scherer2009} the authors concluded that due to the integrability of the isotropic LMG model, ground-state level crossings occur, leading to an ill-defined fidelity metric at zero temperature. To avoid these difficulties, we included in Hamiltonian (\ref{Eq:H-LMG}) two terms linear in the angular momentum components. As it is shown below, this provides a richer model, with interesting properties worth to be explored.


The LMG model~\cite{Lipkin1,Lipkin2,Lipkin3} describes the collective motion of a set of $N$ two-level systems mutually interacting. In this paper, we shall consider the following Hamiltonian
\begin{equation}
	\hat{H}_{\text{LMG}}= \Omega \hat{J}_z + \Omega_x \hat{J}_x  + \frac{\xi_y}{j} \hat{J}_y^2,
\label{Eq:H-LMG}
\end{equation}
where $\hat{J}_{x,y,z}=(1/2)\sum_{n=1}^N \sigma_{x,y,z}^{(n)} $ are the collective pseudo-spin operators given by the sum of the Pauli matrices  $\sigma_{x,y,z}^{(n)} $  for each two-level system $n$, and $j=N/2$, with $j$ coming from the eigenvalue $j(j+1)$ of the total spin operator $\hat{J}^2 =  \hat{J}_x^2 + \hat{J}_y^2 + \hat{J}_z^2$ and thus gives the size of the system. Also, $\Omega$ is the energy difference of the two-level systems, $\Omega_x$ and $\xi_y$ are parameters, and we put $\hbar=1$.


\subsection{Semiclassical analysis}

The classical LMG Hamiltonian is obtained by taking the expectation value of $\hat{H}_{\text{LMG}} /j $ on Bloch coherent states $ |z\rangle=(1+\left|z\right|^{2})^{-j} e^{z \hat{J}_+}|j, -j\rangle$, where $|j, -j\rangle$ is the state with the lowest pseudo-spin projection, 
and $\hat{J}_+$ is the raising operator. 

Defining $z=\tan(\frac{\theta}{2}) e^{-i\phi}$, the classical LMG Hamiltonian has the simple form
\begin{align}
	h_{\text{LMG}}&= \frac{\langle z | \hat{H}_{\text{LMG}} | z \rangle}{j} \nonumber \\
	&= -\Omega \cos\theta + \Omega_x  \sin\theta \cos\phi + \xi_y\sin^2\theta \sin^2\phi.
\label{Eq:H-LMG1}
\end{align}
It can also be written in terms of the canonical variables 
\begin{eqnarray}
	Q = \sqrt{2(1-\cos\theta)} \cos\phi, \nonumber \\
	P = - \sqrt{2(1-\cos\theta)} \sin\phi.
\label{Eq:QP}
\end{eqnarray}
and reads
\begin{align}
 h_{\text{LMG}}(Q,P) = & \frac{\Omega}{2} \left(Q^2 \!+\! P^2\right)-\Omega + \Omega_x Q\sqrt{1 \!-\! \frac{Q^2+P^2}{4}} \nonumber \\
   & + \xi_y P^2 \left(1 \!-\! \frac{Q^2+P^2}{4}\right). 
   \label{Eq:LMGclassical}
\end{align}
Notice that Hamiltonian (\ref{Eq:H-LMG1}) is invariant under the interchange $\phi \leftrightarrow -\phi$, which reflects in Hamiltonian (\ref{Eq:LMGclassical}) in the symmetry under the interchange $P \leftrightarrow -P$. It implies that all stationary points are doubly degenerate if   $\phi \neq 0 \text{~or} \ \pi \,\,(P\neq 0)$.

The allowed domain for the parameters is $\Omega, \Omega_x, \xi_y \in\mathbb{R}$.
To simplify the analysis, we consider only two independent parameters of the system, setting from now on $\Omega=1$ (i.e. we measure the energies in units of $\Omega$). 

\subsubsection{Stationary points}

For the classical Hamiltonian  (\ref{Eq:LMGclassical})  we obtain the equations of motion 
\begin{align}
\dot{Q}=\frac{\partial h_{\text{LMG}}}{\partial P}=\frac{P}{2}\bigg[&2-\xi_{y}(2P^{2}+Q^{2}-4)-\frac{\Omega_{x}Q}{\sqrt{4-(P^{2}+Q^{2})}}\bigg], \nonumber \\
\dot{P}=-\frac{\partial h_{\text{LMG}}}{\partial Q} = \frac{1}{2}\bigg[& \xi_{y}P^{2}Q+\frac{\Omega_{x}Q^{2}}{\sqrt{4-(P^{2}+Q^{2})}} \nonumber \\
& -\Omega_{x}\sqrt{4-(P^{2}+Q^{2})}-2 Q\bigg].
\end{align}

The stationary points correspond to vanishing phase space velocities $(\dot{Q},\dot{P})$ in the region $0\leq \theta \leq \pi$,   $0\leq \phi \leq 2\pi$.  The valid solutions are 

\begin{eqnarray}
\textbf{x}_{1} =(Q_1,P_1) & = \left(\sqrt{2+\frac{2}{\sqrt{\Omega_x^2+1}}},0\right)  \text{ with } \label{x1}\\
(\theta_1,\phi_1) & = \left(\arccos\left(\frac{1}{\sqrt{\Omega_x^2+1}}\right), 0\right) \nonumber \\
\textbf{x}_{2} = (Q_2,P_2) &=  \left(-\sqrt{2-\frac{2}{\sqrt{\Omega_x^2+1}}},0 \right) \text{ with }  \label{x2}\\
(\theta_2,\phi_2) & =\left(\arccos\left(\frac{1}{\sqrt{\Omega_x^2+1}}\right), \pi\right) \nonumber
\end{eqnarray}  
 if  $\xi_y \leq -\frac{\sqrt{1+\Omega_{xc}^2}}{2}$,\\
\begin{eqnarray}
 \textbf{x}_{3} = (Q_3,P_3) & =\left(-\frac{\Omega_x}{\sqrt{\xi_y (2\xi_y-1)}},-\frac{\sqrt{-\Omega_x^2+4 \xi_y^2-1}}{\sqrt{\xi_y (2 \xi_y-1)}}\right) \text{ with }\\
 (\theta_3,\phi_3 & )=\left(\arccos\left(-\frac{1}{2 \xi_y}\right),-\arccos\left(\frac{\Omega_x}{\sqrt{4 \xi_y^2-1}}\right)\right) \text{and} \nonumber \\
 \textbf{x'}_{3} = (Q_3,P_3) & =\left(-\frac{\Omega_x}{\sqrt{\xi_y (2\xi_y-1)}},
\frac{\sqrt{-\Omega_x^2+4 \xi_y^2-1}}{\sqrt{\xi_y (2 \xi_y-1)}}\right) \text{ with } \\
(\theta'_3,\phi'_3) & =\left(\arccos\left(-\frac{1}{2 \xi_y}\right), \arccos\left(\frac{\Omega_x}{\sqrt{4 \xi_y^2-1}}\right)\right) \nonumber
\end{eqnarray}
 if  $\xi_y \geq \frac{\sqrt{1+\Omega_{xc}^2}}{2}$,\\
\begin{eqnarray}
 \textbf{x}_{4} = (Q_4,P_4) & =\left(\frac{\Omega_x}{\sqrt{\xi_y (2\xi_y-1)}},
-\frac{\sqrt{-\Omega_x^2+4 \xi_y^2-1}}{\sqrt{\xi_y (2 \xi_y-1)}}\right) \text{ with}\\
(\theta_4,\phi_4) & =\left(\arccos\left(\frac{1}{2 \xi_y}\right), -\arccos\left(\frac{\Omega_x}{\sqrt{4 \xi_y^2-1}}\right)\right) \text{ and } \nonumber \\
 \textbf{x'}_{4} = (Q'_4,P'_4) & =\left(\frac{\Omega_x}{\sqrt{\xi_y (2\xi_y-1)}},
\frac{\sqrt{-\Omega_x^2+4 \xi_y^2-1}}{\sqrt{\xi_y (2 \xi_y-1)}}\right) \text{ with }\\
(\theta'_4,\phi'_4) & =\left(\arccos\left(\frac{1}{2 \xi_y}\right), \arccos\left(\frac{\Omega_x}{\sqrt{4 \xi_y^2-1}}\right)\right) \nonumber
\end{eqnarray}

\subsubsection{Energy at the stationary points}

The energies $e_i = h_{\text{LMG}}( \textbf{x}_{i})$ associated to each stationary point are listed in Table~\ref{Tablenergies}.

\begin{table}[h]
\caption{\label{Tablenergies} Classical energies at each stationary point.}
\begin{ruledtabular}
\begin{tabular}{cc}
  Point             & Energy          \\
\hline 
 $ \textbf{x}_{1}  $  	& $\sqrt{\Omega_x^2+1}$            \\
$ \textbf{x}_{2} 	$	&$-\sqrt{\Omega_x^2+1}$       \\
 $ \textbf{x}_{3}  $  	& $(1 + \Omega_x^2)/(4 \xi_y) + \xi_y$            \\
 $ \textbf{x}_{4}  $ 	& $(1 + \Omega_x^2)/(4 \xi_y) + \xi_y$  \\
\end{tabular}
\end{ruledtabular}
\end{table}

\begin{figure}[ht]
\begin{tabular}{c c}
\includegraphics[width= 0.5 \columnwidth]{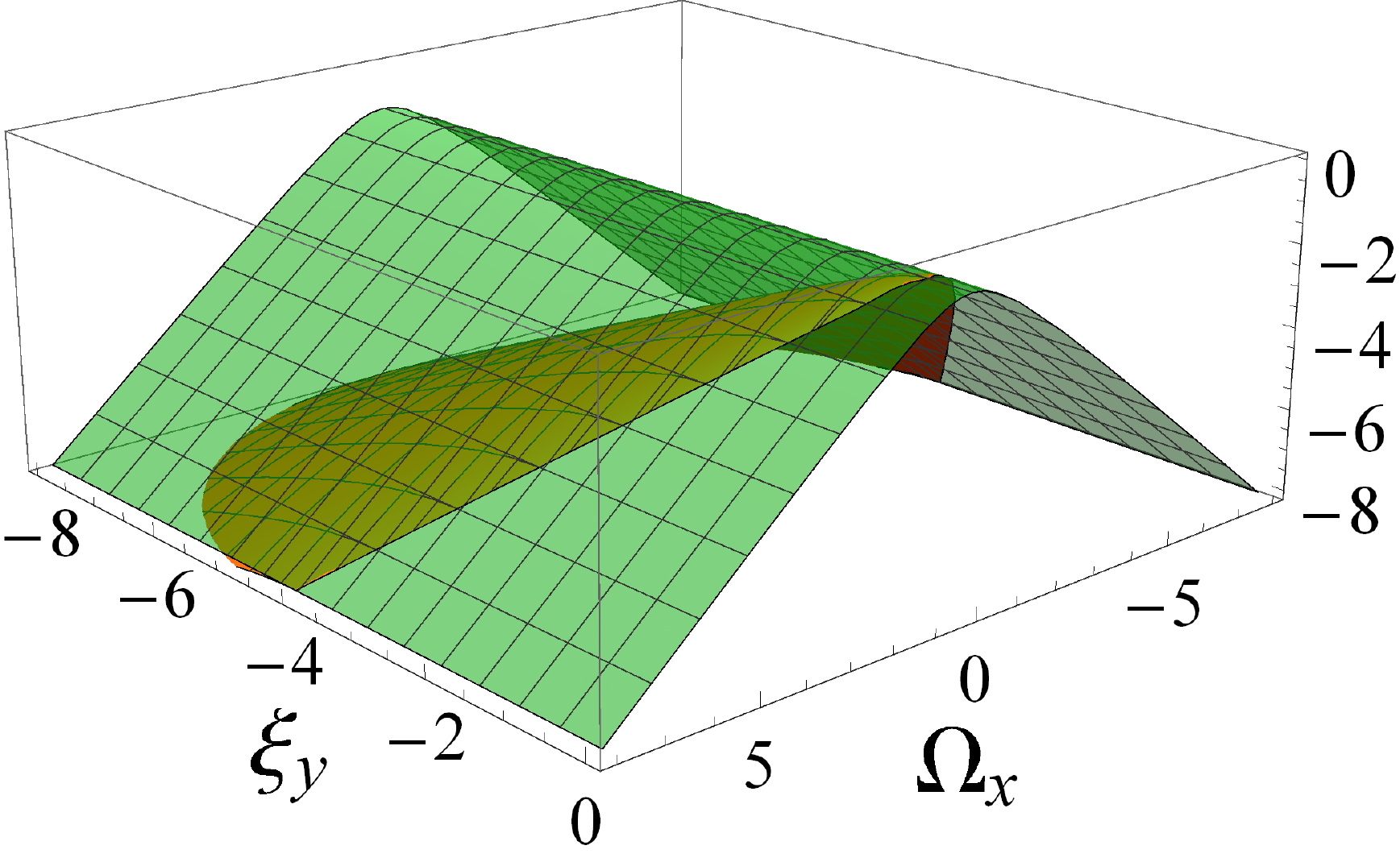}
&  \includegraphics[width= 0.5 \columnwidth]{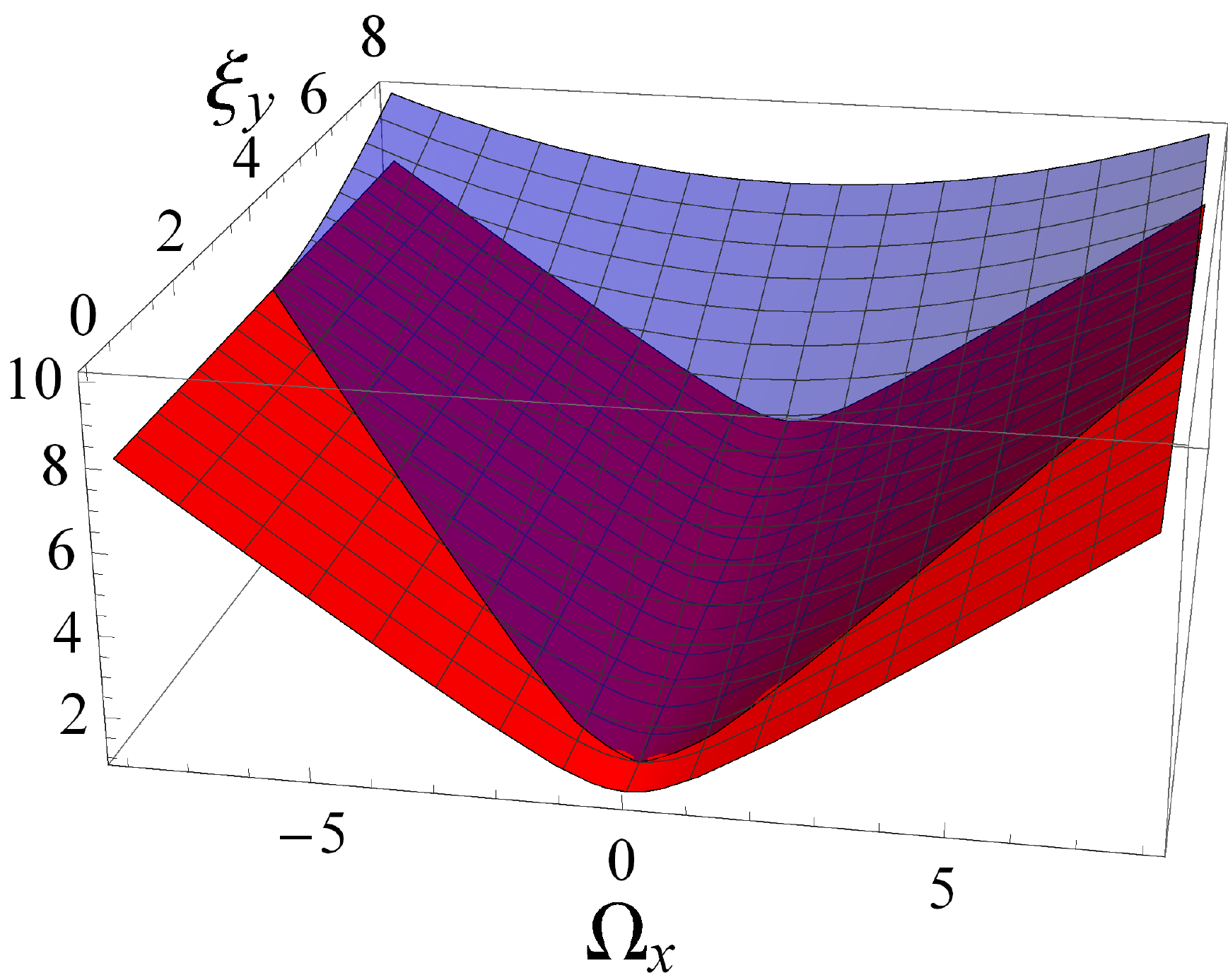}\\
(a) & (b) 
\end{tabular} 
\caption{(a) The two lower energy surfaces $e_2$ (green) and $e_3$ (orange) as functions of the coupling parameters $\Omega_x$ and $\xi_y < 0$.  (b) The two higher energy surfaces $e_1$ (red) and $e_4$ (blue)  as functions of the coupling parameters $\Omega_x$ and $\xi_y > 0$.  
}
 \label{Fig:energy1}
\end{figure}

\begin{figure}[ht]
\begin{tabular}{c c}
\includegraphics[width= 0.49 \columnwidth]{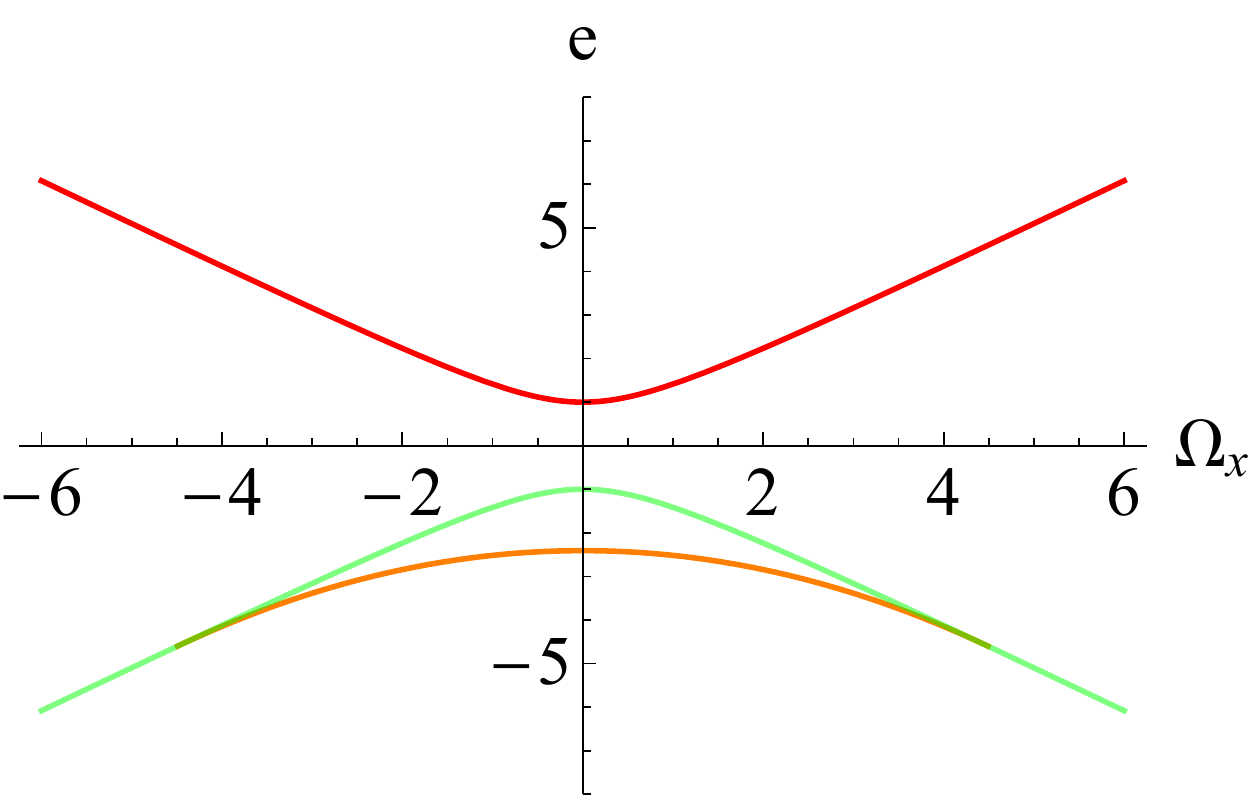}
&  \includegraphics[width= 0.49 \columnwidth]{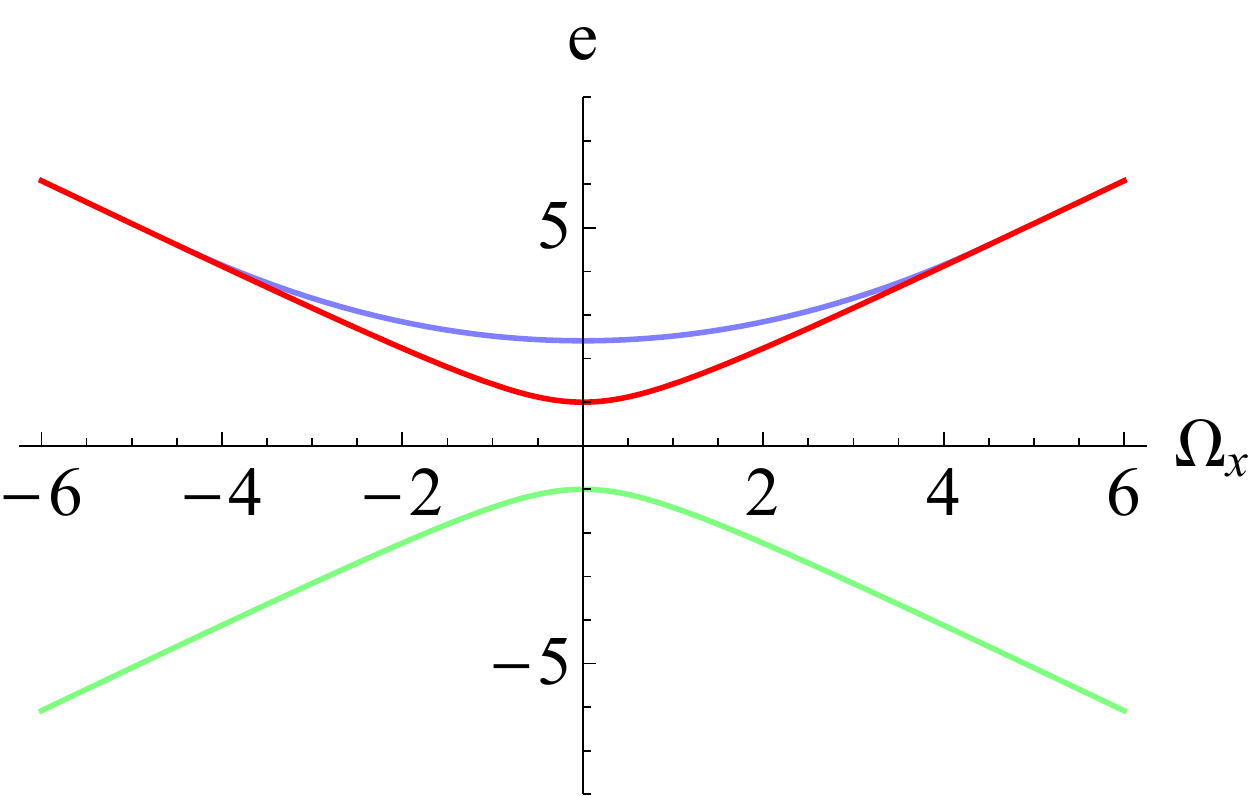}\\
(a) & (b) \\
\end{tabular} 
\caption{(a) Critical energies as a function of the coupling parameter $\Omega_x$ for (a) $\xi_y = -2.3$ and (b) $\xi_y = 2.3$
The color code is $e_1$ (red), $e_2$ (green), $e_3$ (orange) and $e_4$ (blue).  
}
 \label{Fig:energy2}
\end{figure}

These energy surfaces have interesting properties:
\begin{itemize} 
\item They are symmetric under the interchange $\Omega_x \leftrightarrow -\Omega_x$.
\item The energy $e_3$ is only defined in the region  $\xi_y \leq -\frac{\sqrt{1+\Omega_{xc}^2}}{2}$, where it is negative and lower than  $e_2$.
\item The energy $e_4$ is only defined in the region  $\xi_y \geq \frac{\sqrt{1+\Omega_{xc}^2}}{2}$, where it is positive and higher than  $e_1$.
\item The line $\xi_y = -\frac{\sqrt{1+\Omega_{xc}^2}}{2}$ is the {\em separatrix} between two quantum phases. The ground state is described by the coherent state $ \textbf{x}_{2} $ on one side of this line, and by  $ \textbf{x}_{4}  $ on the other side.
A similar situation occurs for the highest energy state for $\xi_y = \frac{\sqrt{1+\Omega_{xc}^2}}{2}$: It is described by $ \textbf{x}_{4}  $ in one phase, and by $ \textbf{x}_{1}  $ on the other phase. These energy surfaces are shown in Fig. \ref {Fig:energy1}.
\item The  energies are shown in Fig.  \ref {Fig:energy2}, as functions of $\Omega_x$ for $\xi_y = -2.3$ in a) and for $\xi_y = 2.3$ in b). In both figures, the color code is red for $e_1$, green for $e_2$, orange for $e_3$, and blue for $e_4$.
\end{itemize}

In what follows, we limit the domain of $\xi_y$ to positive values. Under this condition, the interesting stationary points are located at the top of the energy spectrum. Due to the symmetries in the model, reversing the sign of $\xi_y$ would generate an analogous analysis for the ground state.

\subsubsection{Observables}

\begin{figure}[ht]
\begin{tabular}{c c}
\includegraphics[width= 0.49 \columnwidth]{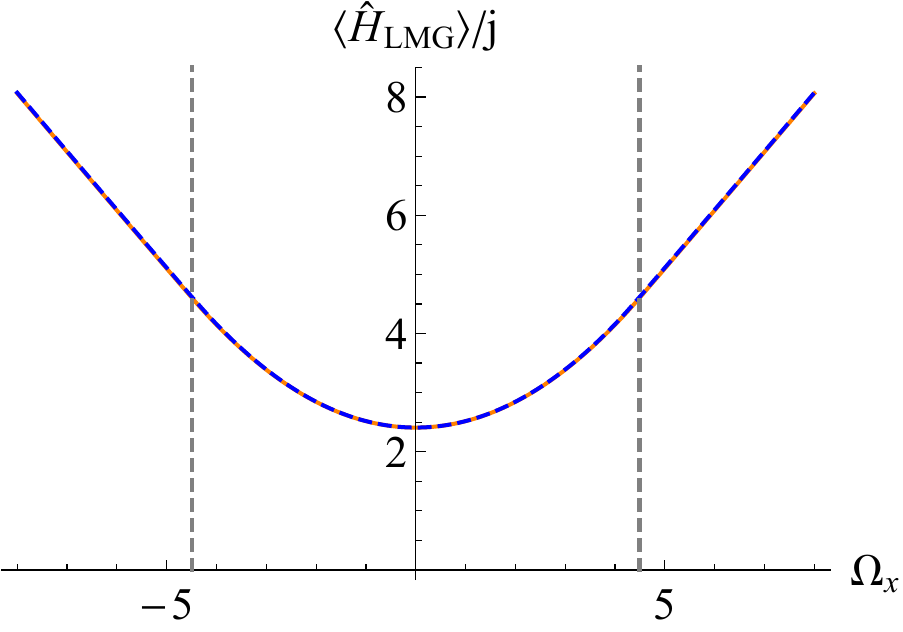} &
\includegraphics[width= 0.49 \columnwidth]{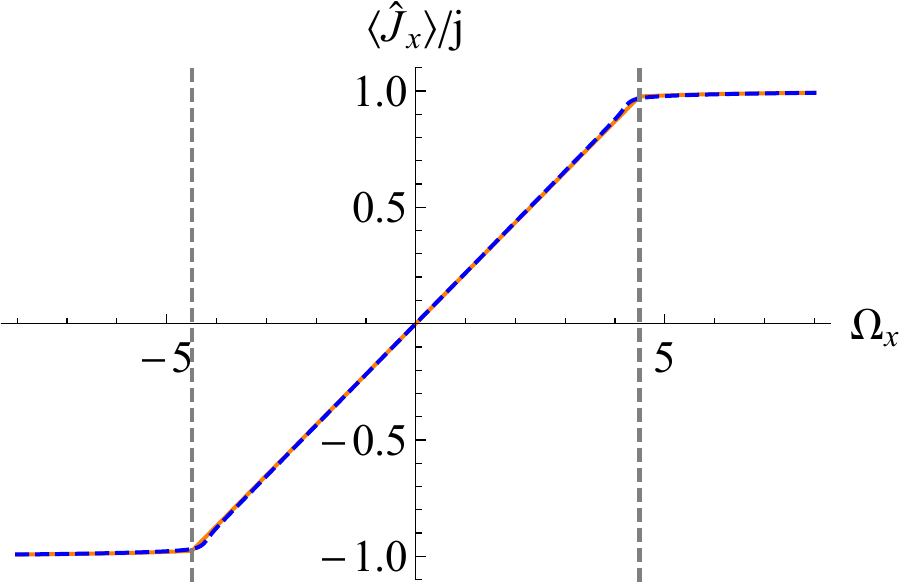} \\
(a) & (b) \\
\includegraphics[width= 0.49 \columnwidth]{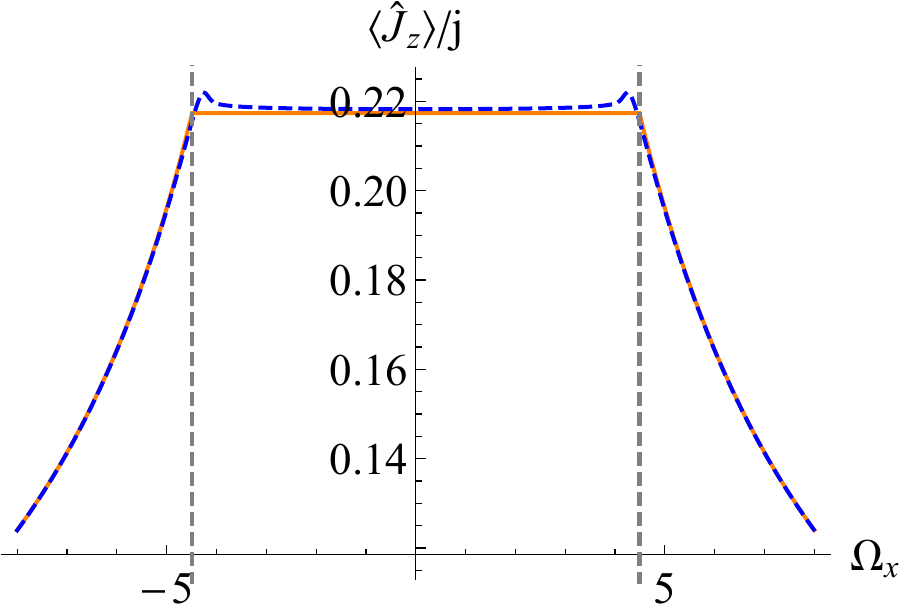} &
\includegraphics[width= 0.49 \columnwidth]{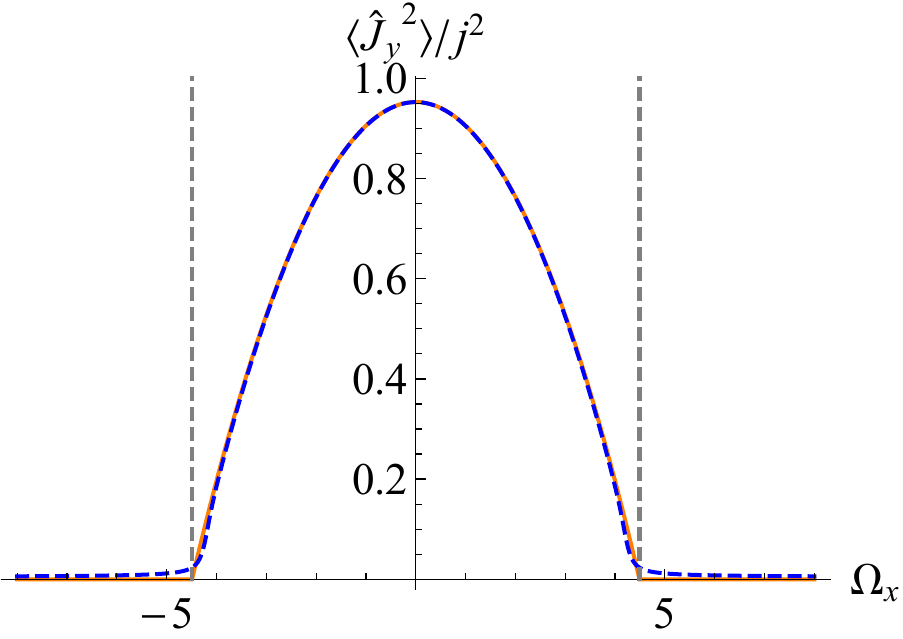} \\
(c) & (d)
\end{tabular} 
\caption{Comparison of the expectation values of different observables taken in the highest energy eigenstate (dashed blue) and in spin coherent states (solid orange) for $j=128$ and $\xi_y=2.3$. }
\label{Fig:NumVsCoh}
\end{figure}

In Fig.~\ref{Fig:NumVsCoh}, we show the expectation values of some observables in spin coherent states for the LMG model, and compare them with the expectation values taken in the highest energy eigenstate, obtained with a numerical diagonalization for $j=128$ and $\xi_y=2.3$. The quantum phase transition manifests as a discontinuity in the derivative of the plots at $\Omega_{xc}=4.490$ for (b), (c), and (d).

\subsubsection{Instabilities and excited state quantum phase transitions}

\begin{figure}[ht]
\begin{tabular}{c}
\includegraphics[width=0.89 \columnwidth]{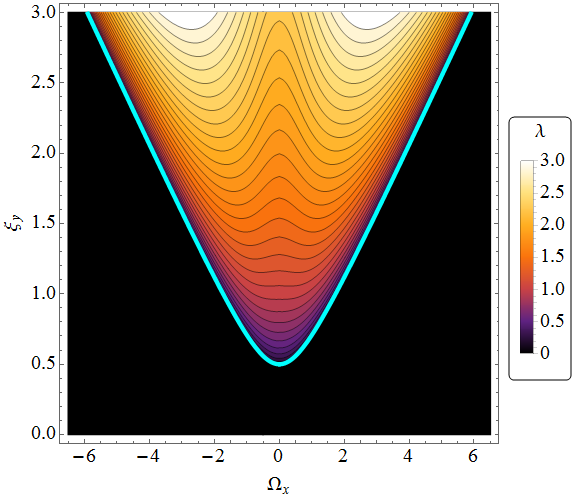}
\end{tabular} 
\caption{Lyapunov exponent for the stationary point $\textbf{x}_{1}$ as a function of the coupling parameters $\Omega_x$ and $\xi_y >0$. The black zone indicates a null Lyapunov exponent.
}
 \label{Fig:LMGLE}
\end{figure}

The Hamiltonian~(\ref{Eq:LMGclassical}) is integrable. The stationary points  $ \textbf{x}_{3}$ and  $ \textbf{x}_{4}$ are regular with Lyapunov exponent equal to zero. When $\xi_y>\frac{\sqrt{1+\Omega_x^2}}{2}$, the point $\textbf{x}_{1}$  is unstable, hyperbolic. The same happens for $\textbf{x}_{2}$ when $\xi_y < -\frac{\sqrt{1+\Omega_x^2}}{2}$. In these cases, they have a positive Lyapunov exponent  given by
\begin{equation}
\lambda = \sqrt{\sqrt{1+\Omega_x^2}\,(2\xi_y-\sqrt{1+\Omega_x^2})}.
\label{Eq:Lambda_LMG}
\end{equation} 
Fig.~\ref{Fig:LMGLE}~ shows the Lyapunov exponent as a function of the parameters $\Omega_x$ and $\xi_y >0$.

\begin{figure}[ht]
\begin{tabular}{c c c}
\includegraphics[width= 0.45 \columnwidth]{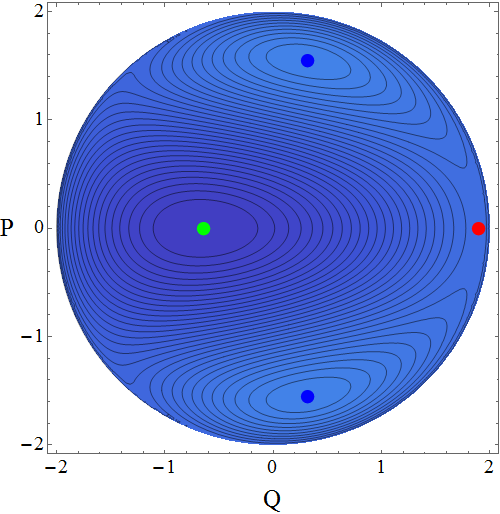} &
\includegraphics[width= 0.45 \columnwidth]{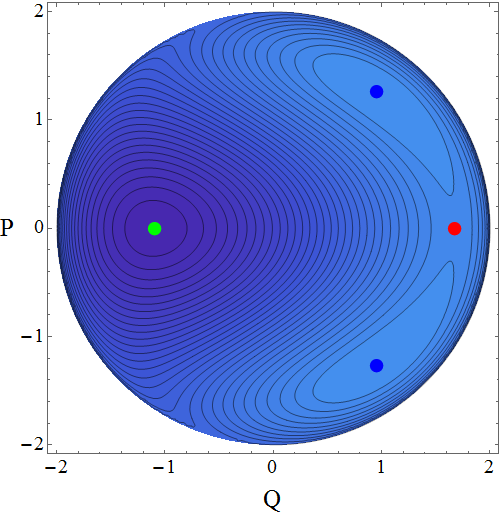} &  \includegraphics[width= 0.07 \columnwidth]{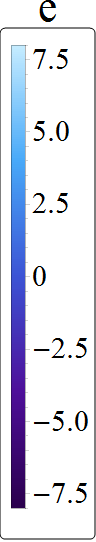} \\
(a) $\Omega_x=0.2\,\Omega_{xc}$ & (b) $\Omega_x=0.6\,\Omega_{xc}$ \\
\includegraphics[width= 0.45 \columnwidth]{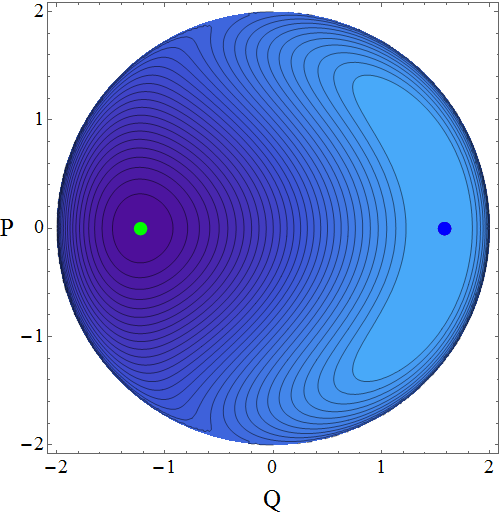} &
\includegraphics[width= 0.45 \columnwidth]{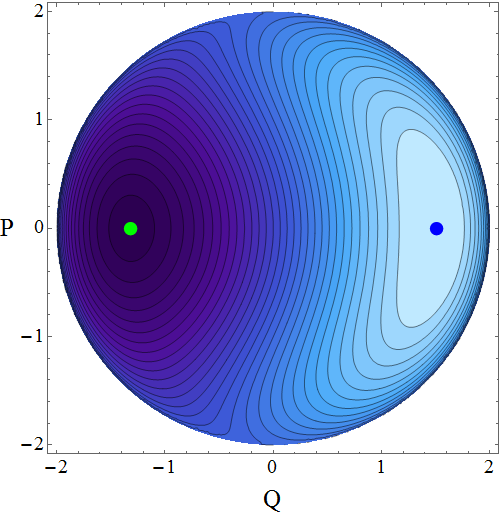} \\
(c) $\Omega_x=\Omega_{xc}$ & (d) $\Omega_x=2\,\Omega_{xc}$  
\end{tabular} 
\caption{Energy surfaces for different values of the parameters $\Omega$,  with $\Omega_{xc}=\sqrt{4 \xi_y^2 -1}$ with $\xi_y=2$.
Green points are stable center points $\textbf{x}_{2}$, the blue ones are unstable center points: $\textbf{x}_{4}$ and $\textbf{x'}_{4}$ in (a) and (b), and in (c) and (d), and the red point is the stationary point with positive Lyapunov exponent $\textbf{x}_{1}$, only present in (a) and (b).
}
 \label{Fig:LMG}
\end{figure}

Moving across the parameter space, it is possible to change the kind of stability of the stationary points of the Hamiltonian. In Fig.~\ref{Fig:LMG}, we can see three points. For positive values of $\xi_y$, the green point corresponds to $\textbf{x}_{2}$, the ground state of the system, which remains a center point. On the other hand, the red point corresponds to $\textbf{x}_{1}$. It is hyperbolic as long as $\Omega_x<\Omega_{xc}$, where $\Omega_{xc}$ satisfies $\xi_y=\frac{\sqrt{1+\Omega_{xc}^2}}{2}$ (Fig.~\ref{Fig:LMG} (a) and (b)), and ends up merging with the two blue points, $\textbf{x}_{4}$ and $\textbf{x'}_{4}$ when $\Omega_x>\Omega_{xc}$, becoming then a maximum (unstable) center point (Fig.~\ref{Fig:LMG} (c) and (d)).

\begin{figure}[ht]
\includegraphics[width=0.9 \columnwidth]{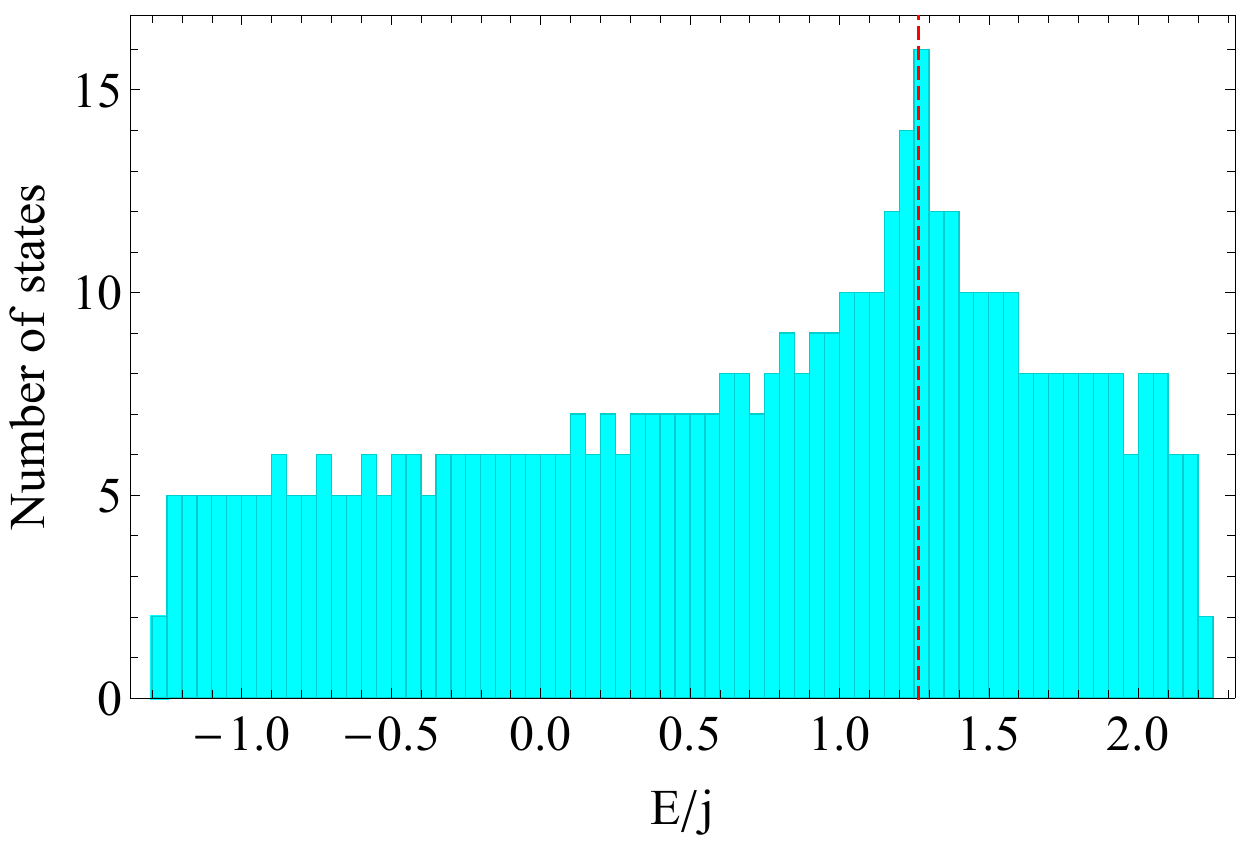}
\caption{Density of states when $\xi_y=2$ and $\Omega_x=0.2\,\Omega_{xc}$ for $j=256$. The red dashed line indicates the classical energy $e_1=\sqrt{1+\Omega_x^2}=1.265$ where the ESQPT takes place.}
\label{Fig:Hist}
\end{figure}

The phase transition appears when $\xi_y=\frac{\sqrt{1+\Omega_x^2}}{2}$.
In the region 
$\xi_y>\frac{\sqrt{1+\Omega_x^2}}{2}$, $\textbf{x}_1$ is saddle point, associated with an ESQPT in the quantum domain. A main signature of ESQPTs is the divergence of the density of states at an energy denoted by $E_{\text{ESQPT}}$. In the mean-field approximation, it has been shown that this energy coincides with the energy of the classical system at the saddle point~\cite{Cejnar2006,Caprio2008}, that is, for the LMG model, 
$E_{\text{ESQPT}}^{\text{LMG}}/ j = h_{\text{LMG}}(\textbf{x}_1) =e_1=\sqrt{1+\Omega_x^2}$. 
In Fig.~\ref{Fig:Hist}, we show the density of states for the LMG model taking $j=256$ and the parameters of Fig.~\ref{Fig:LMG} (a). These parameters correspond to the unstable region where there is a non-zero Lyapunov exponent (see Fig.~\ref{Fig:LMGLE}). The divergence in the density of states associated with the ESQPT is clearly visible at $e_1=1.265$, marked with a vertical red dashed line.

\section{Geometry of the parameter space of the LMG model}
 
 In this section, we study the geometry of the ground and the highest energy states of the LMG model. 
 While the semiclassical analysis performed above employing $SU(2)$ coherent states provides rich and valuable information about the system, its stationary points and its quantum phases, it is not well suited to study the geometry of the parameter space of this system. As shown in Appendix \ref{appCoh}, the metric obtained employing coherent states is ill defined for $\xi_y\leq\frac{\sqrt{1+\Omega_x^2}}{2}$.
 In this case, the coherent states that describe the ground state and the highest energy state are given by $\bf{x}_1$ and $\bf{x}_2$, shown in
Eqs. (\ref{x1}) and (\ref{x2}), respectively. In both cases the states have no dependence in the Hamiltonian parameter $\xi_y$, having null derivates respect to this parameter. 
 To overcome this difficulty, for each of these states, we first compute the QMT and its scalar curvature in the thermodynamic limit via the truncated Holstein-Primakoff transformation~\cite{HolsteinPrimakoff}.  This approximation becomes exact for extremal states in this limit, but has spurious divergences close to a phase transition \cite{Hirsch2013}. 
 We also carry out an exact diagonalization and compare the analytic and numeric results. In what follows, we take $x=\{x^{i}\}=(\Omega_{x},\xi_{y})$ as the adiabatic parameters.

\subsection{Ground state}

We begin our analysis with the ground state, which corresponds via
the classical Hamiltonian~(\ref{Eq:LMGclassical}) to the stationary point $\textbf{x}_2=(\theta_{2},\phi_{2})=\left(\arccos\frac{1}{\sqrt{1+\Omega_{x}^{2}}},\pi \right)$.
The first step is to align the classical pseudospin of the ground
state \break $\vec{J}=(j\sin\theta_{2}\cos\phi_{2},j\sin\theta_{2}\sin\phi_{2},j\cos\theta_{2})$
with the $z$ axis. To do this, we perform a rotation of
the spin operators around the $y$ axis as follows~\cite{Vidal2005}
\begin{equation}
\begin{pmatrix}\hat{J}_{x}\\
\hat{J}_{y}\\
\hat{J}_{z}
\end{pmatrix}=\begin{pmatrix}\cos\theta_{2} & 0 & \sin\theta_{2}\\
0 & 1 & 0\\
-\sin\theta_{2} & 0 & \cos\theta_{2}
\end{pmatrix}\begin{pmatrix}\hat{J}_{x}^{\prime}\\
\hat{J}_{y}^{\prime}\\
\hat{J}_{z}^{\prime}
\end{pmatrix}.
\end{equation}
With this rotation, the Hamiltonian~(\ref{Eq:H-LMG}) takes the form
\begin{equation}
\hat{H}=\sqrt{1+\Omega_{x}^{2}}\hat{J}_{z}^{\prime}+\frac{\xi_{y}}{j}\hat{J}_{y}^{\prime2},\label{Hprimeground}
\end{equation}
which is suitable for applying the Holstein-Primakoff transformation
that maps angular momentum operators into bosonic operators as
\begin{equation}
\hat{J}_{z}^{\prime}=\hat{a}^{\dagger}\hat{a}-j,\,\,\,\,\,\hat{J}_{+}^{\prime}=\hat{a}^{\dagger}(2j-\hat{a}^{\dagger}\hat{a})^{1/2},\,\,\,\,\,\hat{J}_{-}^{\prime}=(2j-\hat{a}^{\dagger}\hat{a})^{1/2}\,\hat{a}.
\end{equation}
It is readily verified that this representation satisfies the $SU(2)$
algebra as long as $[\hat{a},\hat{a}^{\dagger}]=1$. Now, we consider
the thermodynamic limit $j\rightarrow\infty$ and expand the square
roots retaining only the zeroth order term in $1/j$~\cite{GuPRE,Vidal2005}.
The Cartesian components of the angular momentum turn out to be
\begin{equation}
\hat{J}_{x}^{\prime}\simeq\sqrt{\frac{j}{2}}\left(\hat{a}^{\dagger}+\hat{a}\right),\,\,\,\,\,\hat{J}_{y}^{\prime}\simeq-i\sqrt{\frac{j}{2}}\left(\hat{a}^{\dagger}-\hat{a}\right),\,\,\,\,\,\hat{J}_{z}^{\prime}=\hat{a}^{\dagger}\hat{a}-j.
\end{equation}
In principle, we may substitute these expressions into Eq.~(\ref{Hprimeground})
and perform a Bogoliubov transformation to creation and annihilation
operators $(\hat{b},\hat{b}^{\dagger})$ that diagonalize the Hamiltonian.
However, for illustrative purposes, we make first the intermediate
transformation $\hat{Q}=\frac{1}{\sqrt{2}}\left(\hat{a}^{\dagger}+\hat{a}\right)$,
$\hat{P}=\frac{i}{\sqrt{2}}\left(\hat{a}^{\dagger}-\hat{a}\right)$
to find that the Hamiltonian takes the form
\begin{equation}\label{HQPground}
\hat{H}\simeq-j\sqrt{1+\Omega_{x}^{2}}+\left(\frac{\sqrt{1+\Omega_{x}^{2}}+2\xi_{y}}{2}\right)\hat{P}^{2}+\frac{\sqrt{1+\Omega_{x}^{2}}}{2}\hat{Q}^{2}.
\end{equation}
This suggests the use of the following transformation
\begin{align}
\hat{Q}&=\left(\frac{\sqrt{1+\Omega_{x}^{2}}+2\xi_{y}}{4\sqrt{1+\Omega_{x}^{2}}}\right)^{1/4}\left(\hat{b}^{\dagger}+\hat{b}\right), \nonumber \\
\hat{P}&=i\left(\frac{\sqrt{1+\Omega_{x}^{2}}}{4\left(\sqrt{1+\Omega_{x}^{2}}+2\xi_{y}\right)}\right)^{1/4}\left(\hat{b}^{\dagger}-\hat{b}\right),
\end{align}
which will cast the Hamiltonian~(\ref{HQPground}) into the harmonic oscillator
\begin{equation}
\hat{H}\simeq-j\sqrt{1+\Omega_{x}^{2}}+\sqrt{\sqrt{1+\Omega_{x}^{2}}\left(\sqrt{1+\Omega_{x}^{2}}+2\xi_{y}\right)}\left(\hat{b}^{\dagger}\hat{b}+\frac{1}{2}\right),
\end{equation}
where $\omega=\sqrt{\sqrt{1+\Omega_{x}^{2}}\left(\sqrt{1+\Omega_{x}^{2}}+2\xi_{y}\right)}$ is the frequency 
and $E=-j\sqrt{1+\Omega_{x}^{2}}$ is the energy. The QMT can now be calculated with the aid of Eq.~(\ref{QMT}), setting $n=0$, and employing the operators
\begin{align}
\frac{\partial\hat{H}_{LMG}}{\partial\Omega_{x}}= & \hat{J}_{x}=\frac{1}{\sqrt{1+\Omega_{x}^{2}}}\hat{J}_{x}^{\prime}+\frac{\Omega_{x}}{\sqrt{1+\Omega_{x}^{2}}}\hat{J}_{z}^{\prime},\nonumber \\
\frac{\partial\hat{H}_{LMG}}{\partial\xi_{y}}= & \frac{\hat{J}_{y}^{2}}{j}=\frac{\hat{J}_{y}^{\prime2}}{j},
\end{align}
provided that they are expressed in terms of $(\hat{b},\hat{b}^{\dagger})$
to act on the eigenstates of $\hat{H}$. We find the metric components
\begin{align}
g_{11} & =\frac{j}{2\left(\Omega_{x}^{2}+1\right)^{7/4}\sqrt{\sqrt{\Omega_{x}^{2}+1}+2\xi_{y}}} \nonumber \\
&\quad+\frac{\xi_{y}^{2}\Omega_{x}^{2}}{8\left(\Omega_{x}^2+1\right)^{2}\left(\sqrt{\Omega_{x}^{2}+1}+2\xi_{y}\right)^{2}},\nonumber \\
g_{12} & =-\frac{\xi_{y}\Omega_{x}}{8\left(\Omega_{x}^{2}+1\right)\left(\sqrt{\Omega_{x}^{2}+1}+2\xi_{y}\right)^{2}},\nonumber \\
g_{22} & =\frac{1}{8\left(\sqrt{\Omega_{x}^{2}+1}+2\xi_{y}\right)^{2}},\label{metrGround}
\end{align}
and the determinant
\begin{equation}
g=\frac{j}{16\left(\Omega_{x}^{2}+1\right)^{7/4}\left(\sqrt{\Omega_{x}^{2}+1}+2\xi_{y}\right)^{5/2}}.
\end{equation}
Notice that a singularity appears in all the components when $\xi_{y}=-\frac{\sqrt{\Omega_x^2+1}}{2}$. Also, we observe that $g_{11}$ consists of two
terms, one of them is proportional to $j$ and dominant as $j\rightarrow\infty$.
Retaining all the terms and using Eq.~(\ref{scalar}), we find
that the scalar curvature associated to the metric~(\ref{metrGround}) simplifies to
\begin{equation}
R=-4.
\end{equation}
Remarkably, the scalar curvature is constant despite the nontrivial dependence of the metric components on the parameters. Furthermore, there is no sign of the singularity that appeared in the metric, implying that this singularity is not revealed in the parameter space's intrinsic geometry. We must mention that if we had considered only the dominant term in $j$ in $g_{11}$ for the computation of the scalar curvature, the result would have been different; therefore, we chose to retain all the terms. This constant negative curvature signals that the ground state's parameter space possesses a hyperbolic geometry and is isomorphic to the Lobachevskij space~\cite{Alekseevskij}.

\begin{figure}[ht]
\begin{tabular}{c c}
\includegraphics[width= 0.49 \columnwidth]{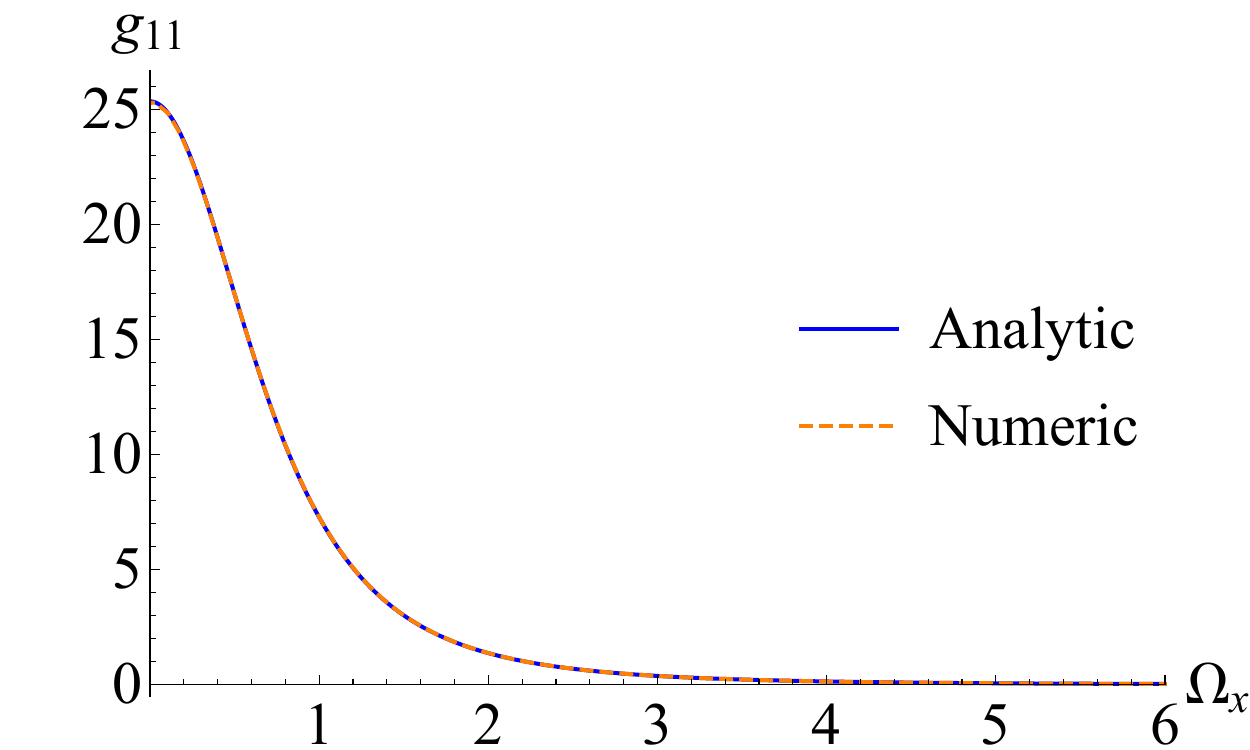} &
\includegraphics[width= 0.49 \columnwidth]{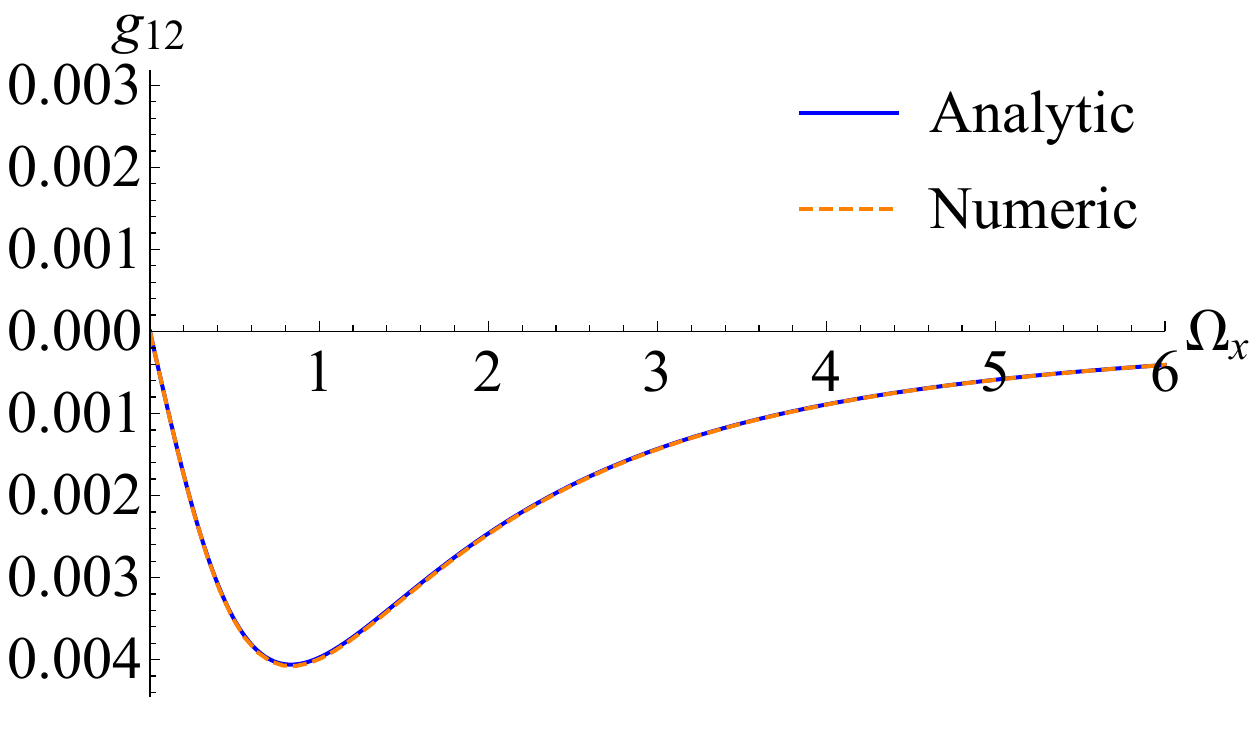} \\
(a) & (b) \\
\includegraphics[width= 0.49 \columnwidth]{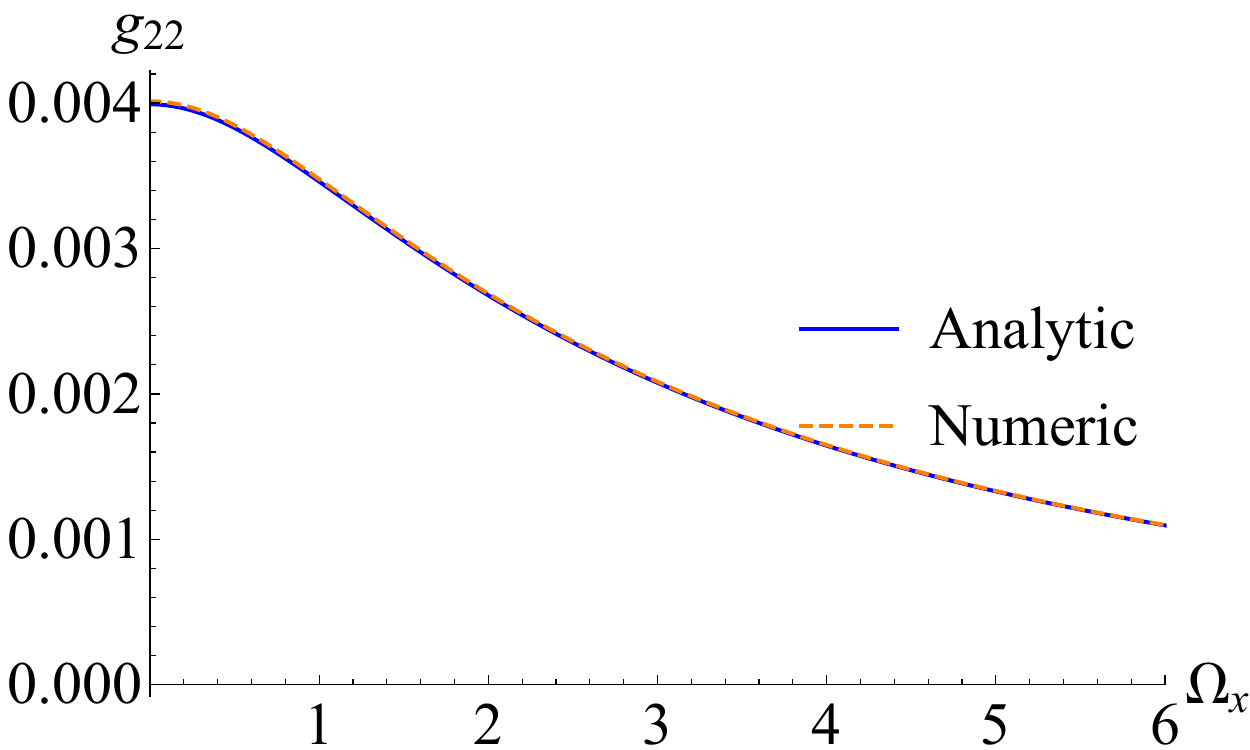} &  
\includegraphics[width= 0.49 \columnwidth]{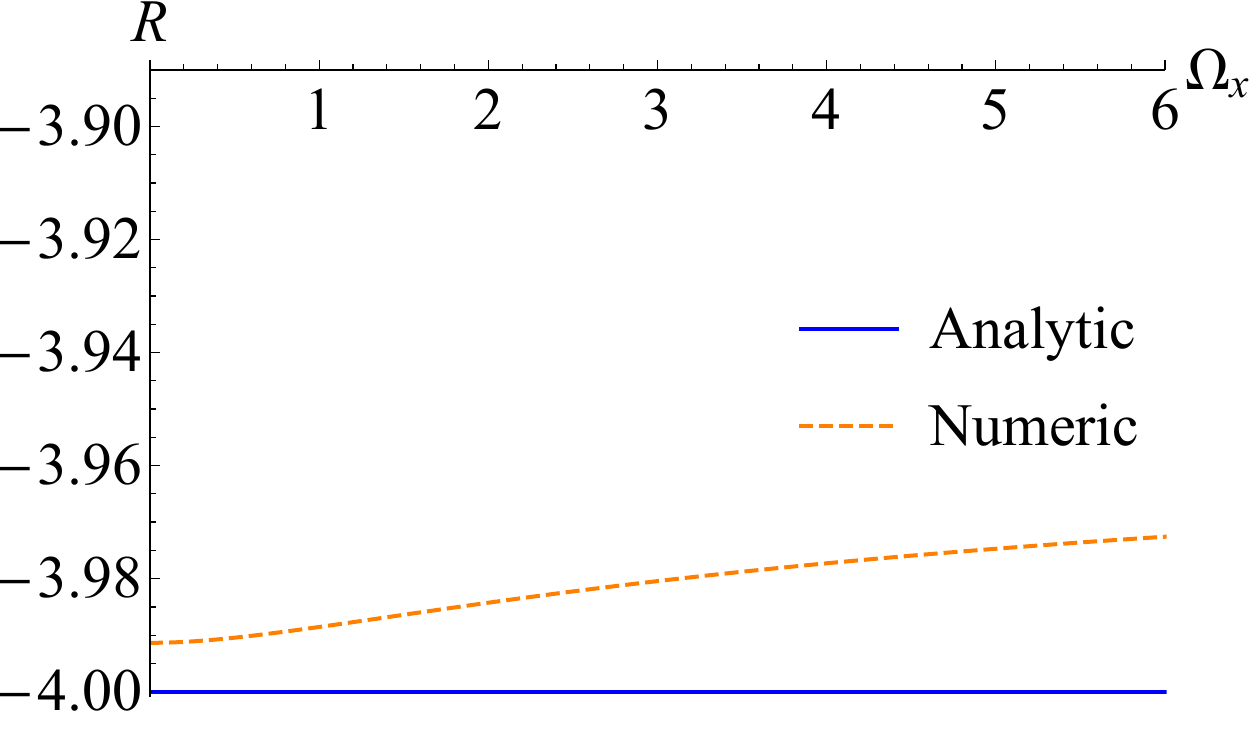} \\
(c) & (d) 
\end{tabular} 
\caption{QMT components and its scalar curvature for the ground state when $j=120$ and $\xi_y=2.3$.}
 \label{Fig:G0}
\end{figure}

\begin{figure}[ht]
\begin{tabular}{c c c}
\includegraphics[width= 0.46 \columnwidth]{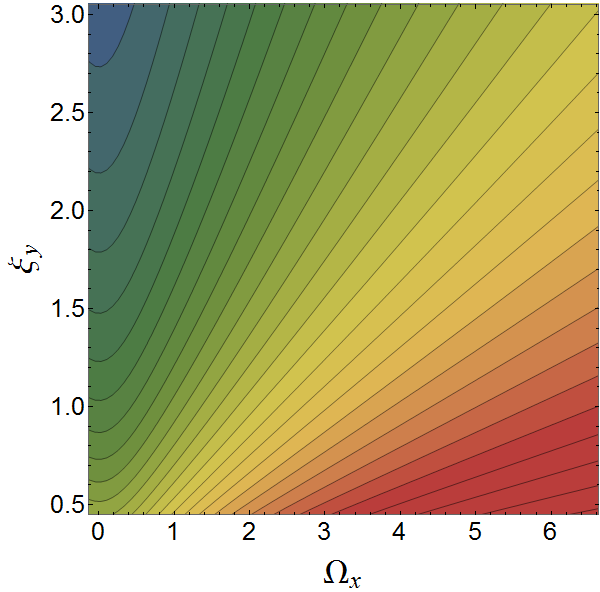} &
\includegraphics[width= 0.46 \columnwidth]{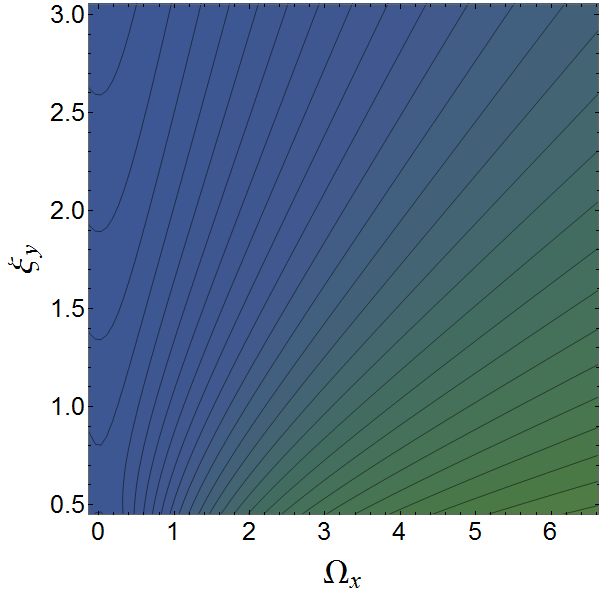} & \includegraphics[width= 0.07 \columnwidth]{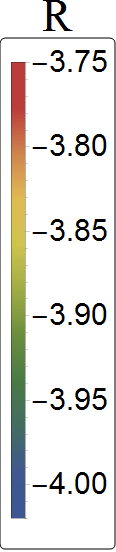}\\
(a) $j=20 $ & (b) $j=50$ \\
\includegraphics[width= 0.46 \columnwidth]{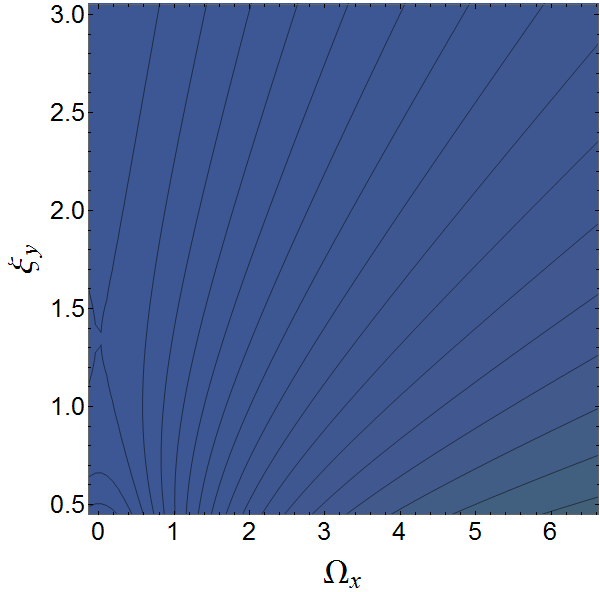} &  
\includegraphics[width= 0.46 \columnwidth]{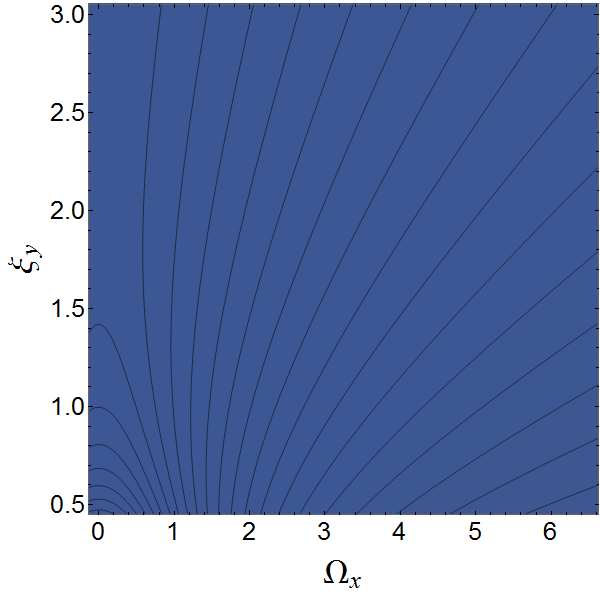} \\
(c) $j=80$ & (d) $j=120$  
\end{tabular} 
\caption{Scalar curvature map for different values of $j$, all in the ground state.}
 \label{Fig:RE0}
\end{figure}

In Fig.~\ref{Fig:G0} we plot the metric components and the scalar curvature for $\xi_y=2.3$ and $j=120$. We see that the numeric results agree well with their analytic counterparts coming from the Holstein-Primakoff approximation. Fig.~\ref{Fig:RE0} shows a map of the scalar curvature for different values of $j$, where we see a tendency to the analytic result as $j$ increases. Notice that the singularity predicted by the metric components~(\ref{metrGround}) is outside the range of the parameters employed in this study ($\xi_{y}\geq 0$).


\subsection{Highest energy state}

We now study the highest energy state for which the QPT occurs.
We divide the parameter space into regions above and below the separatrix, and thus,
treat each phase separately.
The energy of the maximum is
\begin{equation}
\frac{E_{max}}{j}=\begin{cases}
\frac{1+\Omega_{x}^{2}+4\xi_{y}^{2}}{4\xi_{y}} & ,\,\,\Omega_{x}<\Omega_{xc}\\
\sqrt{1+\Omega_{x}^{2}} & ,\,\,\Omega_{x}\geq\Omega_{xc}
\end{cases},
\end{equation}
where $\Omega_{xc}=\sqrt{4\xi_{y}^{2}-1}$ is the critical value of $\Omega_x$
(for a fixed $\xi_{y}$). The energy $E_{max}$ and
its first derivative are continuous as a function of $\Omega_{x}$
at the stationary point $\Omega_{x}=\Omega_{xc}$, whereas its second
derivative presents a discontinuity. This signals a second order QPT.\\

\subsubsection{Symmetric phase}

We begin with the parameter space below the separatrix, i.e. when
$\xi_{y}<\frac{\sqrt{1+\Omega_{x}^{2}}}{2}$, where the procedure
to find the QMT is similar to that of the ground state. The angular
coordinates of the corresponding stationary point (the red one in Fig.~\ref{Fig:LMG}) are $\left(\theta_{1},\phi_{1}\right)=\left(\arccos\left(\frac{1}{\sqrt{1+\Omega_{x}^{2}}}\right),0\right)$,
and the rotation that aligns the classical pseudospin with the $z$
axis is
\begin{equation}
\begin{pmatrix}\hat{J}_{x}\\
\hat{J}_{y}\\
\hat{J}_{z}
\end{pmatrix}=\begin{pmatrix}\cos\theta_{1} & 0 & -\sin\theta_{1}\\
0 & 1 & 0\\
\sin\theta_{1} & 0 & \cos\theta_{1}
\end{pmatrix}\begin{pmatrix}\hat{J}_{x}^{\prime}\\
\hat{J}_{y}^{\prime}\\
\hat{J}_{z}^{\prime}
\end{pmatrix}.
\end{equation}
Making use of the truncated Holstein-Primakoff transformation, we
arrive at the quadratic Hamiltonian
\begin{equation}
\hat{H}\simeq j\sqrt{1+\Omega_{x}^{2}}-\left(\frac{\sqrt{1+\Omega_{x}^{2}}-2\xi_{y}}{2}\right)\hat{P}^{2}-\frac{\sqrt{1+\Omega_{x}^{2}}}{2}\hat{Q}^{2},\label{Hquadred}
\end{equation}
which describes a harmonic oscillator with frequency $\omega=\sqrt{\sqrt{1+\Omega_{x}^{2}}\left(\sqrt{1+\Omega_{x}^{2}}-2\xi_{y}\right)}$ and zero point energy $E=j\sqrt{1+\Omega_{x}^{2}}$. Notice that there is a minus sign in the last two terms of Eq.~(\ref{Hquadred}) due to this stationary point being a maximum of energy. 
The components of the metric are
\begin{align}
g_{11} & =\frac{j}{2\left(\Omega_{x}^{2}+1\right)^{7/4}\sqrt{\sqrt{\Omega_{x}^{2}+1}-2\xi_{y}}} \nonumber \\
&\quad+\frac{\xi_{y}^{2}\Omega_{x}^{2}}{8\left(\Omega_{x}^2+1\right)^{2}\left(\sqrt{\Omega_{x}^{2}+1}-2\xi_{y}\right)^{2}},\nonumber \\
g_{12} & =-\frac{\xi_{y}\Omega_{x}}{8\left(\Omega_{x}^{2}+1\right)\left(\sqrt{\Omega_{x}^{2}+1}-2\xi_{y}\right)^{2}},\nonumber \\
g_{22} & =\frac{1}{8\left(\sqrt{\Omega_{x}^{2}+1}-2\xi_{y}\right)^{2}},\label{QMTsym}
\end{align}
and the determinant is
\begin{equation}
g=\frac{j}{16\left(\Omega_{x}^{2}+1\right)^{7/4}\left(\sqrt{\Omega_{x}^{2}+1}-2\xi_{y}\right)^{5/2}}.
\end{equation}
We can easily identify a singularity in all the components of the
metric and the determinant which occurs at $\xi_{y}=\frac{\sqrt{1+\Omega_{x}^{2}}}{2}$.
This is precisely the stationary point where the QPT takes place,
and confirms the usefulness of the QMT to detect a quantum phase transition.
With the QMT~(\ref{QMTsym}) at hand, we compute its scalar curvature and find
\begin{equation}
R=-4,
\end{equation}
which again means that the underlying geometry is hyperbolic.\\

\subsubsection{Broken symmetry phase}

Above the separatrix, i.e. when $\xi_{y}>\frac{\sqrt{1+\Omega_{x}^{2}}}{2}$,
the angular coordinates of the stationary points (the blue ones in Fig.~\ref{Fig:LMG}) are 
$\left(\theta_{4},\phi_{4}\right)=\left( \arccos\left(\frac{1}{2\xi_{y}}\right),\arccos \left(\frac{\Omega_{x}}{\sqrt{4\xi_{y}^{2}-1}}\right)\right)$,
and the rotation that aligns the classical pseudospin with the $z$
axis is
\begin{equation}
\begin{pmatrix}\hat{J}_{x}\\
\hat{J}_{y}\\
\hat{J}_{z}
\end{pmatrix}=\left(\begin{array}{ccc}
\cos\phi_{4} & -\sin\phi_{4} & 0\\
\sin\phi_{4} & \cos\phi_{4} & 0\\
0 & 0 & 1
\end{array}\right)\left(\begin{array}{ccc}
\cos\theta_{4} & 0 & -\sin\theta_{4}\\
0 & 1 & 0\\
\sin\theta_{4} & 0 & \cos\theta_{4}
\end{array}\right)\begin{pmatrix}\hat{J}_{x}^{\prime}\\
\hat{J}_{y}^{\prime}\\
\hat{J}_{z}^{\prime}
\end{pmatrix}.
\end{equation}
Following similar steps, we find that the quadratic Hamiltonian now
is
\begin{align}
\hat{H}\simeq & j\frac{4\xi_{y}^{2}+\Omega_{x}^{2}+1}{4\xi_{y}}-\frac{\xi_{y}\left(4\xi_{y}^{2}-\Omega_{x}^{2}-1\right)}{4\xi_{y}^{2}-1}\hat{P}^{2} \nonumber \\
&+\frac{\Omega_{x}\sqrt{4\xi_{y}^{2}-\Omega_{x}^{2}-1}}{2\left(4\xi_{y}^{2}-1\right)}\left(\hat{Q}\hat{P}+\hat{P}\hat{Q}\right) \nonumber \\
&-\frac{16\xi_{y}^{4}-8\xi_{y}^{2}+\Omega_{x}^{2}+1}{4\xi_{y}\left(4\xi_{y}^{2}-1\right)}\hat{Q}^{2}.
\label{quadHam}
\end{align}
This Hamiltonian has the form of a generalized harmonic oscillator. In order to remove the crossed term in $\hat{Q}$ and $\hat{P}$, we make the further linear canonical  transformation
\begin{align}
\hat{Q}&=\sqrt{\frac{2\xi_{y}\left(4\xi_{y}^{2}-\Omega_{x}^{2}-1\right)}{4\xi_{y}^{2}-1}}\,\hat{Q}^{\prime}, \nonumber \\
\hat{P}&=\sqrt{\frac{4\xi_{y}^{2}-1}{2\xi_{y}\left(4\xi_{y}^{2}-\Omega_{x}^{2}-1\right)}}\,\hat{P}^{\prime}+\frac{\Omega_{x}}{\sqrt{2\xi_{y}\left(4\xi_{y}^{2}-1\right)}}\hat{Q}^{\prime}.
\end{align}
The Hamiltonian then also turns into a harmonic oscillator, except for a sign,
\begin{equation}
\hat{H}=j\frac{4\xi_{y}^{2}+\Omega_{x}^{2}+1}{4\xi_{y}}-\frac{\hat{P}^{\prime2}}{2}-\frac{\omega^{2}}{2}\hat{Q}^{\prime2},
\end{equation}
with frequency $\omega=\sqrt{4\xi_{y}^{2}-\Omega_{x}^{2}-1}$ and
zero point energy $E=j\frac{4\xi_{y}^{2}+\Omega_{x}^{2}+1}{4\xi_{y}}$.
The expressions for the metric components are cumbersome; however, it can be seen that all of them are singular at points on the separatrix. Finally, the computation of the scalar curvature via Eq.~(\ref{scalar}) also yields a complicated and not illuminating expression.

\begin{figure}[ht]
\begin{tabular}{c c}
\includegraphics[width= 0.49 \columnwidth]{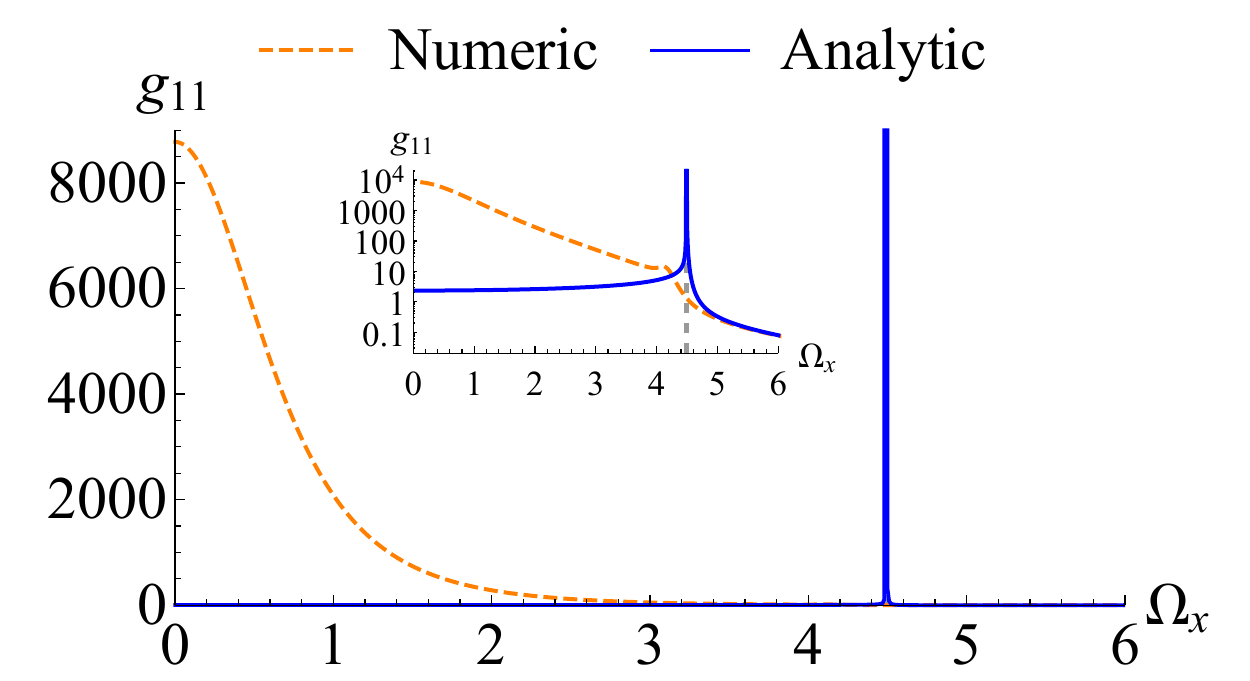} & \includegraphics[width= 0.49 \columnwidth]{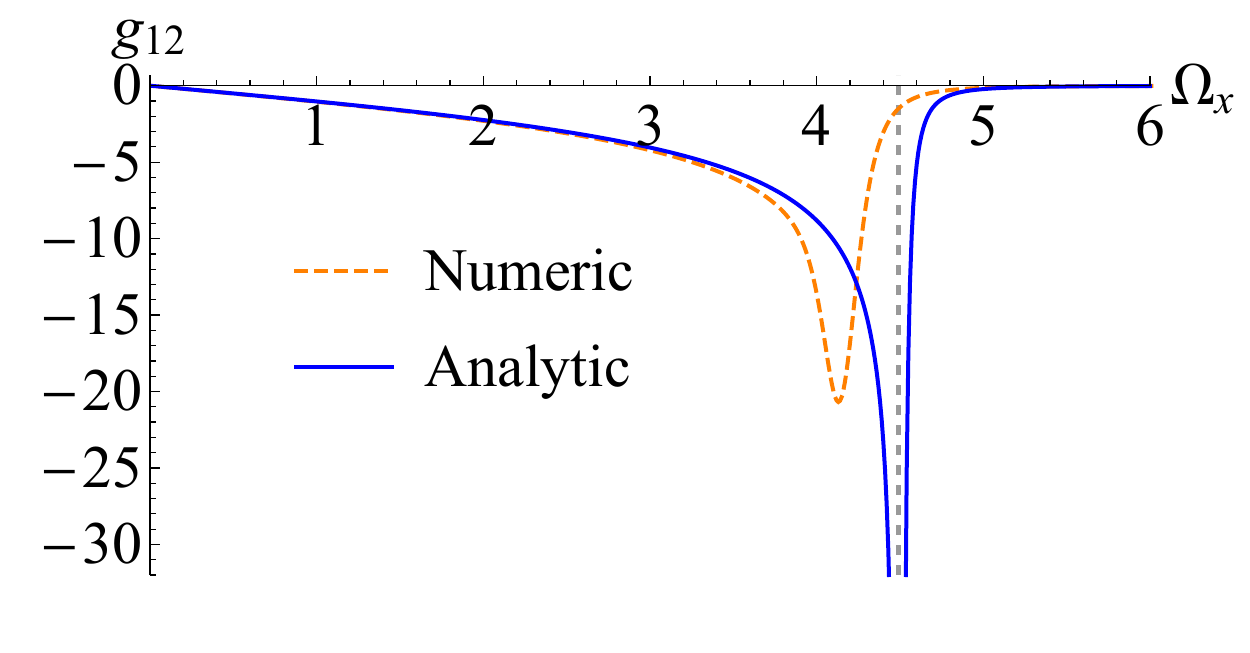} \\
(a) $g_{11}$ & (b) $g_{12}$ \\
\includegraphics[width= 0.49 \columnwidth]{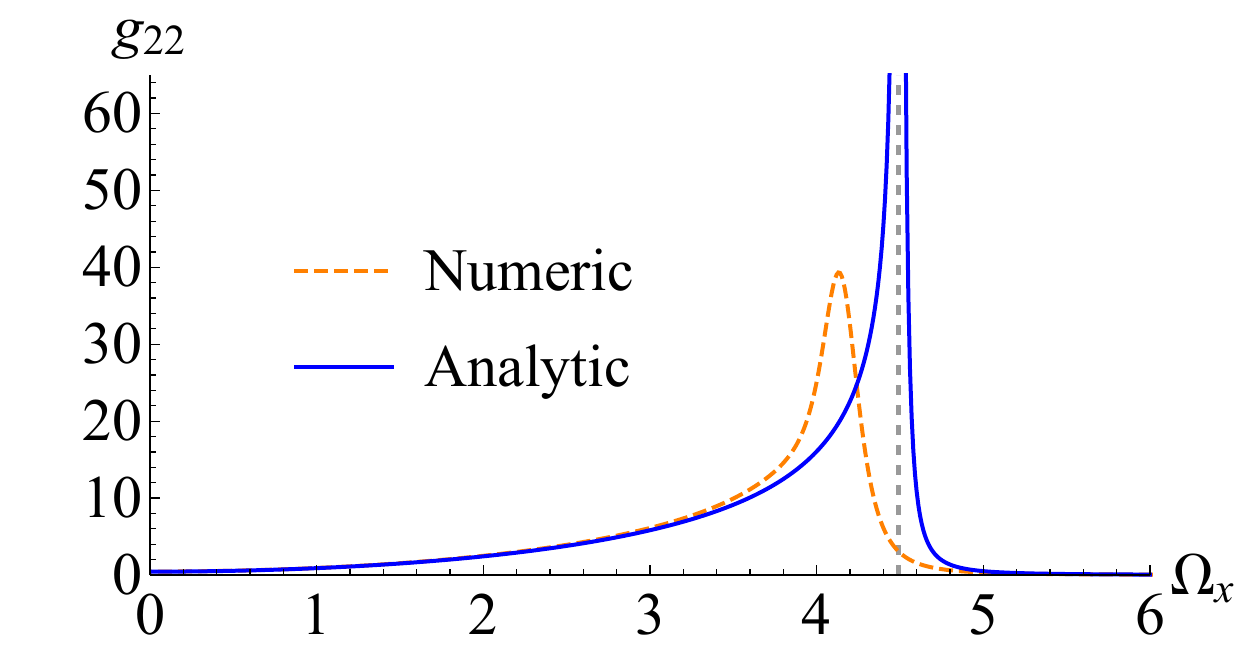} & \includegraphics[width= 0.49 \columnwidth]{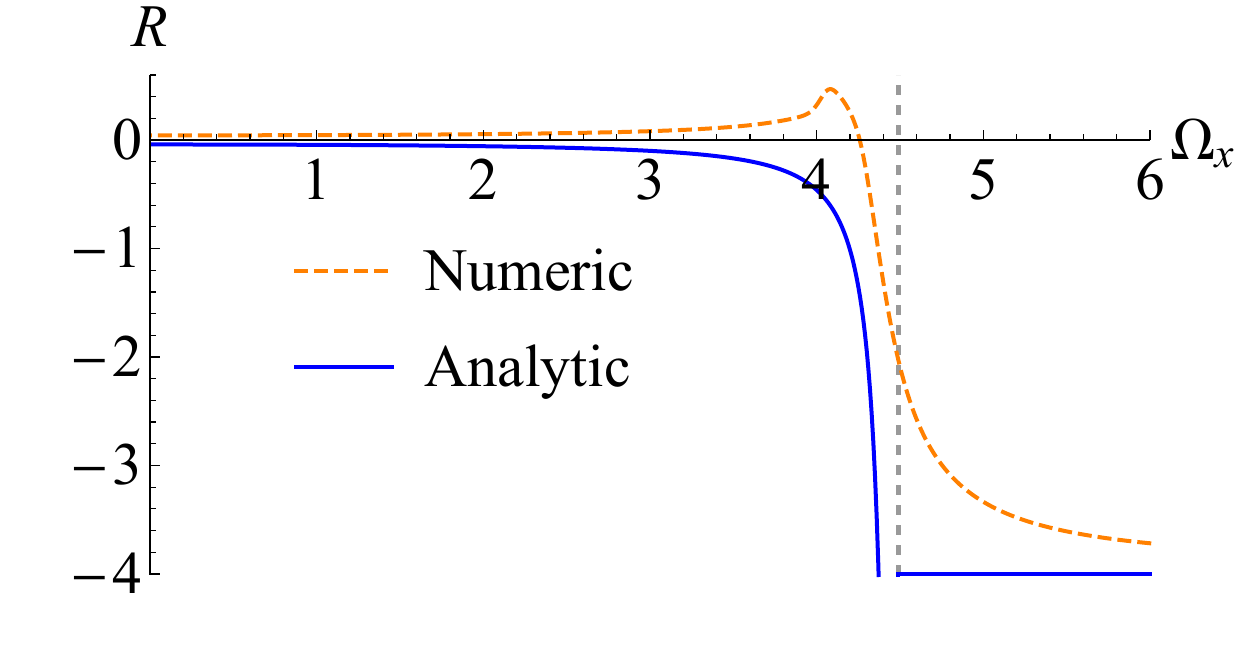} \\
(c) $g_{22}$ & (d) $R$
\end{tabular}
\caption{Comparison of the numeric QMT components and the scalar curvature with the analytic results for $\xi_y=2.3$ and $j=96$ for the highest energy state. The inset shows the $g_{11}$ component in logarithmic scale. The agreement is excellent except near to the QPT (dashed gray). Notice in the plots of $g_{11}$the difference between the analytical and numerical curves as $\Omega_x \rightarrow 0$.}
\label{Fig:CutVs}
\end{figure}

In Fig.~\ref{Fig:CutVs}, we show the QMT and its scalar curvature and compare them with their numeric counterparts. A good agreement is observed between the analytic and numeric results. 
The exception is the $g_{11}$ component, which exhibits a maximum in the numerical calculations at $\Omega_x =0$ which is not predicted by the Holstein-Primakoff approximation. It seems to be related with the crossover in the projection of $\hat{J}_x$ when $\Omega_x$ changes sign, as can be seen in Fig.~\ref{Fig:NumVsCoh}, which is clearly detected by $g_{11}$ but does not affect the scalar curvature, which approaches to zero in the broken phase both in the analytic and numerical analysis. The numerical evaluation of the metric
demands high precision, due to the near degeneracy of the
eigenvalues. To handle this issue specially designed computational
techniques are required \cite{doi:10.1137/1.9780898718027.ch27}.

The broken phase exhibits a Berry curvature, whose only component is given by
\begin{equation}
F_{12}=-\frac{2j+1}{4\xi_{y}^{2}\sqrt{4\xi_{y}^{2}-\Omega_{x}^{2}-1}}+\frac{16\xi_{y}^{2}-\Omega_{x}^{2}+1}{16\xi_{y}^{3}\left(4\xi_{y}^{2}-\Omega_{x}^{2}-1\right)}.
\end{equation}
We observe, as expected that the Berry curvature is also singular at the separatrix. On the other hand, the numeric analysis yields a zero Berry curvature. 
This discrepancy is a consequence of the rotation performed to align the highest energy state with the classical pseudospin, which introduced the crossed term in $\hat Q$ and $\hat P$ in Eq. (\ref{quadHam}). In the numerical calculations we would obtain a no vanishing Berry curvature rotating the collective pseudo-spin operators in Hamiltonian (\ref{Eq:H-LMG}), to have a linear term in $J_y$, as has been done in \cite{Cui,Yuan}.

\begin{figure}[ht]
\begin{tabular}{c c}
\includegraphics[width= 0.49 \columnwidth]{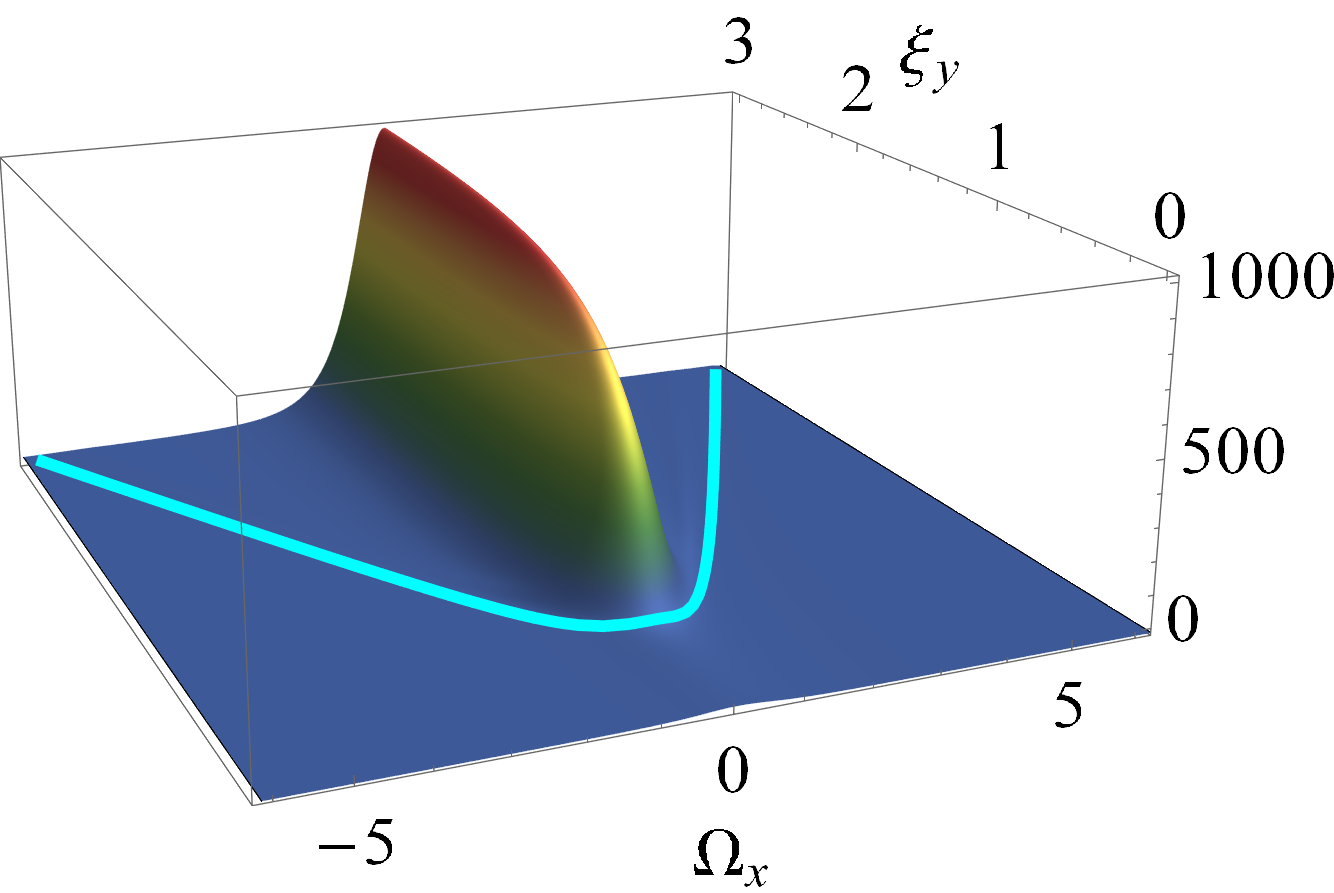} &
\includegraphics[width= 0.49 \columnwidth]{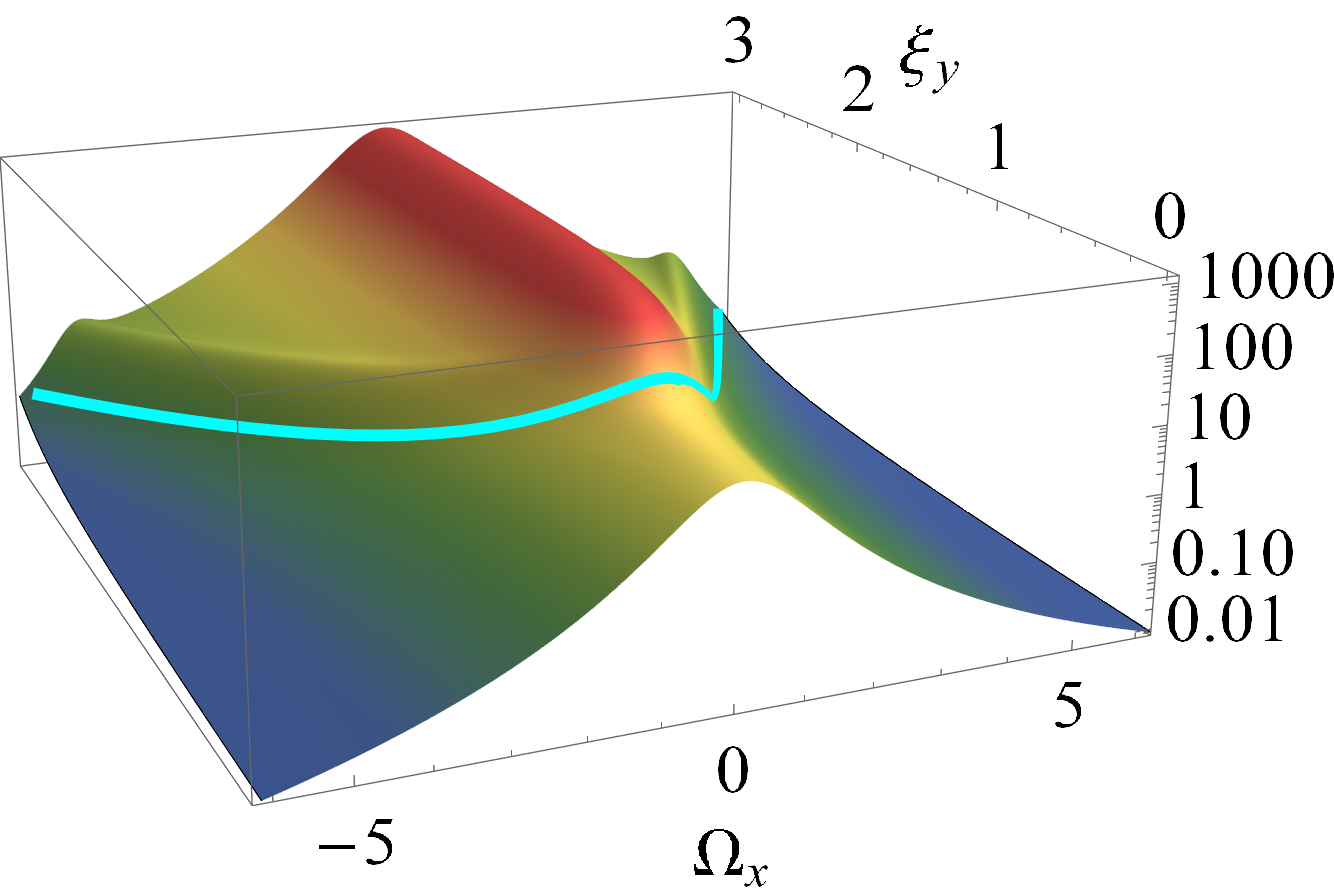} \\
(a) $g_{11}$ & (b) $g_{11}$ in log scale \\
\includegraphics[width= 0.49 \columnwidth]{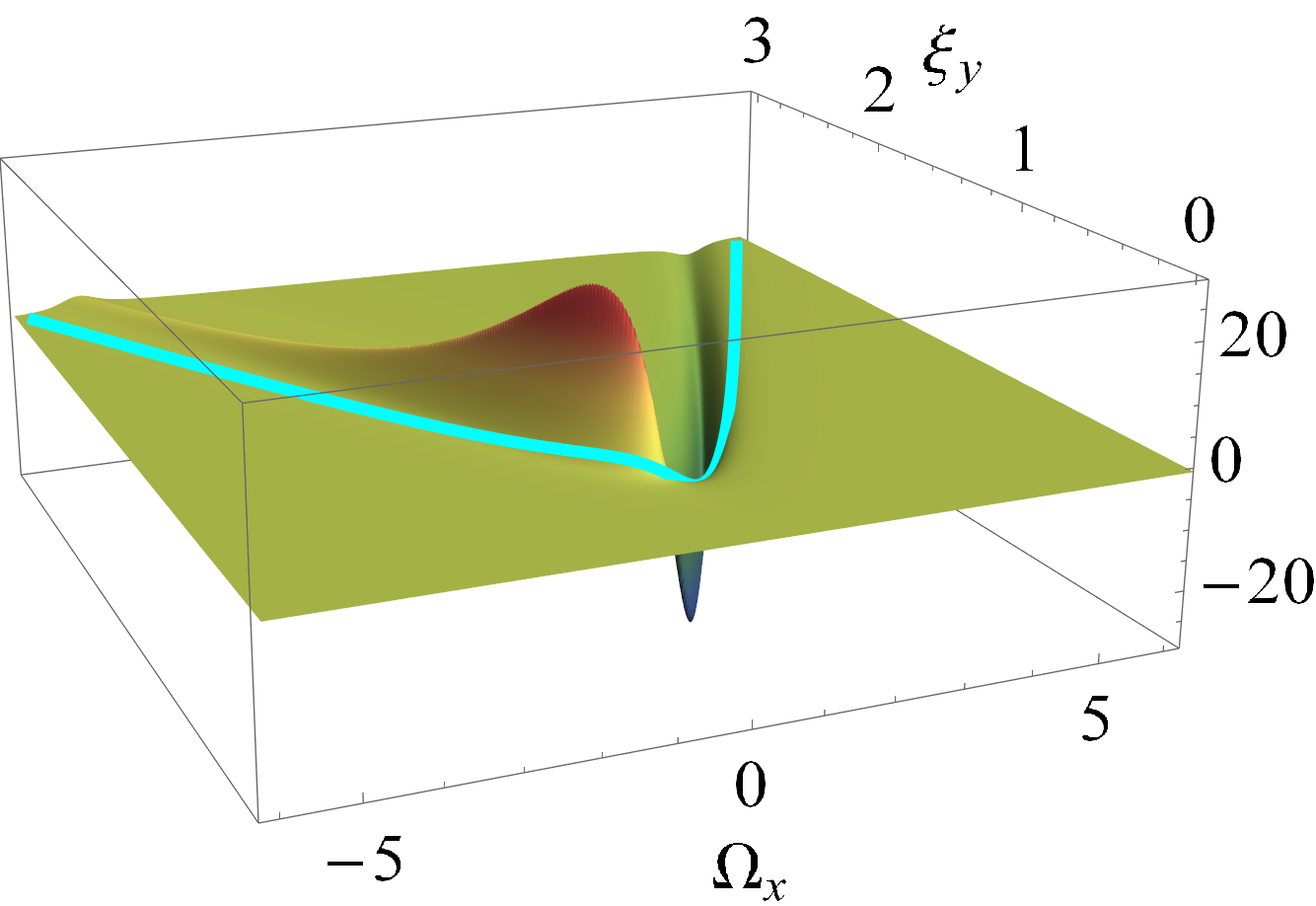} &  
\includegraphics[width= 0.49 \columnwidth]{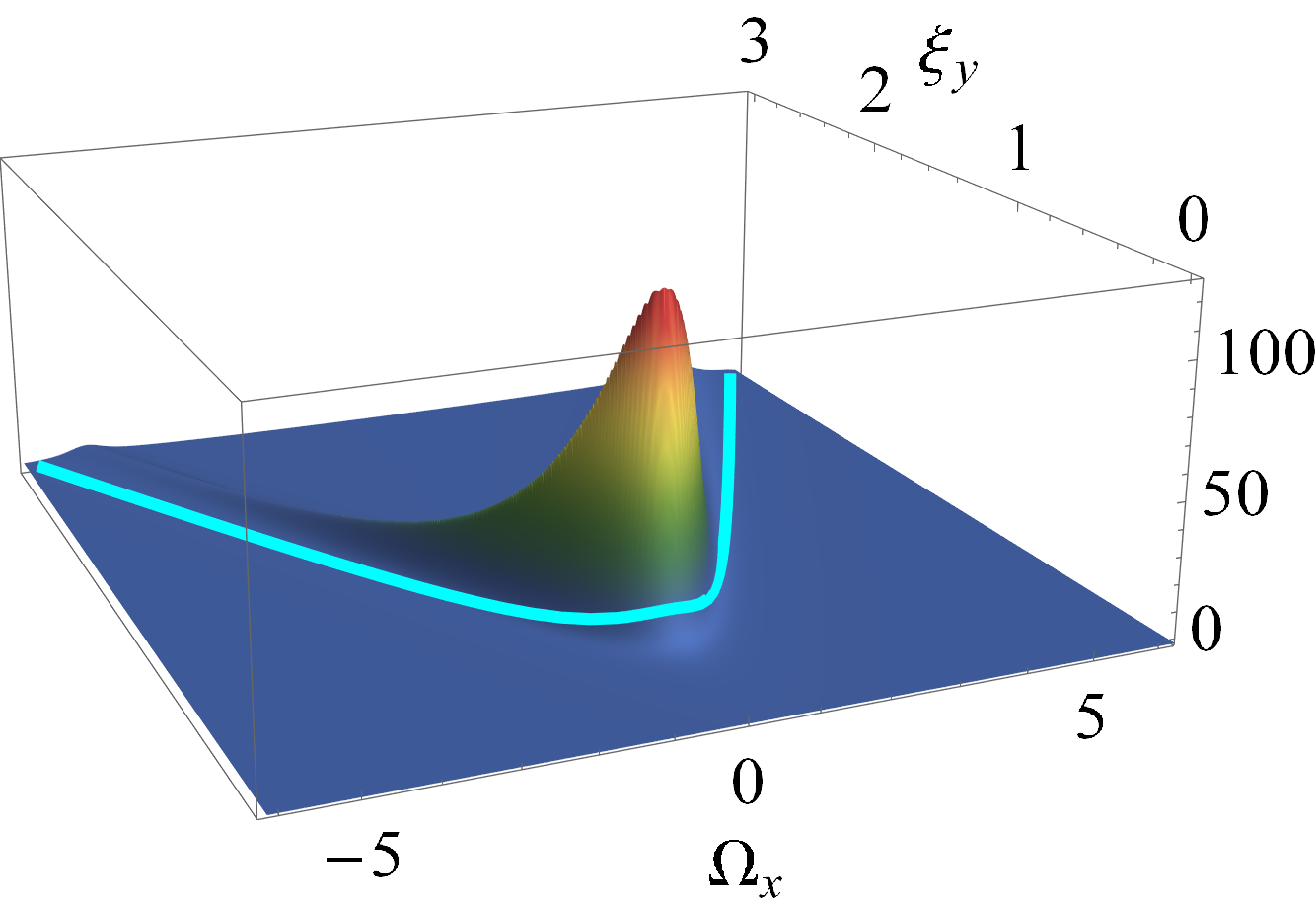} \\
(c) $g_{12}$ & (d) $g_{22}$  \\
\includegraphics[width= 0.49 \columnwidth]{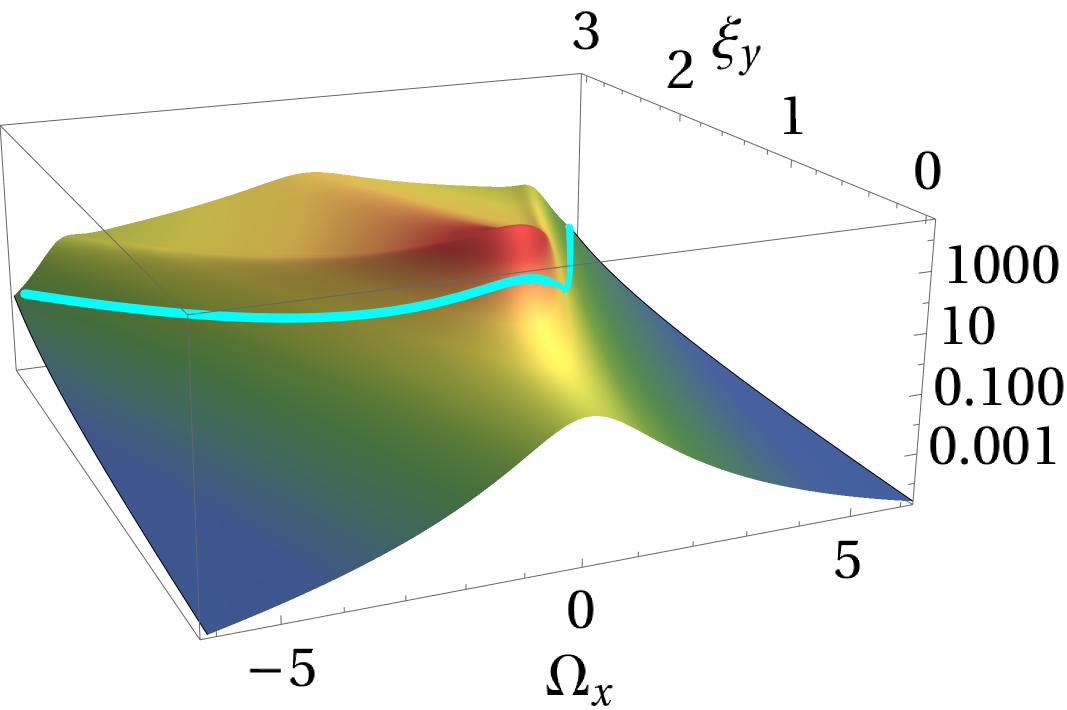} &  
\includegraphics[width= 0.49 \columnwidth]{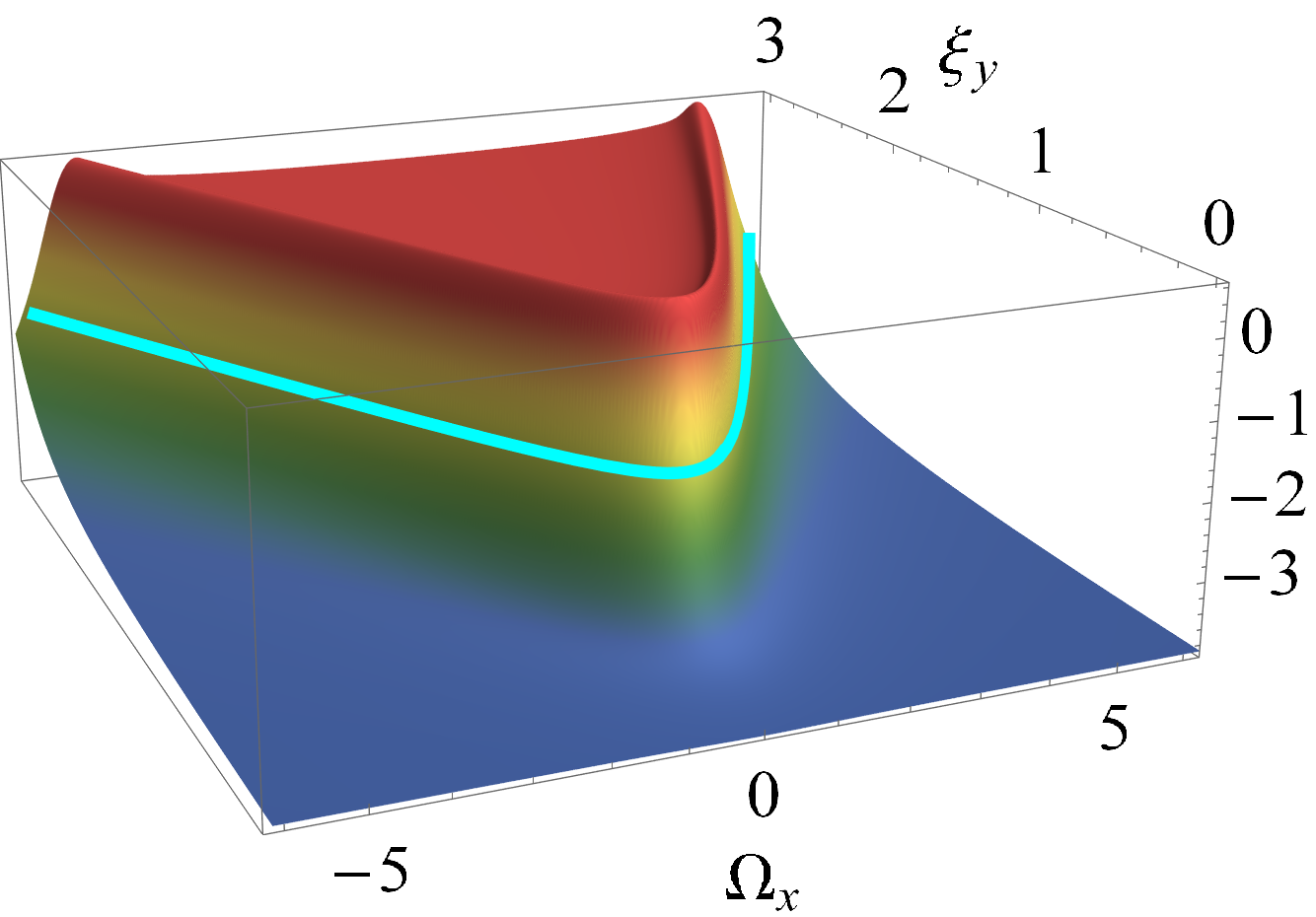} \\
(e) $g$  in log scale  & (f) $R$   
\end{tabular}
\caption{QMT components and scalar curvature for the highest energy state with $j=32$. The cyan line is the separatrix given in Fig.~\ref{Fig:LMG} (a) when $\xi_y=\frac{\sqrt{1+\Omega_x^2}}{2}$.}
 \label{Fig:REc23D}
\end{figure}

\begin{figure}[ht]
\begin{tabular}{c c}
\includegraphics[width= 0.49 \columnwidth]{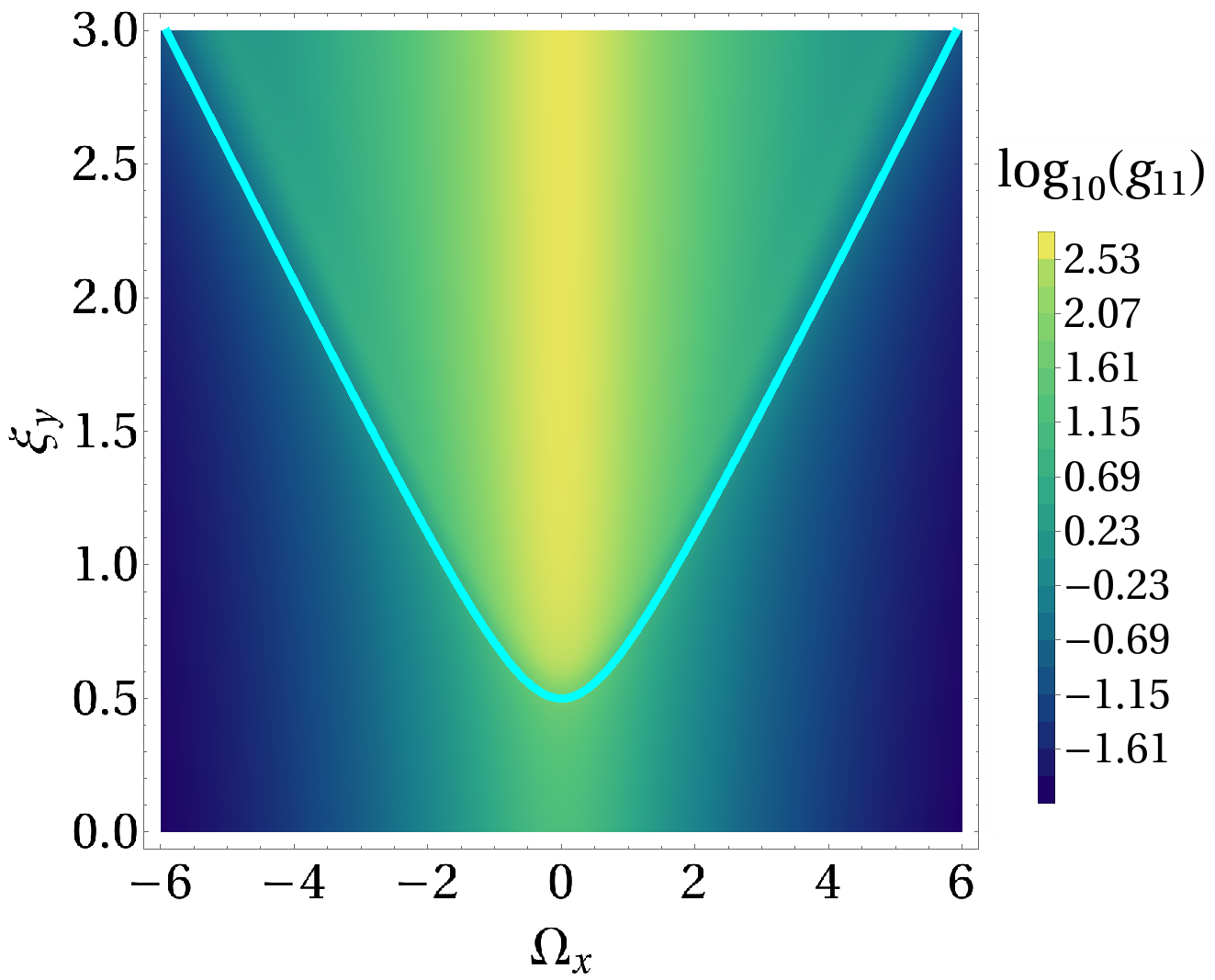} & \includegraphics[width= 0.49 \columnwidth]{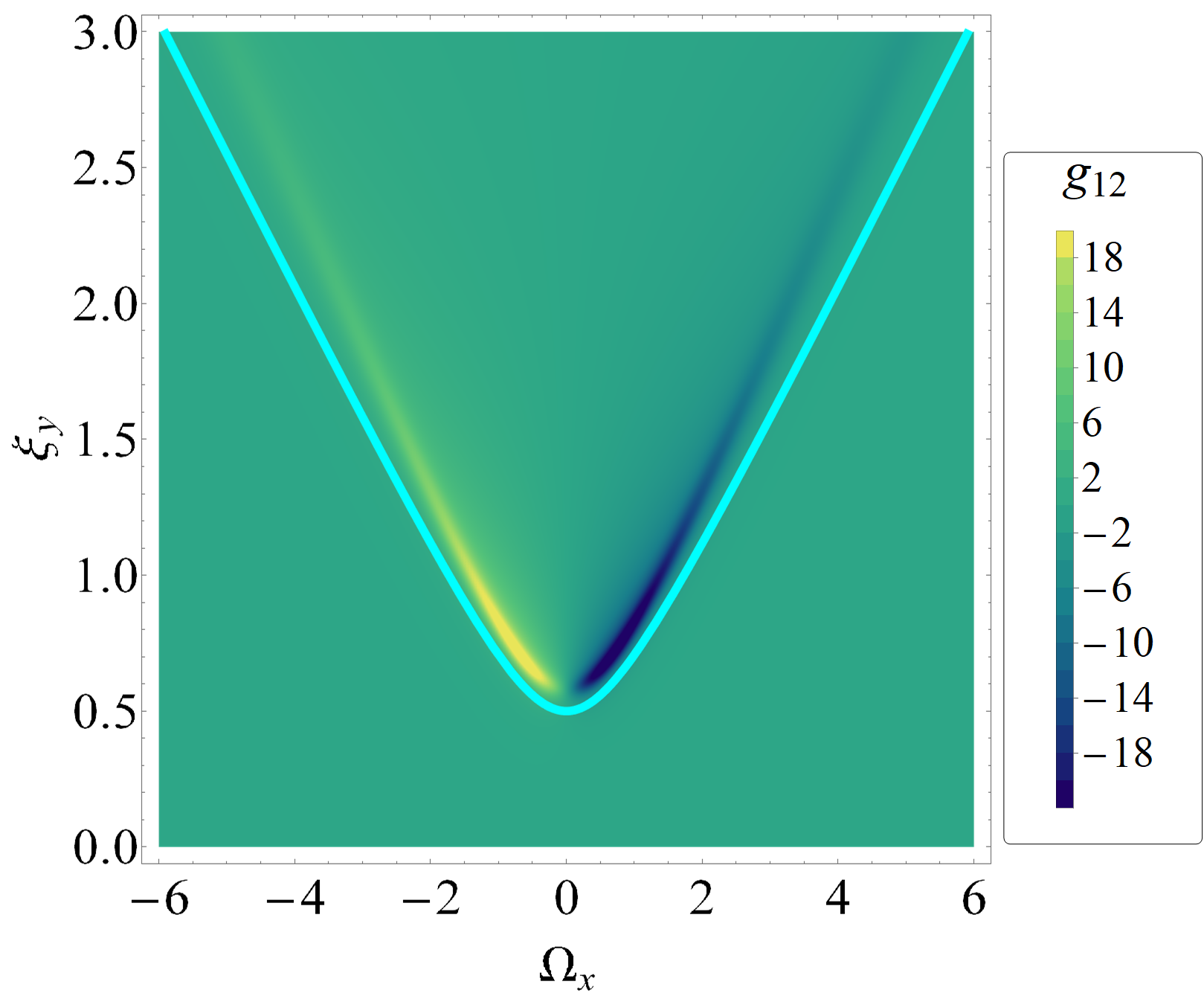} \\
(a) $g_{11}$ & (b) $g_{12}$ \\
\includegraphics[width= 0.49 \columnwidth]{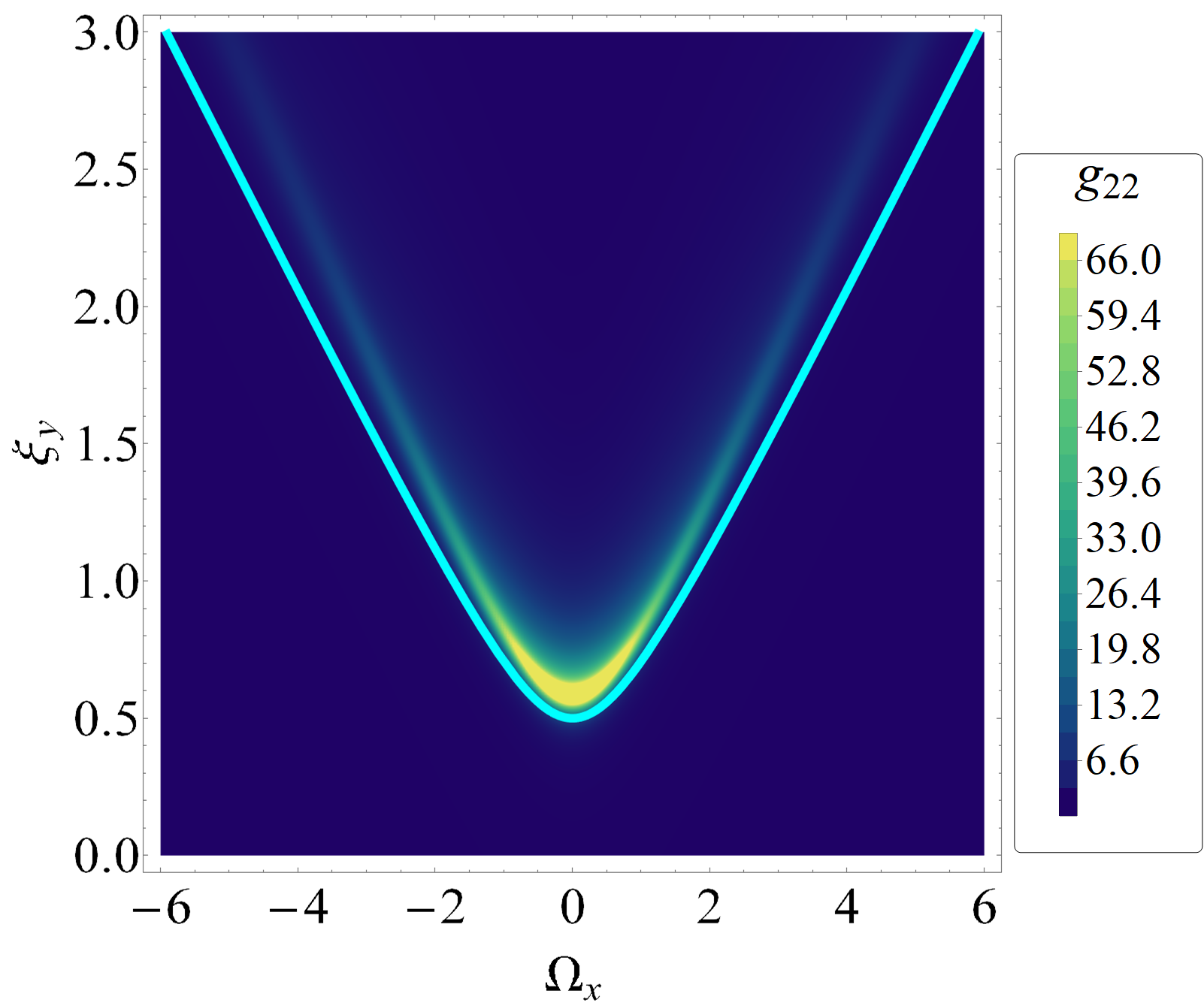} & \includegraphics[width= 0.49 \columnwidth]{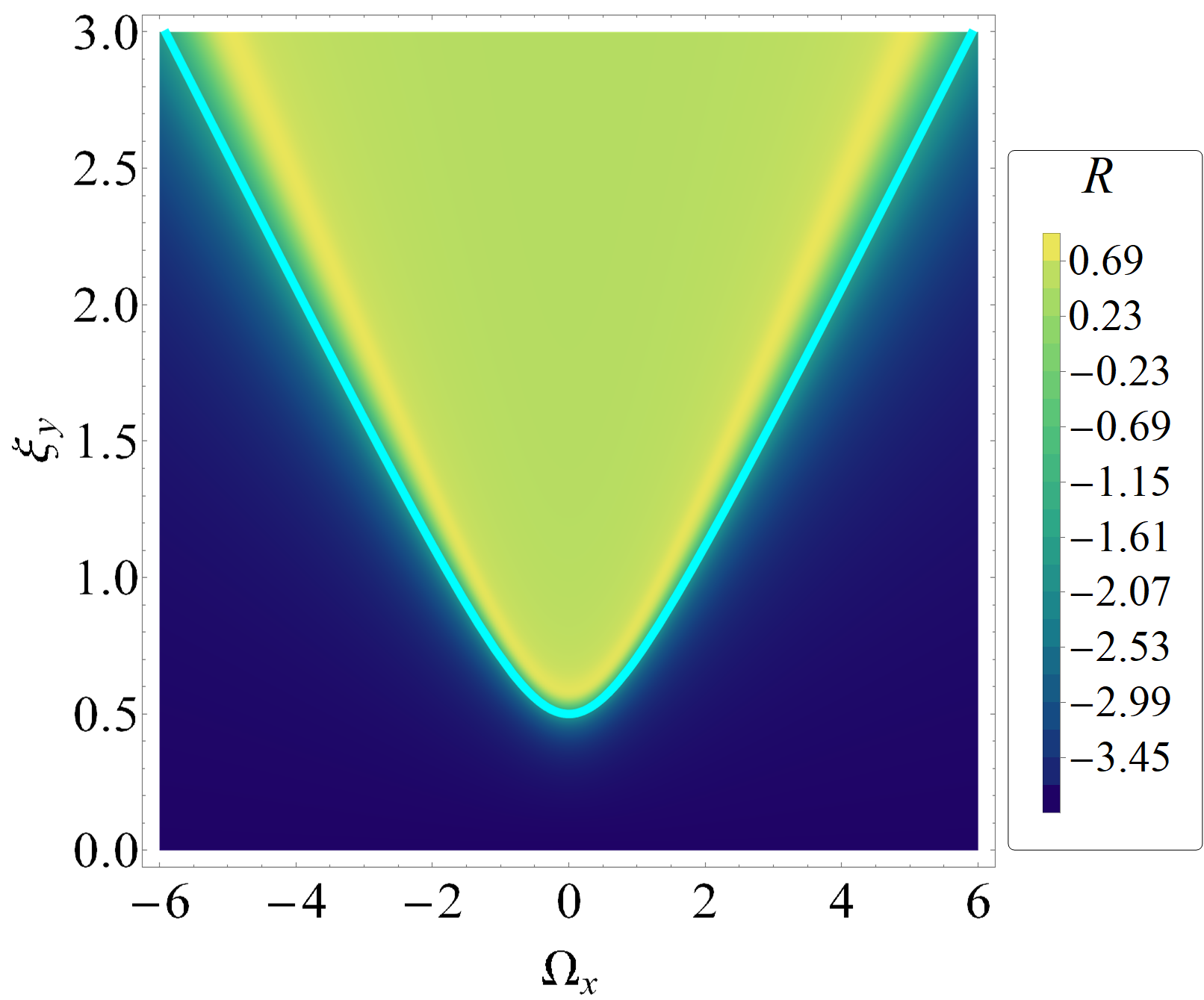} \\
(c) $g_{22}$ & (d) $R$  
\end{tabular} 
\caption{Maps of the QMT components and the scalar curvature for the highest energy state with $j=32$.}
\label{Fig:DP}
\end{figure}

In Fig.~\ref{Fig:REc23D}, the plots of the numeric QMT components and the scalar curvature for the highest energy state are shown in 3D, while Fig.~\ref{Fig:DP} contains their maps. We see that in the regular region where the Lyapunov exponent vanishes, the scalar curvature has a value around -4, just as the Holstein-Primakoff calculation predicts. However, near the region with a positive Lyapunov exponent, the scalar curvature begins to grow, passes the separatrix, takes small positive values, reaches a maximum until it descends to near zero values. Moreover, as the size of the system increases, the peak gets closer to the separatrix, and the transition region between the asymptotic values $-4$ on one side and $0$ on the other becomes thinner. 

How can we interpret the change of sign near the separatrix? Does it mean that there is a change in the topology of the parameter space and that it switches between a closed and an open shape? In order to answer this, we must recall that a fundamental quantity in the study of two-dimensional surfaces is the Gaussian curvature $K$, which is related to the scalar curvature as $K=R/2$. It is defined as the product of the two principal curvatures, $\kappa_1$ and $\kappa_2$, which quantify the bending of the surface along each direction~\cite{Spivak}. 
If both principal curvatures, $\kappa_1$ and $\kappa_2$, have the same sign, then the Gaussian curvature is positive and the local geometry is spherical. On the contrary, when both principal curvatures have opposite signs, the Gaussian curvature is negative and the surface is locally hyperbolic. A change in topology would imply that the metric becomes singular at some point, which means that its determinant should vanish there \cite{Hawking1978}. Since we do not observe that the metric determinant be equal to zero (see Fig. \ref{Fig:REc23D}), we conclude that a change in topology  
does not take place. Instead,
there is only a change in sign in one of the two principal curvatures, $\kappa_1$ or $\kappa_2$, producing a local change between dome-like and saddle-like shapes.

\begin{figure}[ht]
\begin{tabular}{c c}
\includegraphics[width= 0.49 \columnwidth]{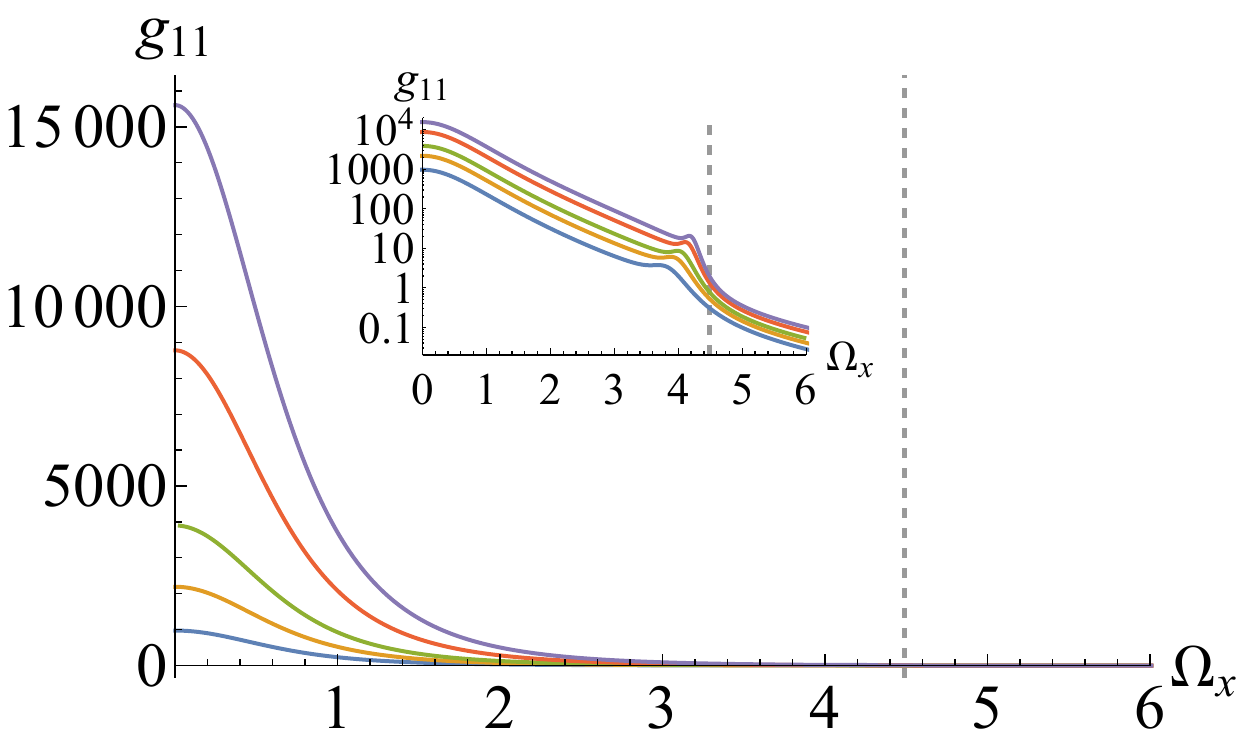} & \includegraphics[width= 0.49 \columnwidth]{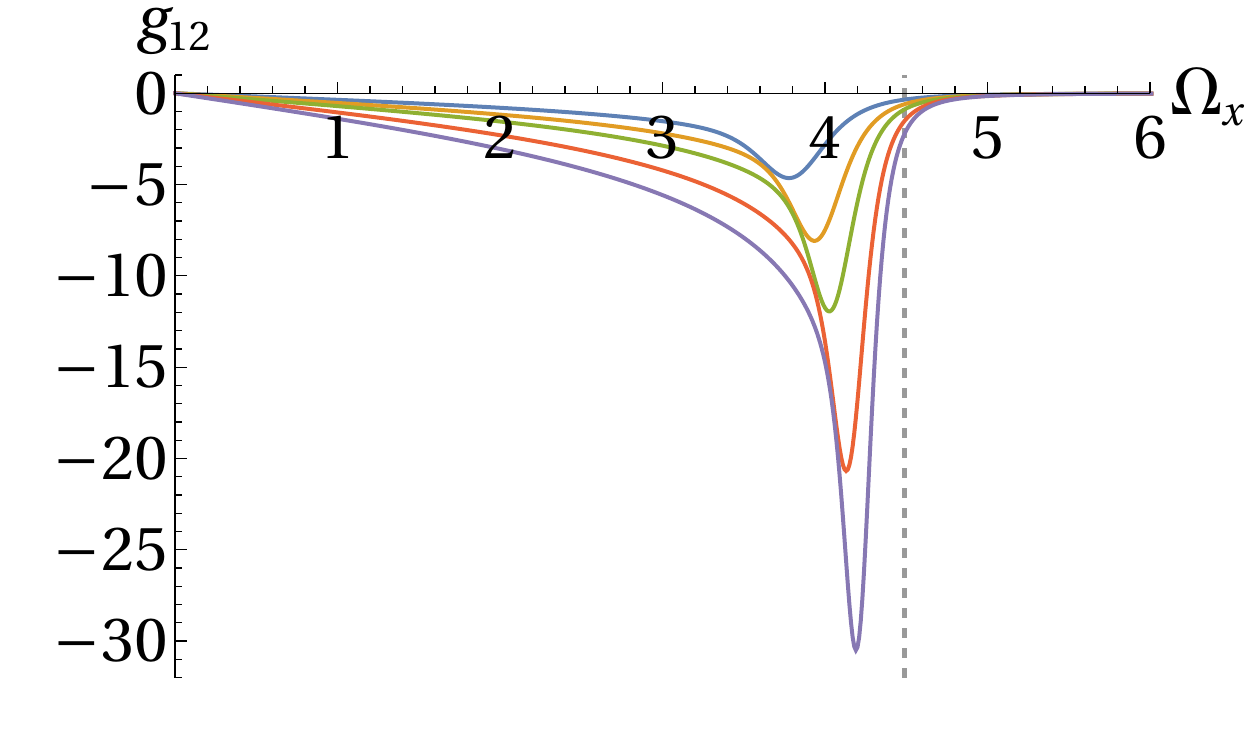} \\
(a) $g_{11}$ & (b) $g_{12}$ \\
\includegraphics[width= 0.49 \columnwidth]{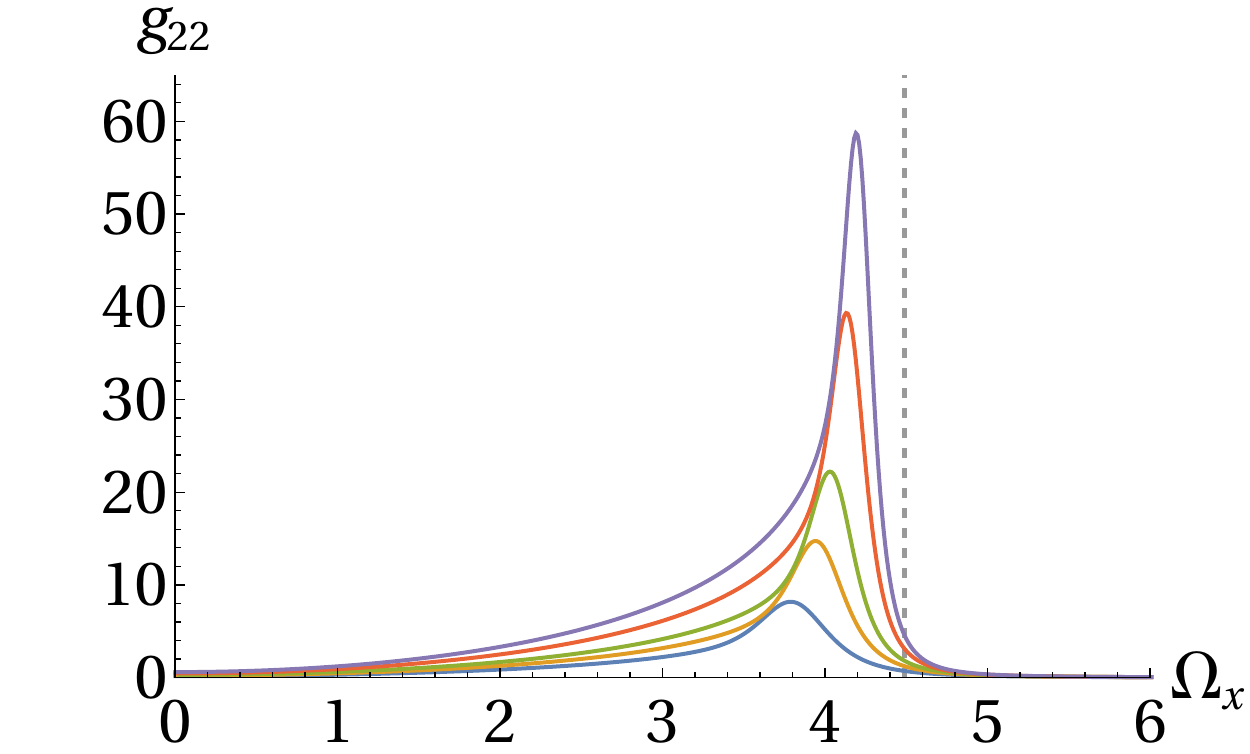} & \includegraphics[width= 0.49 \columnwidth]{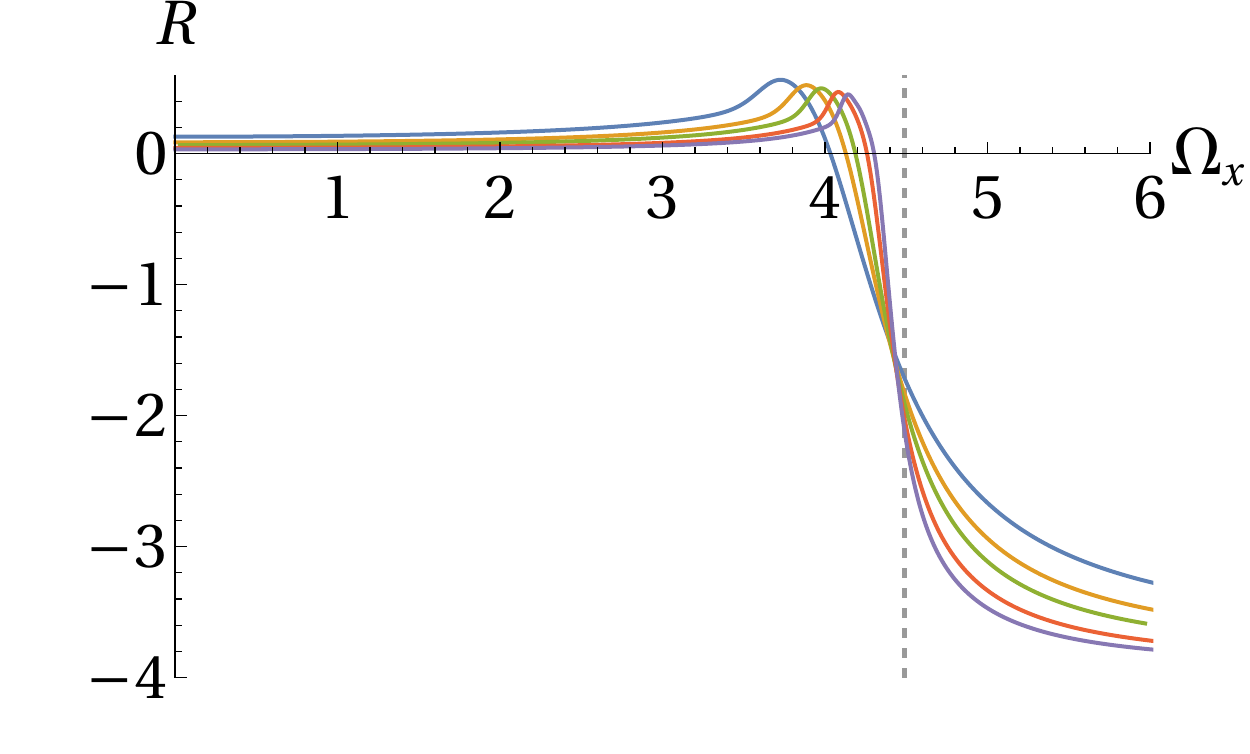} \\
(c) $g_{22}$ & (d) $R$
\end{tabular}
\caption{QMT components and scalar curvature for $\xi_y=2.3$ and $j=32,48,64,96,128$. The inset shows the $g_{11}$ component in logarithmic scale.}
\label{Fig:Cut}
\end{figure}

In Fig~\ref{Fig:Cut}, the plots of the numeric QMT and its scalar curvature are shown for various $j$'s when $\xi_y=2.3$. We can see how the peaks in the metric components and the curvature get closer to the critical value $\Omega_{xc}=4.490$ as $j$ increases. Also, the peaks become sharper, indicating that in the thermodynamic limit $j\to \infty$, the curvature's maximum will be identified with the change of sign of the curvature, indicating these two features as distinctive of the phase transition. In the next section, we carry out an analysis to extract more relevant information about this behavior.


\subsection{Analysis of the peaks}

In this section, we analyze the (positive or negative) peaks that are present in the numeric QMT components and the scalar curvature. 

\subsubsection{$\xi_y=2.3$}

\begin{figure}[ht]
\begin{tabular}{c c}
\includegraphics[width= 0.49 \columnwidth]{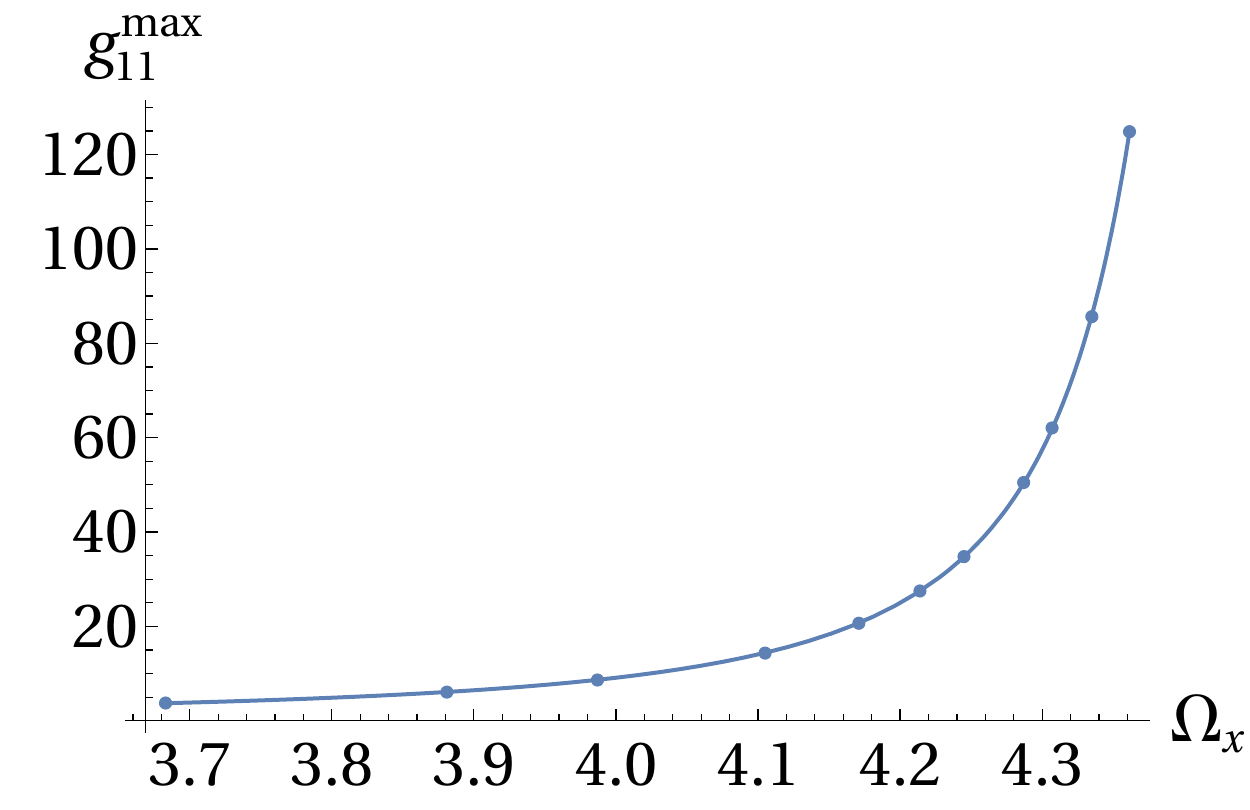} & \includegraphics[width= 0.49 \columnwidth]{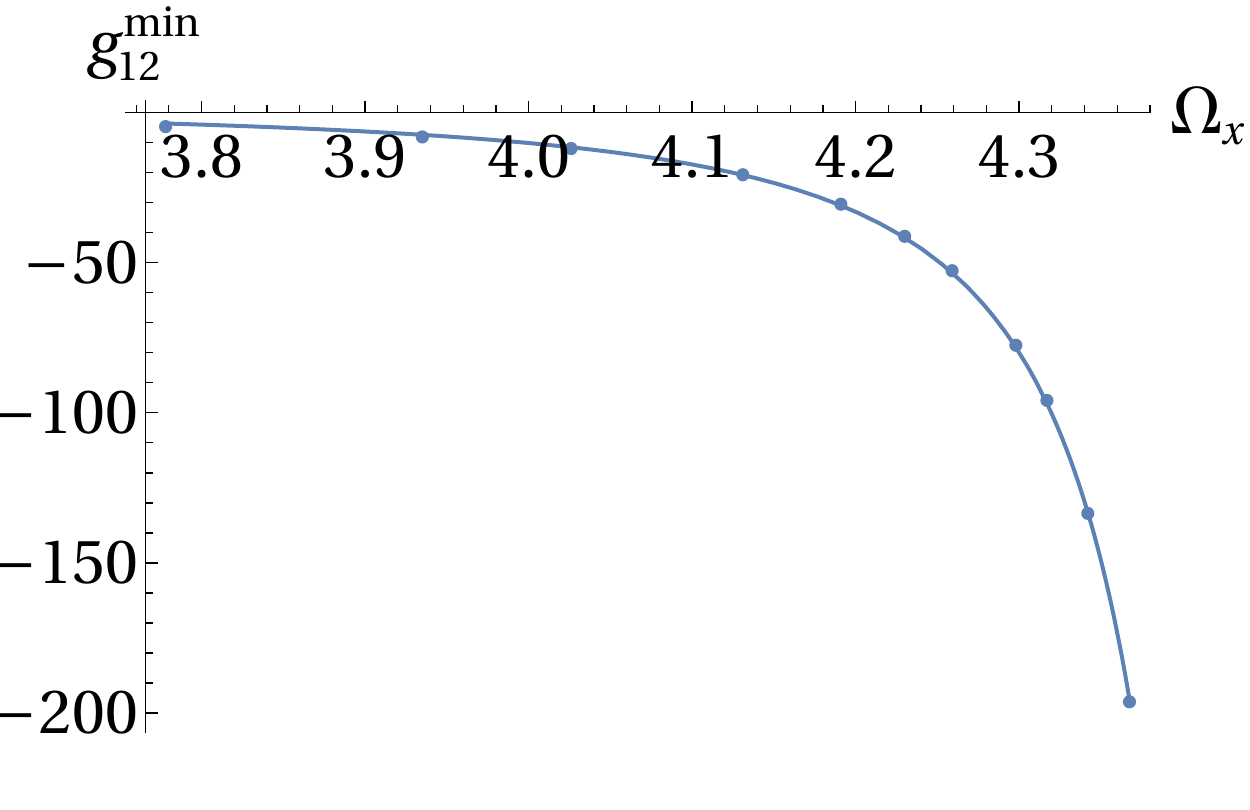} \\
(a) & (b) \\
\includegraphics[width= 0.49 \columnwidth]{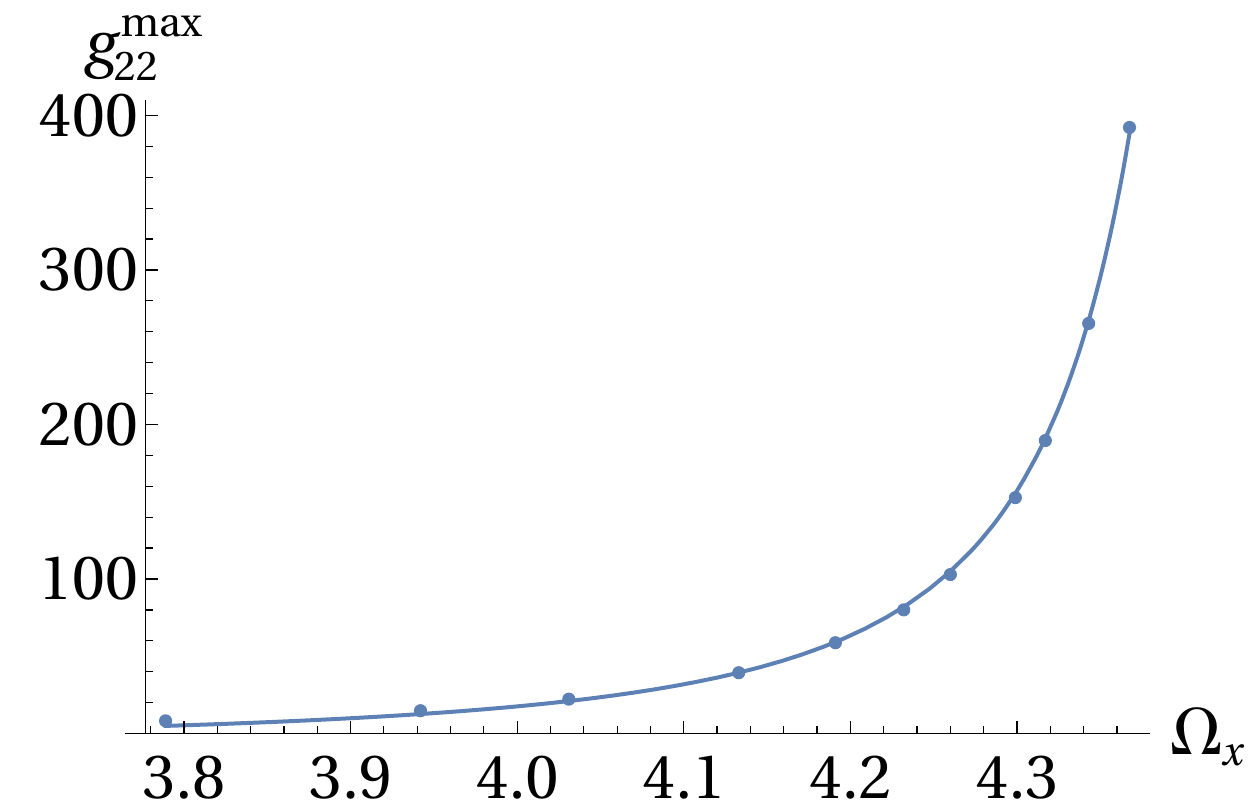} & \includegraphics[width= 0.49 \columnwidth]{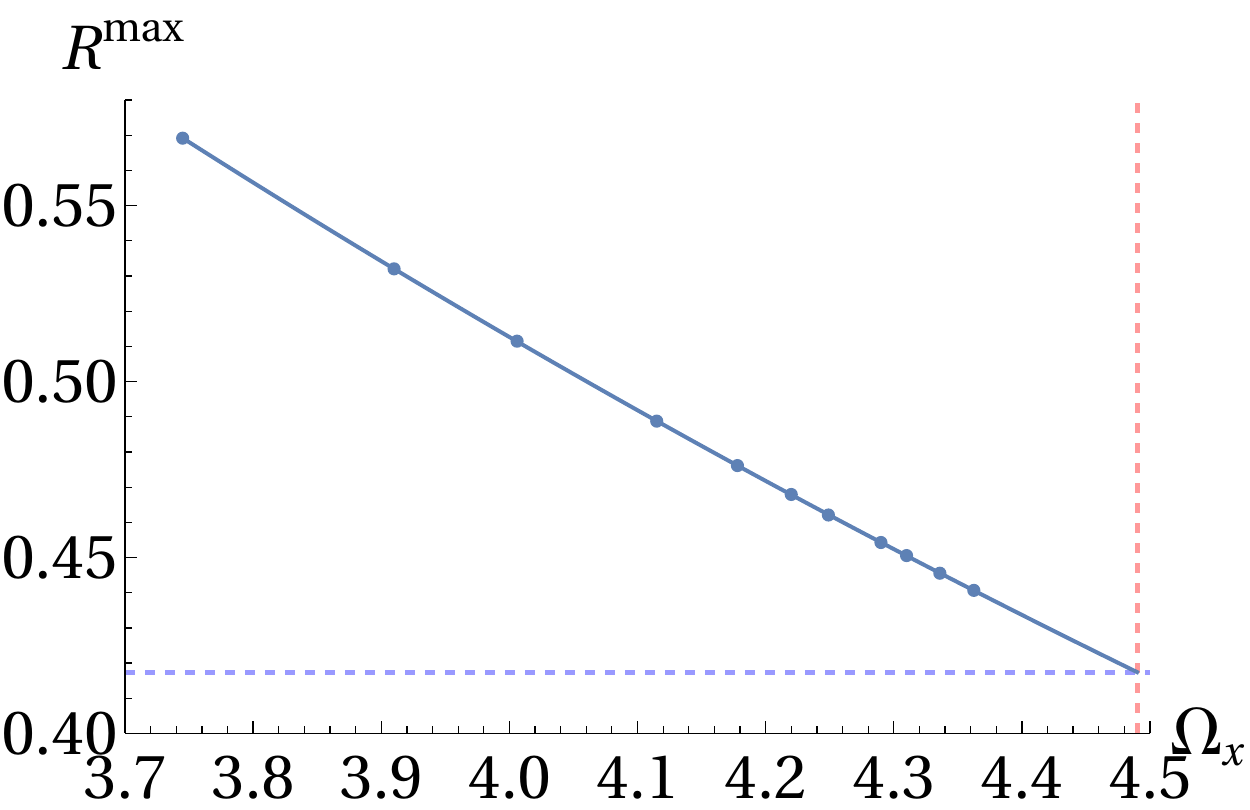} \\
(c) & (d)
\end{tabular}
\caption{Behavior of the maximum of each metric component and scalar curvature with respect to $\Omega_x$ for $\xi_y=2.3$. The points correspond to $j=32, 48, 64, 96, 128, 160, 192, 256, 300, 384, 512$.}
\label{Fig:PeaksOmegax}
\end{figure}

In Fig.~\ref{Fig:PeaksOmegax}, we plot the extrema (maximum or minimum) of every metric component and of the scalar curvature as a function of $\Omega_x$ for $\xi_y=2.3$ and increasing values of $j$. The functions used to fit the data are:
\begin{align}
g_{11}^{\rm max}&=0.582+\frac{2.051}{(\Omega_x-4.490)^2}, \nonumber \\
g_{12}^{\rm min}&=2.249-\frac{2.966}{(\Omega_x-4.490)^2}, \nonumber \\
g_{22}^{\rm max}&=-7.170+\frac{5.931}{(\Omega_x-4.490)^2}, \nonumber \\
R^{\rm max}&=-0.083+2.470\,\mathrm{e}^{-0.356\Omega_x}.
\end{align}

We see that at the stationary point, $\Omega_{xc}=4.490$, the metric components are singular, just as the analytic formulas~(\ref{QMTsym}) predict. On the other hand, the maximum of the scalar curvature at the stationary point takes the value $0.416$. This is an indication that the maximum of the scalar curvature persists in the thermodynamic limit. 

\begin{figure}[ht]
\begin{tabular}{c c}
\includegraphics[width= 0.49 \columnwidth]{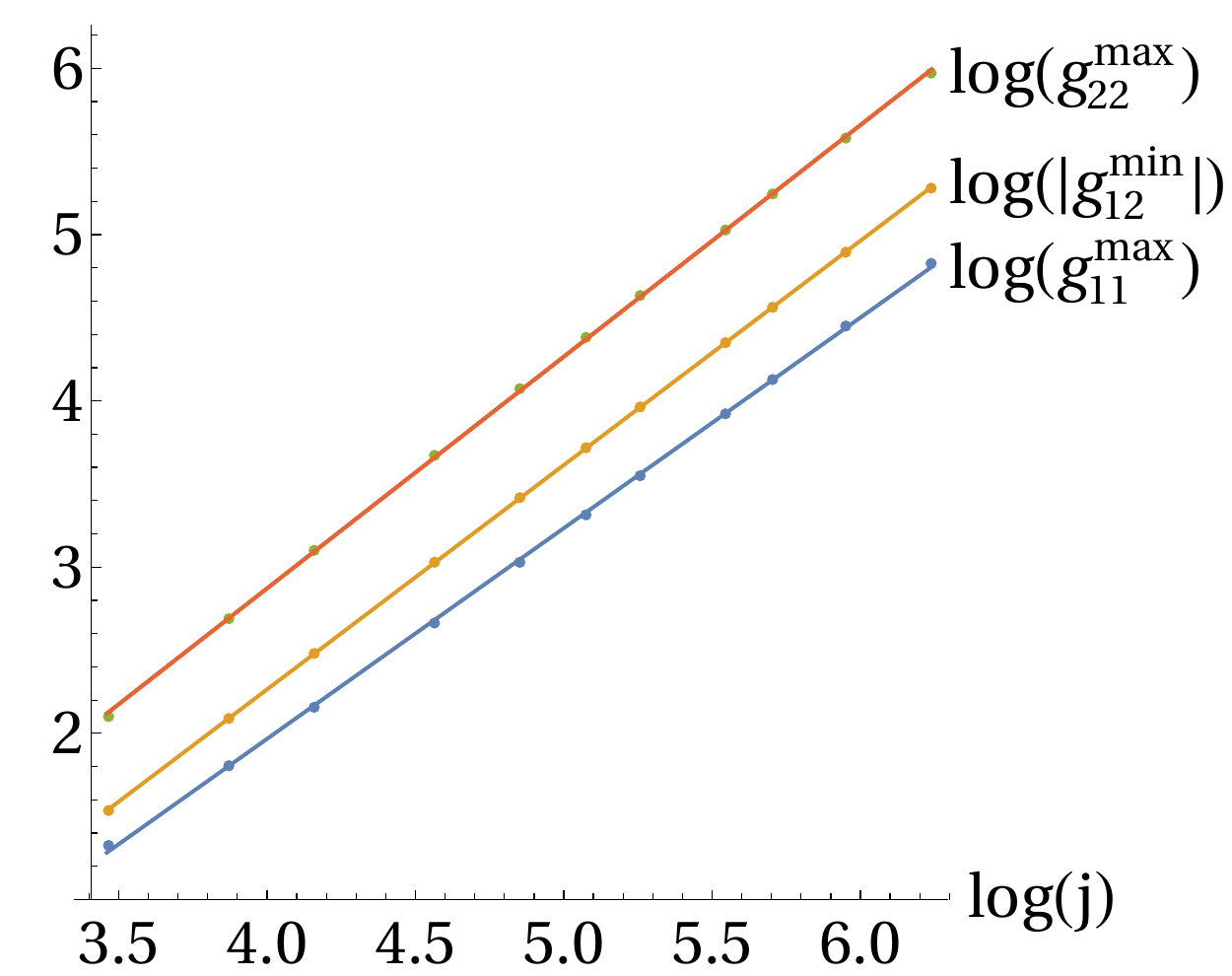} & \includegraphics[width= 0.49 \columnwidth]{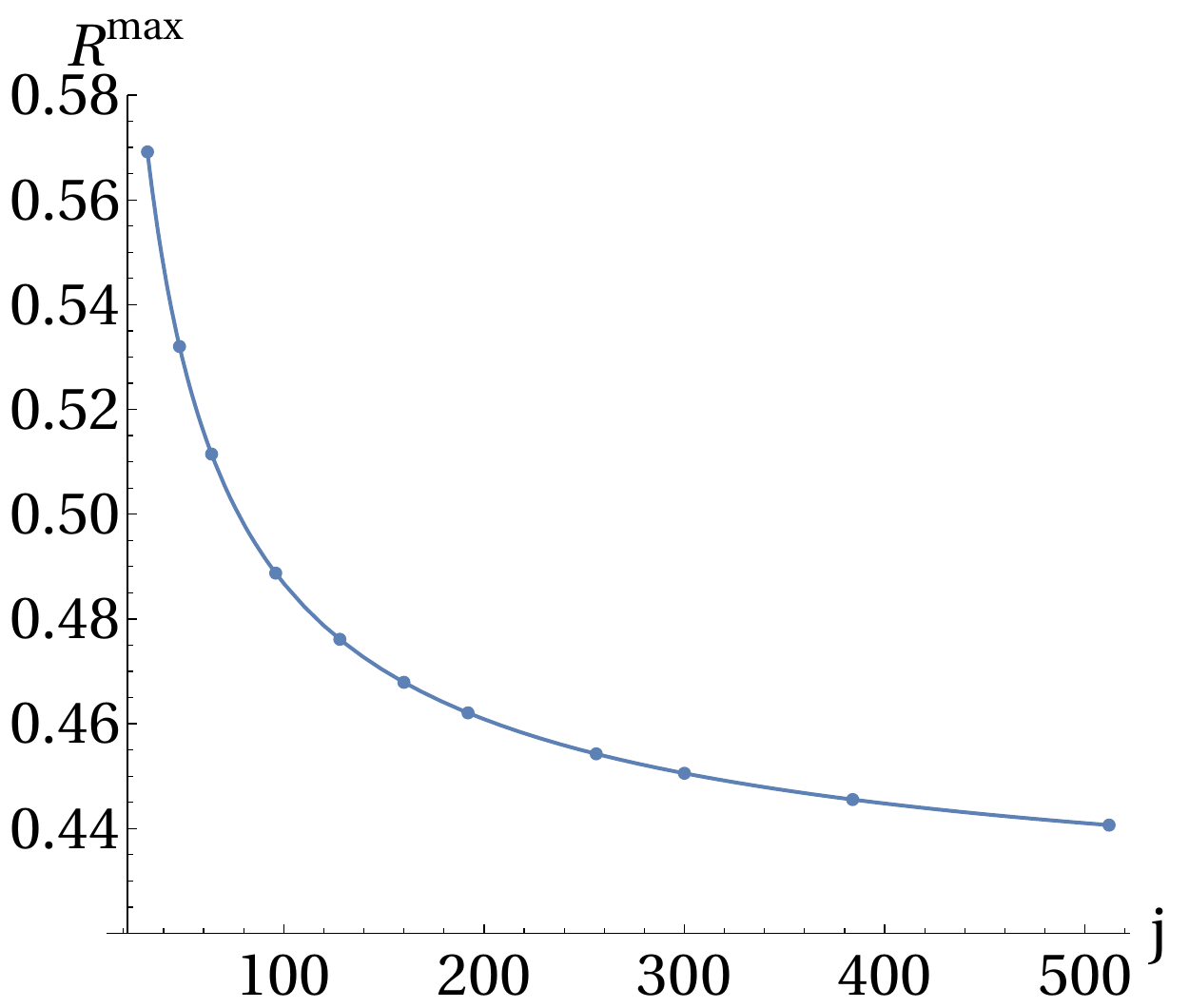} \\
(a) Maxima of $|g_{ij}|$. & (b) $R^{\rm max}$.
\end{tabular}
\caption{Maximum of each metric component and the scalar curvature with respect to $j$ for $\xi_y=2.3$.}
\label{Fig:Peaksj}
\end{figure}

Next, in Fig.~\ref{Fig:Peaksj}, we show the peaks of QMT components and the scalar curvature as functions of $j$. In this case, the functions used to fit the data are:
\begin{align}
\log(g_{11}^{\rm max})&=-3.102+1.267\log(j), \nonumber \\
\log(g_{12}^{\rm min})&=-3.134+1.349\log(j), \nonumber \\
\log(g_{22}^{\rm max})&=-2.702+1.394\log(j), \nonumber \\
R^{\rm max}&=0.418+\frac{1.563}{(j-0.913)^{0.680}}.
\end{align}

In the thermodynamic limit, $j\rightarrow\infty$, the metric components diverge, whereas the scalar curvature approaches $0.418$. This confirms that $R$ is not singular across the QPT, although it has a discontinuity as the tendency in Fig.~\ref{Fig:Cut} shows. As a consequence, it can be said that the singularity that appears in the metric is removable and it is not a true singularity of the parameter space.

\subsubsection{$\Omega_x=0$}

\begin{figure}[ht]
\begin{tabular}{c c}
\includegraphics[width= 0.49 \columnwidth]{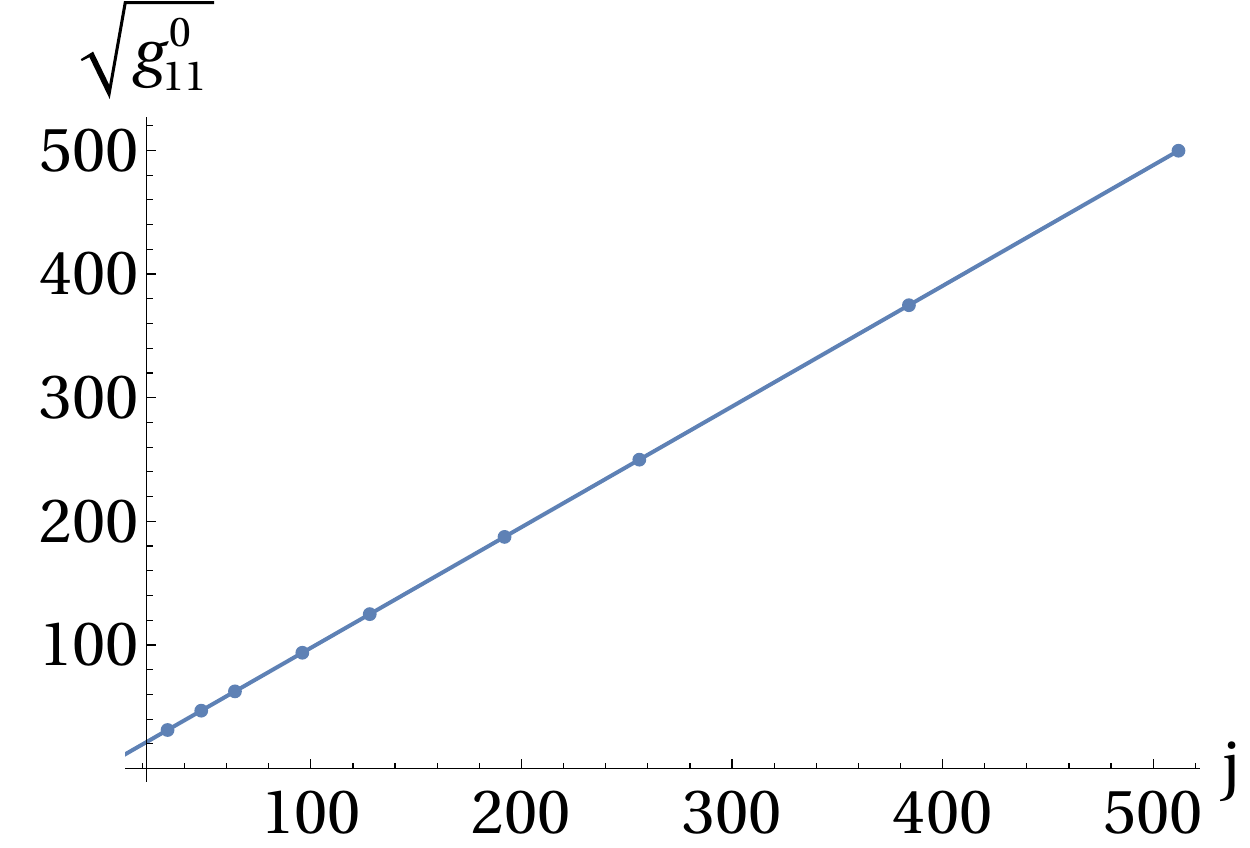} & \includegraphics[width= 0.49 \columnwidth]{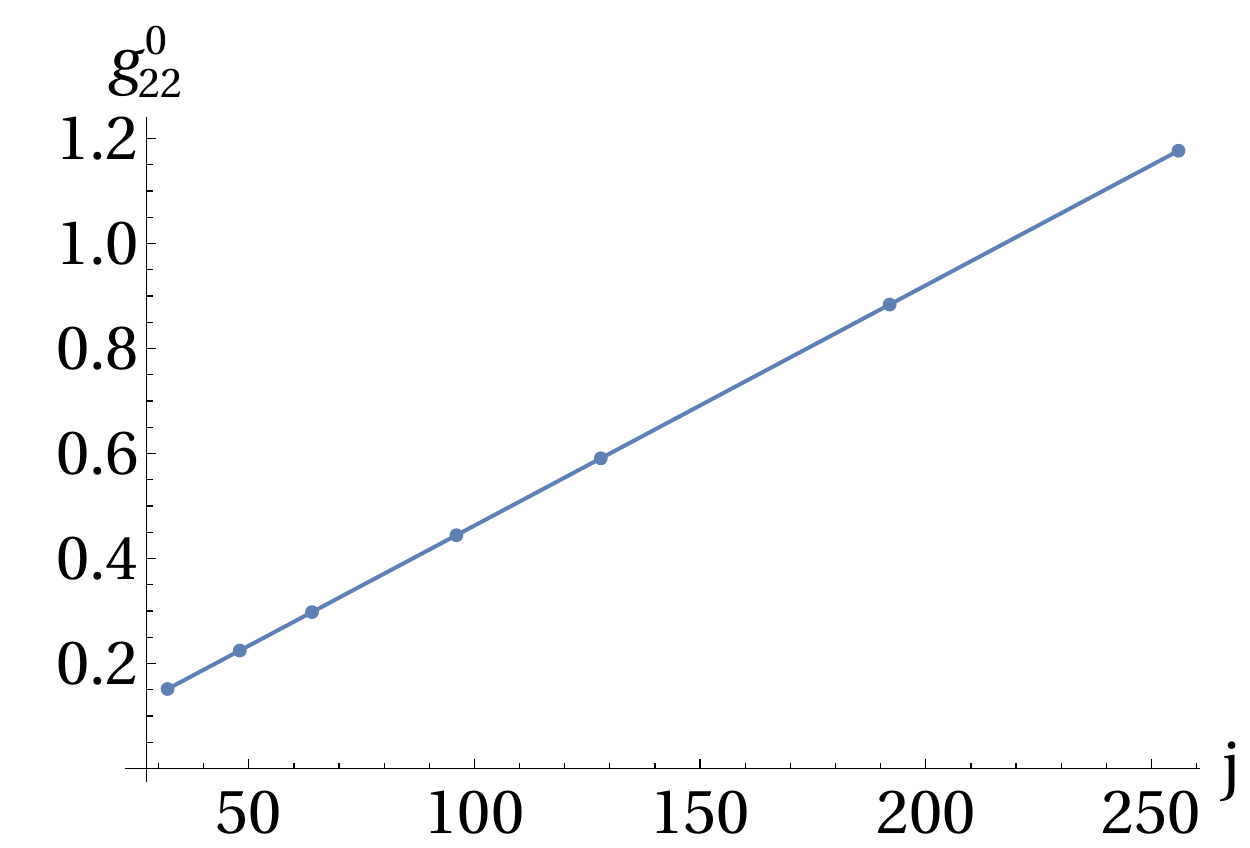} \\
(a) & (b) \\
\includegraphics[width= 0.49 \columnwidth]{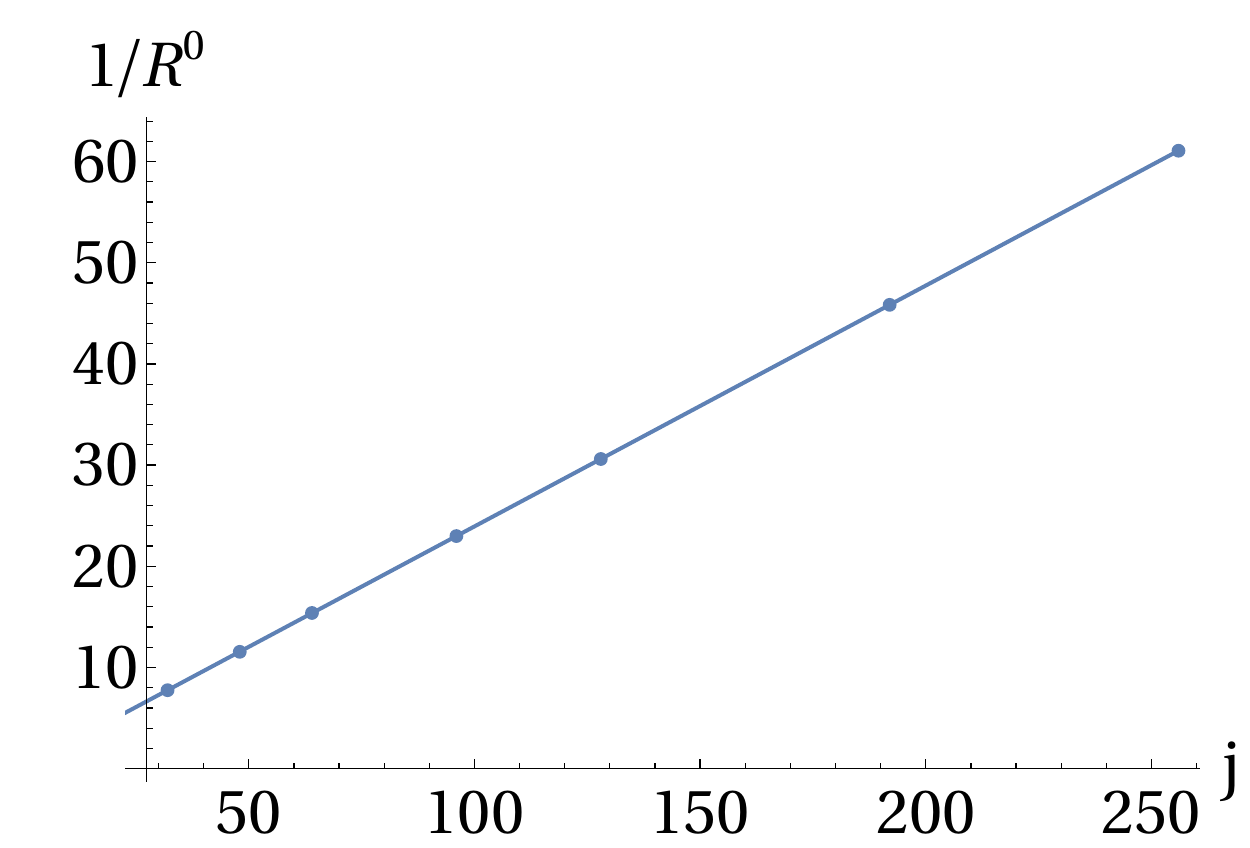} &  \\
(c)
\end{tabular}
\caption{Plots of $g_{11}$, $g_{22}$ and $R$ as functions of $j$ for $\Omega_x=0$ and $\xi_y=2.3$.}
\label{Fig:Peaks0}
\end{figure}

Now, we analyze the value of the metric components $g_{11}$ and $g_{22}$, as well as the scalar curvature when $\Omega_x=0$.  The plots are shown in Fig.~\ref{Fig:Peaks0} and the functions that fit the data are:
\begin{align}
g_{11}^0&=(-0.013+0.976 j)^2, \nonumber \\
g_{22}^0&=0.005+0.005j, \nonumber \\
R^0&=\frac{1}{0.131+0.238j}.
\end{align}
Notice that $g_{12}$ does not appear because its value at $\Omega_x=0$ is zero. Remarkably, the scalar curvature goes to $0$ as $j\rightarrow\infty$, which is precisely the prediction of the coherent-state approach of Appendix~\ref{appCoh} (see Eq.~\ref{scalarCoh}).

\subsubsection{ $\xi_y=0.5$ }

\begin{figure}[ht]
	\begin{tabular}{c c}
		\includegraphics[width= 0.49 \columnwidth]{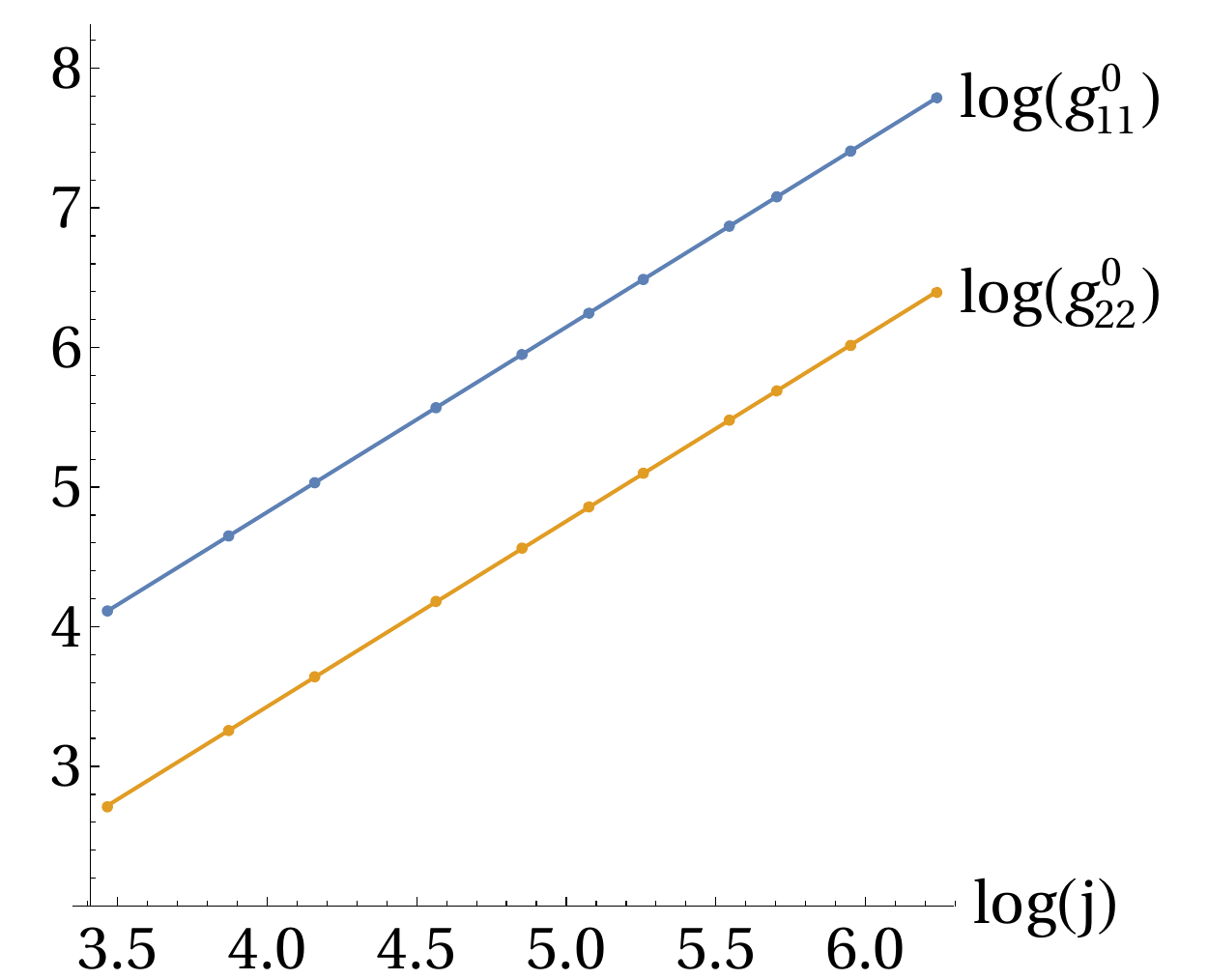} & \includegraphics[width= 0.49 \columnwidth]{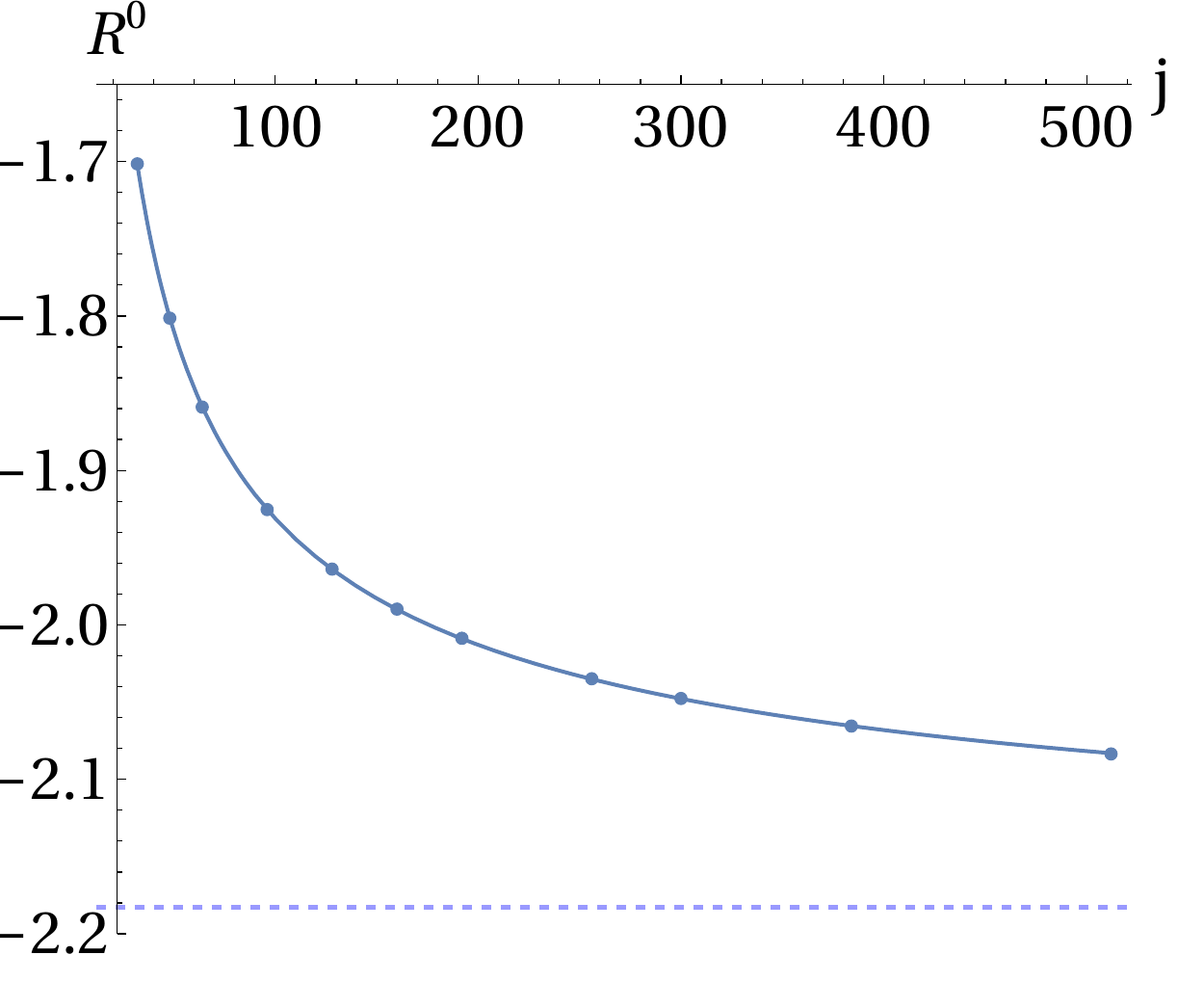} \\
		(a) Maxima of $g_{ij}$. & (b) $R^{0}$.
	\end{tabular}
	\caption{Maximum of each metric component and the scalar curvature with respect to $j$ for $\xi_y=0.5$.}
	\label{Fig:Peaksj05}
\end{figure}

Finally, in Fig.~\ref{Fig:Peaksj05}, we show the peaks of the metric components $g_{11}$ and $g_{22}$, as well as the scalar curvature for $\xi_y=0.5$ as functions of $j$. This case is particularly interesting since, at $\xi_y=0.5$, the peaks of the QMT and the scalar curvature occur at $\Omega_x=0$, regardless of the value of $j$. The functions used to fit the data in Fig.~\ref{Fig:Peaksj05} are:
\begin{align}
	\log(g_{11}^{0})&=-0.480 + 1.325\log(j), \nonumber \\
	\log(g_{22}^{0})&=-1.881 + 1.327\log(j), \nonumber \\
	R^{0}&=-2.183+\frac{3.430}{(j^2-6.206)^{0.284}}.
\end{align}

In this case, in the thermodynamic limit, $j\rightarrow\infty$, the metric components also diverge, whereas the scalar curvature approaches $-2.183$. This is a limiting point, over the separatrix,  where the scalar curvature takes an intermediate value between the asymptotic ones in the two phases ($0$ and $-4$).  Once again, this confirms that $R$ is not singular across the QPT, there is not a true singularity at this point in the parameter space.


\section{Conclusions}

In this paper, we have studied the geometry of the parameter space of a LMG model. The LMG model that we chose has a non-degenerate metric across the parameter space, as opposed to that considered in Ref.~\cite{SarkarPRE2012}, where the quantum metric was obtained only for the symmetric phase of the model, and the other phase was not studied due to the vanishing metric determinant. This model allowed us to compute the scalar curvature in both phases. 

We first introduced the QMT and the scalar curvature, which contain the relevant geometric information. Then, we carried out a classical mean-field analysis employing Bloch coherent states. This analysis allowed the determination of the Hamiltonian stationary points and their classification according to their stability properties. Two points were of relevance in our work: the ground state and the highest energy state. The first state did not show any singular behavior in the parameter space under consideration ($\xi_y>0$). On the other hand, the second state exhibited a QPT, and as a result, the QMT components turned out singular. Both states were studied from an analytic and numeric perspective. The analytic treatment was performed under the truncated Holstein-Primakoff transformation, which yields a quadratic approximation to the system near the chosen state. The numeric computation used the original Hamiltonian and employed exact diagonalization to extract the QMT and then numerical differentiation over a mesh to find the scalar curvature.

The analytic and numeric results match for the ground state. On the other hand, the $g_{11}$ component of the numeric QMT for the highest energy state shows a peak for small values of $\Omega_x$, which the analytic counterpart does not possess. The reason for this behavior may be related to the fact that the classical system is not invariant under $Q\rightarrow-Q$, i.e., it has a broken symmetry, as can be seen in Eq.~\ref{Eq:LMGclassical} and is related to the crossover in the projection of $\hat J_x$ for the change of sign of $\Omega_x$,  indicated in  Fig.~\ref{Fig:NumVsCoh}. In contrast to this, the analytic computation is carried out under the truncated Holstein-Primakoff approximation, which yields a quadratic Hamiltonian possessing the $Q\rightarrow-Q$ symmetry. Another difference between the analytic and numeric results is the appearance of a Berry curvature in the analytic case, which also detects the QPT but is not present in the numeric case. The same Berry curvature could be obtained in the numerical calculations rotating the collective pseudo-spin operators in the Hamiltonian (\ref{Eq:H-LMG}).
 
For the highest energy state, we find that far from the separatrix, both the analytic and numeric results show that the scalar curvature takes the value $-4$ in the symmetric phase. In the broken symmetry phase, the scalar curvature tends to zero in the thermodynamic limit $j \rightarrow \infty$. In this limit, the metric components diverge over the separatrix, whereas the scalar curvature $R$ always takes finite values across the QPT in the numerical calculations. This contrasts with its singular behavior predicted employing the truncated Holstein-Primakoff, which becomes unreliable in close vicinity of the phase transition. We conclude that the singularity appearing in the metric is removable and is not a true singularity of the parameter space. However, in terms of the scalar curvature, the second-order QPT is indicated as a sudden change of sign.


\acknowledgments

We acknowledge the support of the Computation Center - ICN, in particular of Enrique Palacios, Luciano D\'\i az, and Eduardo Murrieta.
Daniel Guti\'errez-Ruiz is supported by a CONACyT Ph.D. scholarship No. 332577. Diego Gonzalez was partially supported by a DGAPA-UNAM postdoctoral fellowship and by Consejo Nacional de Ciencia y Tecnolog\'ia (CONACyT), M\'exico, Grant No. A1-S-7701. Jorge Ch\'avez is supported by a DGAPA-UNAM postdoctoral fellowship. This work was partially supported by DGAPA-PAPIIT Grants No. IN103919 and IN104020.

\appendix

\section{Bloch coherent states}\label{appCoh}

The $SU(2)$ coherent states are parameterized by the complex number $z=\tan\frac{\theta}{2}\,{\rm e}^{-i \phi}$ and are given by
\begin{equation}
|z\rangle=\frac{{\rm e}^{z\hat{J}_+}|j,-j\rangle}{(1+|z|^2)^j}=\sum_{m=-j}^{j} c_{m}^{(j)}|j,m\rangle, 
\end{equation}
where
\begin{equation}
c_{m}^{(j)}:=\binom{2j}{j+m}^{1/2} \sin^{j+m}\frac{\theta}{2} \cos^{j-m}\frac{\theta}{2}\,{\rm e}^{-i (j+m)\phi}.
\end{equation}

The angles on the Bloch sphere are functions of the parameters, i.e., $\theta=\theta(x)$ and $\phi=\phi(x)$, with $x=(\Omega_x,\xi_y)$. The coherent state of the maximum has the coordinates:
\begin{itemize}
\item $(\theta_4,\phi_4)=\left(\arccos\left(-\frac{1}{2\xi_y}\right),\arccos\left(-\frac{\Omega_x}{\sqrt{4\xi_y^2-1}}\right)\right)$ for $\xi_y>\frac{\sqrt{1+\Omega_x^2}}{2}$
\item $(\theta_1,\phi_1)=\left(\arccos\left(-\frac{1}{\sqrt{1+\Omega_x^2}}\right),0\right)$ for $\xi_y\leq\frac{\sqrt{1+\Omega_x^2}}{2}$
\end{itemize}

Therefore, the classical angular momentum vector is given by $\vec{J}=j(\sin\theta\cos\phi,\sin\theta\sin\phi,\cos\theta)$. 

The quantum geometric tensor for the coherent state $|z\rangle$ is given by
\begin{align}
Q^{(z)}_{ij}&=\langle\partial_i z|\partial_j z\rangle-\langle\partial_i z|z\rangle\langle z|\partial_j z\rangle \nonumber \\
&=\sum_{m=-j}^{j}\partial_i c_{m}^{*(j)} \partial_j c_{m}^{(j)}-\sum_{m=-j}^{j}\partial_i c_{m}^{*(j)} c_{m}^{(j)}\sum_{n=-j}^{j}c_{n}^{*(j)}\partial_j c_{n}^{(j)}.
\end{align}

\begin{figure}[ht]
\begin{tabular}{c c}
\includegraphics[width= 0.49 \columnwidth]{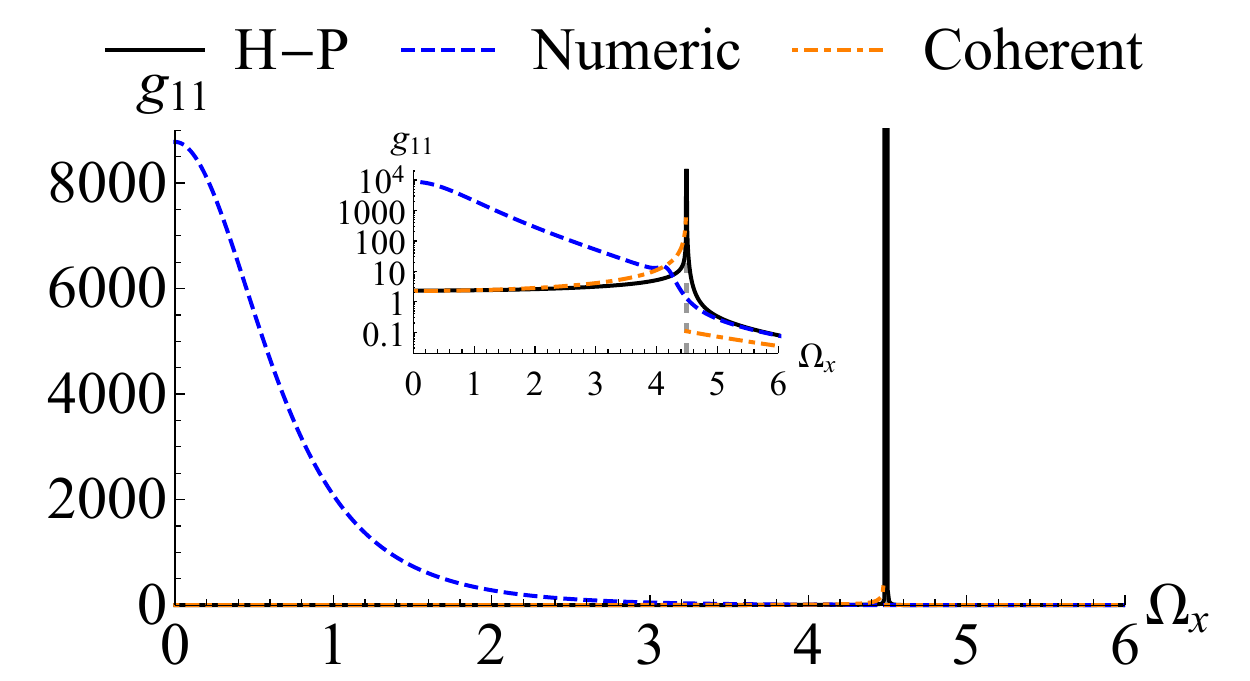} &
\includegraphics[width= 0.49 \columnwidth]{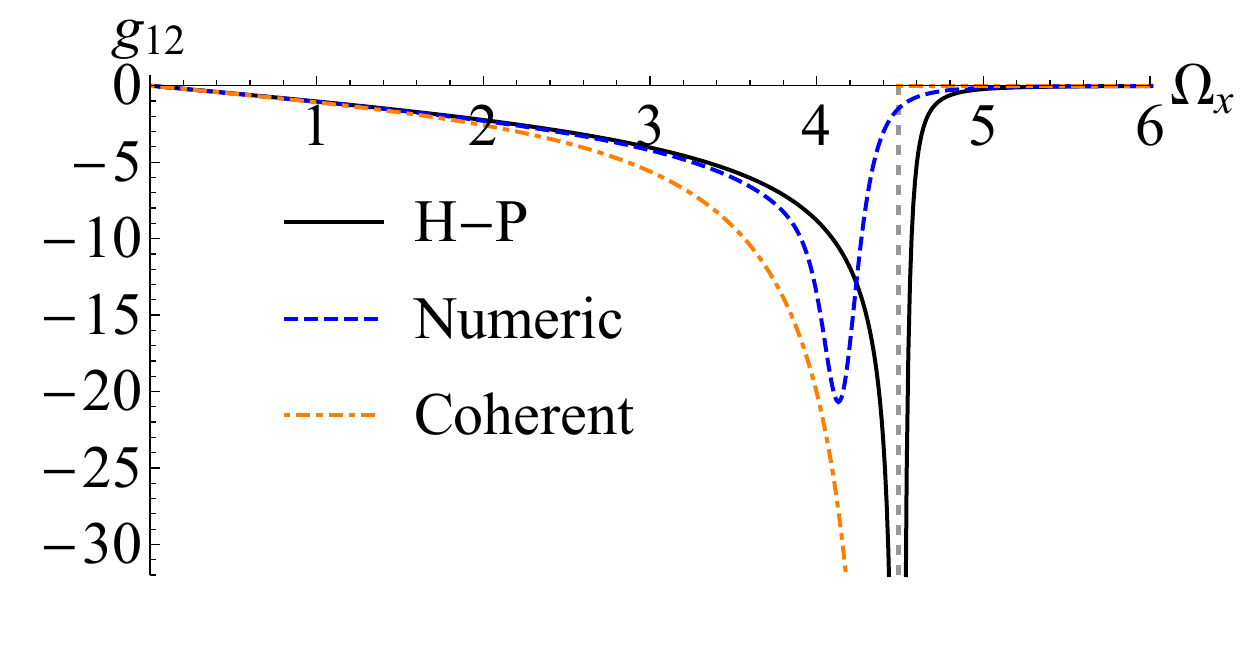} \\
(a) & (b) \\
\includegraphics[width= 0.49 \columnwidth]{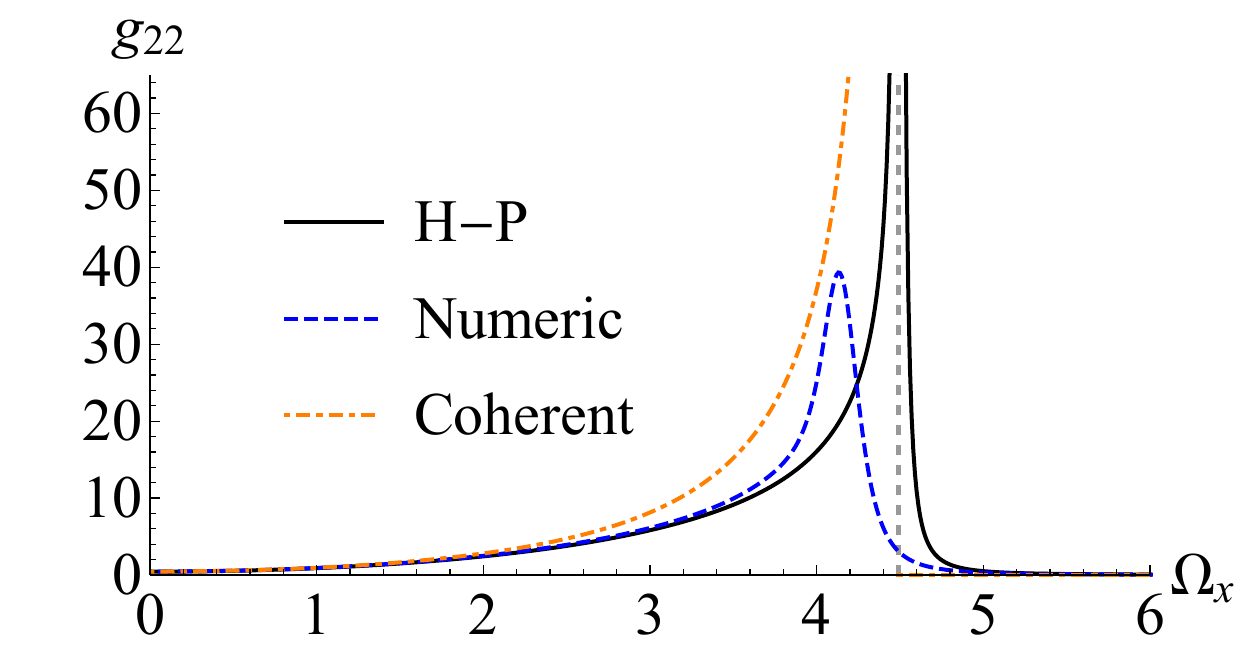} &
\includegraphics[width= 0.49 \columnwidth]{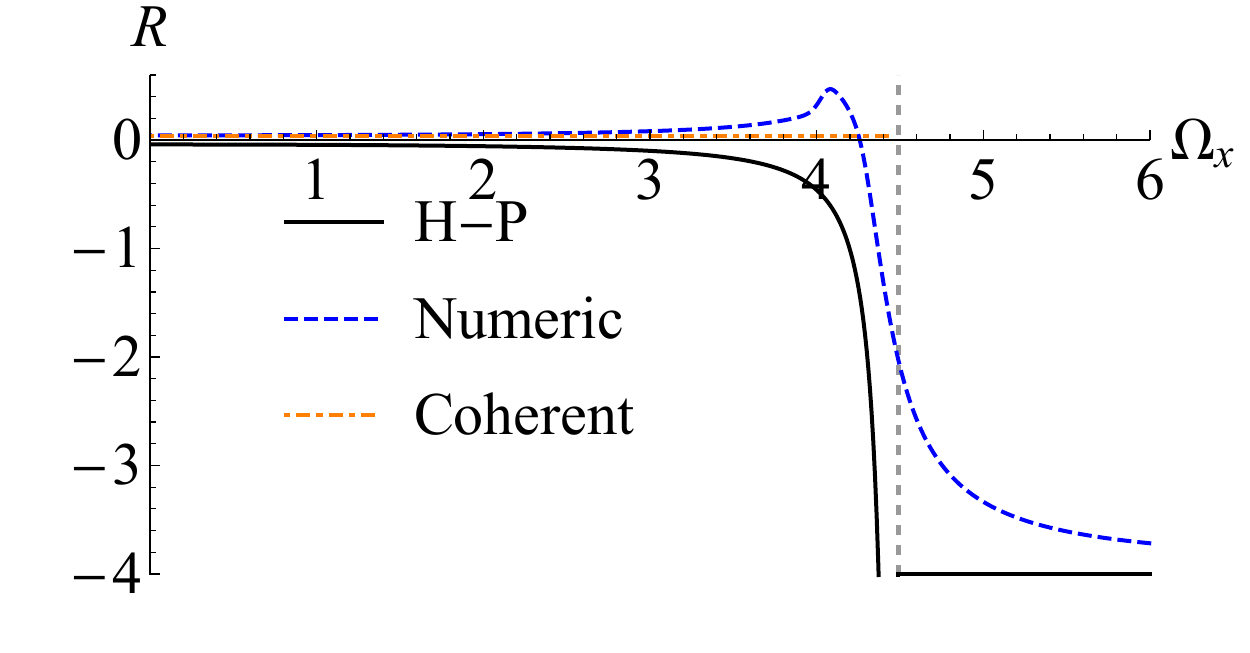} \\
(c) & (d)
\end{tabular} 
\caption{Comparison of the QMT and $R$ obtained with the truncated Holstein-Primakoff approximation (solid black), coherent states (dot-dashed orange), and exact diagonalization (dashed blue). We fixed $j=96$ and $\xi_y=2.3$.}
\label{Fig:gCoh}
\end{figure}

With this at hand, we find the metric tensor, its determinant, and the Berry curvature for both phases.\\
-  for $\xi_y>\frac{\sqrt{1+\Omega_x^2}}{2}$\
\begin{align}
F^{(z)}_{12}&=-\frac{j}{2}\,\frac{1}{\xi_y^2 \sqrt{4\xi_y^2-\Omega_x^2-1}}, \nonumber\\
g^{(z)}_{11}&=\frac{j}{2}\,\frac{4\xi_y^2-1}{4\xi_y^2 (4\xi_y^2-\Omega_x^2-1)}, \nonumber\\
g^{(z)}_{12}&=-\frac{j}{2}\,\frac{\Omega_x}{\xi_y (4\xi_y^2-\Omega_x^2-1)}, \nonumber\\
g^{(z)}_{22}&=\frac{j}{2}\,\frac{\Omega_x^2+1}{\xi_y^2 (4\xi_y^2-\Omega_x^2-1)}, \nonumber\\
g^{(z)}&=\frac{j^2}{16\xi_y^4(4\xi_y^2-\Omega_x^2-1)}.
\end{align}
- for $\xi_y\leq\frac{\sqrt{1+\Omega_x^2}}{2}$
\begin{align}
F^{(z)}_{12}&=0, \nonumber\\
g^{(z)}_{11}&=\frac{j}{2}\,\frac{1}{(\Omega_x^2+1)^2}, \nonumber\\
g^{(z)}_{12}&=0, \nonumber\\
g^{(z)}_{22}&=0, \nonumber\\
g^{(z)}&=0.
\end{align}
The scalar curvature is only defined above the separatrix (since the metric is invertible in that region), and is given by
\begin{equation}\label{scalarCoh}
R=4/j.
\end{equation}
In Fig.~\ref{Fig:gCoh}, we plot the QMT and its scalar curvature obtained through the Bloch coherent states and compare the results with those coming from the truncated Holstein-Primakoff transformation, as well as those coming from the exact diagonalization of the Hamiltonian.

%
%
%

\end{document}